\documentclass[a4paper,oneside]{amsart}

\usepackage{amsmath,amssymb,epsf}
\usepackage{amsfonts,amsbsy,amscd,stmaryrd}
\usepackage{latexsym,amsopn}
\usepackage{amsthm}
\usepackage{geometry}
\usepackage{setspace}
\usepackage{graphicx}
\usepackage{color}
\definecolor{Blue}{rgb}{0.,0.,1.}
\definecolor{Red}{rgb}{1.,0.,0.}

\newcounter{smallarabics}
\newenvironment{arabicenumerate}
{\begin{list}{{\normalfont\textrm{(\arabic{smallarabics})}}}
  {\usecounter{smallarabics}\setlength{\itemindent}{0cm}
   \setlength{\leftmargin}{5ex}\setlength{\labelwidth}{4ex}
   \setlength{\topsep}{0.75\parsep}\setlength{\partopsep}{0ex}
   \setlength{\itemsep}{0ex}}}
{\end{list}}

\newcounter{smallroman}

\newenvironment{nouppercase}{%
  \renewcommand{\uppercasenonmath}[1]{}}{}

\newcommand{\ben}{\begin{arabicenumerate}}  
\newcommand{\een}{\end{arabicenumerate}}

\def\init{\setcounter{equation}{0}}

\newtheorem{theoreme}{Theorem }[section]
\newtheorem{assumption}{Hypothesis}[section]
\newtheorem{proposition}[theoreme]{Proposition}
\newtheorem{lemma}[theoreme]{Lemma}
\newtheorem{definition}[theoreme]{Definition}
\newtheorem{corollary}[theoreme]{Corollary}
\newtheorem{remark}[theoreme]{Remark}
\newtheorem{example}[theoreme]{Example}
\newcommand{\beq}{\begin{equation}}
\newcommand{\eeq}{\end{equation}}

\newcommand{\bex}{\begin{example}}
\newcommand{\eex}{\end{example}}
\def\bel{\begin{lemma}}
\def\eel{\end{lemma}}
\def\bet{\begin{theoreme}}
\def\eet{\end{theoreme}}
\def\bed{\begin{definition}}
\def\eed{\end{definition}}
\def\ber{\begin{remark}}
\def\eer{\end{remark}}

\def\rr{{\mathbb R}}
\def\zz{{\mathbb Z}}
\def\cc{{\mathbb C}}
\def\nn{{\mathbb N}}

\def\part{{\rm par}}

\def\wlim{{\rm w-}\lim}

\def\bar{\overline}

\def\cinf{C^\infty}
\def\c0inf{C_0^\infty}

\def\proof{
\noindent{\bf Proof.}\ \ }

\def\tg{{\rm tg}}

\usepackage{calrsfs}
\DeclareMathAlphabet{\pazocal}{OMS}{zplm}{m}{n}

\def\cV{{\pazocal V}}

\def\cN{{\pazocal N}}

\def\CCR{{\rm CCR}}

\def\wf{{\rm WF}}

\def\mod{{\rm mod}}

\def\i{{\rm i}}

\def\Dom{{\rm Dom}}

\def\Ker{{\rm Ker}}

\newcommand{\qeds}{\qed\medskip}

\def\Int{\rm Int}
\def \p{ \partial}

\def\12{\frac{1}{2}}
\def\14{\frac{1}{4}}
\def\x{\langle x \rangle}

\def\supp{{\rm supp}}

\def\e{{\rm e}}
\def\D{D}

\def\Ran{{\rm Ran}}

\def\bbbone{{\mathchoice {\rm 1\mskip-4mu l} {\rm 1\mskip-4mu l}
{\rm 1\mskip-4.5mu l} {\rm 1\mskip-5mu l}}}
\newcommand{\one}{\boldsymbol{1}}
\def\cH{{\pazocal H}}

\def\coinf{C_0^\infty}

\def\Int{\rm Int}

\def\12{\frac{1}{2}}
\def\x{\langle x \rangle}

\def\supp{{\rm supp}}

\def\e{{\rm e}}

\def\Ran{{\rm Ran}}

\def\Diff{{\rm Diff}}

\def\bep{\begin{proposition}}
\def\eep{\end{proposition}}

\def\Op{{\rm Op}^{\rm w}}

\newcommand{\mat}[4]{\left(\begin{array}{cc}#1 &#2  \\ #3 &#4 \end{array}\right)}
\newcommand{\lin}[2]{\left(\begin{array}{c}#1 \\#2\end{array}\right)}

\def\CARal{{\rm C\hskip 0.25 em \hbox{\raise 1.72 ex 
\hbox{$\scriptscriptstyle\rm al$}\kern -0.57 em A}R}}

\def\otimesal{\mathop{\hbox{\raise 1.5 ex
  \hbox{$\scriptscriptstyle\rm al$}
\kern -0.92 em \hbox{$\otimes$}}}}
\def\oplusal{\mathop{\hbox{\raise 1.5 ex
  \hbox{$\scriptscriptstyle\rm al$}
\kern -0.92 em \hbox{$\oplus$}}}}
\def\Gammal{\hbox{\raise 1.68 ex 
\hbox{$\scriptscriptstyle\rm al$}\kern -0.50 em $\Gamma$}}
\def\Bal{\hbox{\raise 1.68 ex 
\hbox{$\scriptscriptstyle\rm  al$}\kern -0.50 em $B$}}
\def\CARal{{\rm C\hskip 0.25 em \hbox{\raise 1.72 ex 
\hbox{$\scriptscriptstyle\rm al$}\kern -0.57 em A}R}}

\def\cE{\pazocal{E}}
\def\kE{\pazocal{E}}

\renewcommand{\Int}{\, \lrcorner \,}
\newcommand{\tnI}{\, \llcorner \,}

\DeclareMathAlphabet{\mathpzc}{OT1}{pzc}{m}{it}
\newcommand{\bra}{\langle} 
\newcommand{\ket}{\rangle}

\DeclareSymbolFont{boldoperators}{OT1}{cmr}{bx}{n}
\SetSymbolFont{boldoperators}{bold}{OT1}{cmr}{bx}{n}
\edef\wbar{\unexpanded{\protect\mathaccentV{bar}}\number\symboldoperators16}

\def\bA{\wbar{A}}
\def\bds{\wbar{d}^{\Sig}}
\def\bdelta{\wbar{\delta}}
\def\bdeltas{{\wbar{\delta}^{\Sig}}}
\def\bF{\wbar{F}}

\def\bds{\wbar{d}_{\Sig}}
\def\bdelta{\wbar{\delta}}
\def\bdeltas{{\wbar{\delta}_{\Sig}}}
\def\bF{\wbar{F}}
\def\bardel{\wbar{d}}
\def\bardels{{\bdeltas}}
\def\bards{\wbar{d}_{\Sig}}
\def\tf{\tilde{f}}
\def\tg{\tilde{g}}

\def\kasd{K_{\Sig}^{\dag}}

\def\Sig{{\scriptscriptstyle\Sigma} }
\def\pback{{\scriptscriptstyle *}}

\def\sD{1}
\def\sDo{{}}
\def\sP{{\scriptscriptstyle P}}
\def\sQ{0}
\def\sR{0}
\def\sV{{\scriptscriptstyle V}}
\makeatletter
\newcommand*{\defeq}{\mathrel{\rlap{%
                     \raisebox{0.3ex}{$\m@th\cdot$}}%
                     \raisebox{-0.3ex}{$\m@th\cdot$}}%
                     =}
                     
\makeatother
\makeatletter
\newcommand*{\eqdef}{=\mathrel{\rlap{%
                     \raisebox{0.3ex}{$\m@th\cdot$}}%
                     \raisebox{-0.3ex}{$\m@th\cdot$}}%
                     }
\makeatother

\def\Texp{{\rm Texp}}

\newcommand{\pdo}[2]{ \Psi^{#1}(\Sigma; {#2})}
\newcommand{\pdoscal}[2]{\Psi_{\rm scal}^{#1}(\Sigma; {#2})}

\newcommand{\tpdo}[2]{C^{\infty}(\rr, \Psi^{#1}(\Sigma; {#2}))}

\newcommand{\tpdos}[3]{C^{\infty}(\rr, \Psi^{#1}(\Sigma; {#2}, {#3}))}
\def\sc{{\rm sc}}

\def\Proj{\Pi}

\def\gs{{\scriptscriptstyle >\!}}
\def\ls{{\scriptscriptstyle <\!}}
\DeclareMathAlphabet{\mathpzc}{OT1}{pzc}{m}{it}
\def\killing{\mathpzc{k}}
\def\gi{{\rm (g.i.)}}
\def\pos{{\rm (pos)}}
\def\musc{(\mu{\rm sc})}

\def\tarrow{\stackrel{{}_\sim}{\rightarrow}}
 \newcommand{\co}[1]{T^{*} #1}
\newcommand{\cooo}[1]{T^{*} #1\backslash \{0\}}
\def\Op{{\rm Op}}
\def\WF{{\rm WF}}
\def\fg{\mathfrak{g}}
\def\vara{{\rm\textsl{a}}}
\def\varb{{\rm\textsl{b}}}

\begin{document}
\title[Hadamard states for the Yang-Mills equation on curved spacetime]{\Large Hadamard states for the linearized Yang-Mills equation on curved spacetime}
\author{C. G\'erard }
\address{Universit\'e Paris-Sud XI, D\'epartement de Math\'ematiques, 91405 Orsay Cedex, France}
\email{christian.gerard@math.u-psud.fr}
\author{M. Wrochna}
\address{Universit\'e Joseph Fourier (Grenoble 1), Institut Fourier, UMR 5582 CNRS, BP 74 38402 Saint-Martin
d'H\`eres Cedex, France}
\email{michal.wrochna@ujf-grenoble.fr}
%\address{D\'epartement de Math\'ematiques, Universit\'e de Paris XI, 91405 Orsay Cedex France}

\begin{abstract}We construct Hadamard states for the Yang-Mills equation linearized around a smooth, space-compact background solution. We assume the spacetime is globally hyperbolic and its Cauchy surface is compact or equal $\rr^d$. 

We first consider the case when the spacetime is ultra-static, but the background solution depends on time. By methods of pseudodifferential calculus we construct a parametrix for the associated vectorial Klein-Gordon equation. We then obtain Hadamard two-point functions in the gauge theory, acting on Cauchy data. A key role is played by classes of pseudodifferential operators that contain microlocal or spectral type low-energy cutoffs.

The general problem is reduced to the ultra-static spacetime case using an extension of the deformation argument of Fulling, Narcowich and Wald.

As an aside, we derive a correspondence between Hadamard states and parametrices for the Cauchy problem in ordinary quantum field theory.
\end{abstract}

\begin{nouppercase}
\maketitle
\end{nouppercase}

\section{Introduction}\init\label{sec:intro}\init

The construction of a sufficiently explicit parametrix for the Klein-Gordon is essential in Quantum Field Theory on curved spacetime, where two-point functions of physically admissible states (\emph{Hadamard states}) are required to be distributions  with a specified wave front set. By using methods of pseudodifferential calculus it is possible to control at the same time the propagation of singularities and the additional properties of the parametrix, which are needed to treat physical conditions such as positivity (or purity) of states. As shown in the scalar case in \cite{junker,GW} for a large class of spacetimes, this allows to construct a large class of Hadamard states.
% and to parametrize them by symplectic transformations that preserve the wave front set.

The generalization to gauge theories poses difficulties which are due to two main obstacles. 

First of all, the equations of motions are given by a \emph{non-hyperbolic} differential operator $P$. This is usually coped with by identifying the space of solutions of $P$ with a quotient $\cV_\sP$ of subspaces of solutions of some hyperbolic operator $D_1$. Although one is essentially reduced to constructing two-point functions for $D_1$, one has to make sure that their restriction to $\cV_\sP$ is well defined. This entails a compatibility condition that will be termed \emph{gauge-invariance}.

Secondly, the hyperbolic operator $D_1$ is formally self-adjoint w.r.t. a hermitian product which is typically \emph{non-positive} on fibers. This results in a conflict between the Hadamard condition and positivity of states for $D_1$. Although one can still expect positivity to hold on the subspace $\cV_\sP$, it is not obvious  how this can be controlled.

An additional difficulty are {\em infrared problems}, which are inherent to any massless theory, but  have also their special incarnations in the context of gauge-invariance and positivity on $\cV_\sP$.

In the present paper we study those issues in the case of the Yang-Mills equation, linearized around a (possibly non-vanishing) background solution $\wbar{A}$.

\subsection*{Framework for gauge theories} We work (when possible) in the abstract framework for gauge theories proposed recently by Hack \& Schenkel \cite{HS}. More precisely, we consider its simplified version, in which the classical theory is determined by:
\begin{enumerate}
\item two vector bundles $V_0$, $V_1$ over a globally hyperbolic manifold $(M,g)$, both equipped with a hermitian structure,
\item a formally self-adjoint operator $P\in\Diff(M;V_1)$, which accounts for the equations of motion,
\item a non-zero operator $K\in\Diff(M;V_0,V_1)$ s.t. $PK=0$, which accounts for gauge transformations $u\to u+Kf$.
\end{enumerate}
We then assume $D_1\defeq P+ K K^*$ is hyperbolic and define the physical space by identifying solutions of $P$ with those solutions of $D_1$ which satisfy the additional constraint $K^*u=0$ (cf. Sect. \ref{sec:classical} for precise definitions). The latter is often called \emph{subsidiary condition} in the physics literature, we will thus term this approach the \emph{subsidiary condition framework}\footnote{Because we are working in a purely algebraic setting, the terminology is rather ambiguous. We refer the interested reader to \cite{derezinski} for a review on the flat case that explains the terminology used in the physics literature.}. The version we consider applies to the Maxwell and Yang-Mills equations, $K$ being then the covariant differential $\wbar{d}$ (note however that for other gauge theories one would have to use the more extended version from \cite{HS}).

\subsection*{Hadamard two-point functions} In our framework, a pair of operators $\lambda^\pm_1:\Gamma_{\rm c}(M;V_1)\to\Gamma(M;V_1)$ induces  two-point functions\footnote{We work with complex fields rather than with real ones, therefore it is natural to speak of a \emph{pair} of two-point functions, cf. \cite{hollands,GW,wrothesis}. It should be noted that the real and complex approaches are equivalent, see for instance \cite{GW} for the bosonic case.} of a Hadamard state on the phase space of $P$ if it satisfies
\beq\label{eq:pseucov}
D_1\lambda^\pm_1=\lambda^\pm_1 D_1 =0, \quad \lambda^+_1-\lambda^-_1=\i G_1,
\eeq
where $G_1$ is the causal propagator of $D_1$ and if moreover:
\[
\begin{aligned}
\musc &\quad  \wf'(\lambda^\pm_1)\subset\cN^\pm\times\cN^\pm,\\
\gi & \quad(\lambda^\pm_1)^*=\lambda^\pm_1 \mbox{ \ and \ } \lambda^\pm_1:\Ran\,K\to\Ran\,K,\\
\pos & \quad  \lambda^\pm_1  \geq 0 \mbox{ \ on \ } \Ker\,K^*.
\end{aligned}
\]
Condition $\musc$ is just the same as the Hadamard condition in ordinary (i.e., hyperbolic) field theory. What differs is the non-trivial requirement of gauge-invariance $\gi$. Moreover, positivity $\pos$ is no longer required to hold on all test sections, but on a specified subspace instead.

%Pairs of operators $\lambda^\pm_1:\cD(M;V_1)\to\cD'(M;V_1)$ that satisfy (\ref{eq:pseucov}) will be called \emph{pseudo-covariances} in order to avoid confusion with actual two-point functions, which are a priori defined only on the phase space of $P$ (and satisfy positivity).

\subsection*{Main results} Our main result is the construction of Hadamard states  for the Yang-Mills equation linearized around a smooth background solution $\wbar{A}$, under various assumptions on $\wbar{A}$ and the spacetime $(M,g)$.  
Let us first formulate some hypotheses.
\subsubsection{Spacetimes}
\begin{assumption}\label{as:spacetime}
 $(M, g)$ is a globally hyperbolic spacetime with a Cauchy surface $\Sigma$ diffeomorphic either to $\rr^{d}$ for $d\geq 3$,  or to a compact, parallelizable manifold.
 \end{assumption}
 \begin{assumption}\label{as:metric}
If $\Sigma= \rr^{d}$, $h_{ij}(x)dx^{i}dx^{j}$ is  a smooth Riemannian metric on $\Sigma$ such that:
  \[
c^{-1}\one\leq [h_{ij}(x)]\leq  c\one, \  c>0, \ |\p^{\alpha}_{x}h_{ij}(x)|\leq C_{\alpha}, \ \forall \alpha\in \nn^{d}, \ x\in \rr^{d}.
\]
\end{assumption}
\subsubsection{Background Yang Mills connections}
\begin{assumption}\label{as:group}
 $G$ is a linear Lie group with compact Lie algebra $\mathfrak{g}$.   \end{assumption}
 We consider the trivial principal bundle   $(M\times G, M, G)$ and the associated trivial vector bundle $(M\times \mathfrak{g}, M, \mathfrak{g})$.
Using the horizontal connection on  $M\times G$, a connection on $M\times\mathfrak{g}$ can be identified with a section $\wbar{A}$ of the bundle $T^{*}M\times \mathfrak{g}$, i.e. with a Lie algebra valued $1-$form $\wbar{A}$.

\begin{assumption}\label{as:background}
 If $\Sigma= \rr^{d}$,  $\wbar{A}$ is  a smooth global solution of the non-linear Yang-Mills equation (\ref{eq:nlYM}) on $\rr_{t}\times \Sigma$ such that
 \[
 \begin{array}{rl}
i)& \wbar{A}\mbox{ is in the {\em temporal gauge} i.e. }\wbar{A}_{t}=0,\\[2mm]
 ii)&|\p^{\alpha}_{x}\wbar{A}_{\Sig}(t,x)|\leq C_{\alpha}, \mbox{ locally uniformly in }t,\\[2mm]
 iii)&|\p^{\alpha}_{x}{\wbar{\delta}_{\Sig}}\wbar{F}_{\Sig}(0, x)|\leq C_{\alpha}\langle x\rangle^{-1}, \ |\p^{\alpha}_{x} \wbar{F}_{t}(0,x)|\leq C_{\alpha}\langle x\rangle^{-2}\ 
 \alpha\in \nn^{d}, \ x\in \rr^{d},
\end{array}
\]
where   the components $\wbar{A}_{\Sig}$, $\wbar{A}_{t}$,  $\wbar{F}_{\Sig}$, $\wbar{F}_{t}$ of $\wbar{A}$ and the curvature $\wbar{F}= \bardel \wbar{A}$ are defined in \ref{sss:ident}.
\end{assumption}

Our first theorem deals with ultra-static background metrics and background solutions $\wbar{A}$ satisfying conditions near infinity in the case $\Sigma= \rr^{d}$.

\begin{theoreme}\label{maintheo2}
Let us assume Hypotheses \ref{as:spacetime}, \ref{as:group} and  if $\Sigma= \rr^{d}$  also Hypotheses \ref{as:metric}, \ref{as:background}. Let 
$g= - dt^{2}+ h_{ij}(x)dx^{i}dx^{j}$ on $M =  \rr_{t}\times\Sigma$.  
Then there exist quasi-free Hadamard states  for the linearized Yang Mills equation on $(M, g)$ around $\wbar{A}$.
\end{theoreme}

Our next theorem covers the  general case, with a {\em space-compact} background solution $\wbar{A}$. We will deduce it from Thm. \ref{maintheo2} by a deformation argument explained in  Subsect. \ref{ss:fnw}. This deformation relies on the global solvability of the {\em non-linear} Yang-Mills equation, which requires that $\dim M\leq 4$.
\begin{theoreme}\label{maintheo1}
Let us assume Hypotheses \ref{as:spacetime}, \ref{as:group} and $\dim M\leq 4$. 
 
 Let $\wbar{A}\in \cE^{1}_{\rm sc}(M)\otimes \mathfrak{g }$ a smooth, space-compact solution of the non-linear Yang-Mills equation  (\ref{eq:nlYM})  on $(M, g)$. Then there exist quasi-free Hadamard states  for the linearized Yang Mills equation around $\wbar{A}$.
 \end{theoreme}

Let us emphasize that the case $\wbar{A}\neq 0$ differs substantially from the case of a vanishing background solution (or of an abelian gauge group), as was so far assumed in other works on Hadamard states. Indeed, if $\wbar{A}\neq0$ then the deformation argument cannot be used to reduce the problem to the situation when $(M,g)$ is ultra-static \emph{and} the coefficients of $D_1$, $P$ do not depend on time. 

As further explained in Subsect. \ref{ss:fnw}, the difficulty comes from the fact that the background $\wbar{A}$ must be a solution of the non-linear Yang-Mills equation and therefore cannot be arbitrarily deformed.  This is our main motivation for considering the case of a \emph{time-dependent} Klein-Gordon operator $D_1$ on an ultra-static spacetime.

\subsection*{Known results}
In the literature, other constructions were already considered in the special case of the Maxwell equations or Yang-Mills linearized around $\wbar{A}=0$. 

In these cases the deformation argument yields a time-independent problem, and it is possible to use arguments from spectral theory at least if the Cauchy surface $\Sigma$ has special properties that make the infrared problems less serious. For the Maxwell equations, this strategy was employed in \cite{FP} for $\Sigma$ compact with vanishing first cohomology group (extending some earlier results of \cite{furlani0}), and in \cite{FS} for $\Sigma$ subject to an `absence of zero resonances' condition for the Laplace-Beltrami operator on $1$-forms. This condition appears to be more general but similar in nature to our assumptions, as it involves the behaviour of $\Sigma$ at infinity\footnote{  The two methods are difficult to compare: in \cite{FS} the infrared problem amounts to an obstruction to invertibility of the Laplacian, whereas in our approach the Laplacian is effectively replaced by an invertible operator and an infrared problem occurs in attempts of restoring gauge-invariance.}. The Yang-Mills equation with $\wbar{A}=0$ was considered in \cite{hollands2} (in the BRST framework) for $\Sigma$ compact with vanishing first cohomology group.

Another approach was studied in \cite{DS} on asymptotically flat spacetimes, where the use of spectral theory arguments is made possible by considering a characteristic Cauchy problem.

\subsection*{Summary of the construction}Let us summarize the strategy adopted in the paper. 

The construction of the parametrix by pseudodifferential calculus is a generalization of the arguments used in \cite{GW} in the scalar case. As an output, we obtain Hadamard two-point functions $\lambda^\pm_1$ that satisfy $\gi$ only `modulo smooth terms'. Moreover, they are positive on some subspace (the space of `purely spatial' $1$-forms on $M$) that needs not to coincide with $\Ker K^*$.

To solve this,  we work with quantities on a fixed  Cauchy surface $\Sigma$. We define a Cauchy-surface analogue $K_\Sig$ of the operator $K$, and deduce that the Cauchy-surface version of the phase space for $P$ can be expressed as a quotient $\Ker K^\dag_\Sig / \Ran K_\Sig$ (where $^\dag$ is the {\em symplectic adjoint}, defined in (\ref{eq:defkdag})).

Next, we argue that gauge-invariance can be obtained by modifying $\lambda^\pm_1$ with the help of a projection $\Proj$ that maps to a complement of $\Ran K_\Sig$. The whole task that remains then is to show that:
\begin{itemize}
\item The range of $\Proj$ is a space on which $\lambda^\pm_1$ is positive (after restricting to the phase space of $P$).
\item The modification of $\lambda_{1}^{\pm}$ does not affect $\musc$.
\end{itemize} 
Both tasks are unfortunately made difficult by infrared problems. For example, the projection $\Proj$ can contain terms such as $(\wbar{\delta}_\Sig\wbar{d}_\Sig)^{-1}\wbar{\delta}_\Sig$ (see Subsect.\ref{ss:refproj}), whose definition is already ambiguous, not to mention boundedness between Sobolev spaces of appropriate order.

One way we deal with such problems is to use a  {\em Hardy's inequality} on $\rr^d$ for the Hodge Laplacian on $0$-forms. 

The essential novelty is the systematic use of two classes of pseudodifferential operators
\[
\Psi^p_{\rm as}(\Sigma;V_\alpha,V_\beta), \quad \Psi^p_{\rm reg}(\Sigma;V_\alpha,V_\beta),
\]
that contain infrared regularizations of different type --- either a simple `microlocal' cutoff in the low frequencies (for the $\Psi^p_{\rm as}$ class), or in addition to that a `spectral' cutoff  (the $\Psi^p_{\rm reg}$ class), defined using (functions of) some elliptic self-adjoint operators. Moreover, the norm of the regularization is controlled by a parameter $R$ that can be chosen arbitrarily large. This allows to obtain \emph{exact inverses} in situations where standard pseudodifferential calculus gives only inverses modulo regularizing remainders. Using this method, we first construct a reference projection, establish its boundedness as an operator between appropriate (weighted) Sobolev spaces, and then perturb it in order to finally get the positivity.

\subsection*{Auxiliary results} Beside of what is of direct interest for Maxwell and Yang-Mills fields, let us mention some auxiliary results obtained in the present work.

First of all, in the context of ordinary field theory (without gauge), we derive a direct relation between (bosonic) Hadamard two-point functions and parametrices that satisfy certain special properties (Subsect. \ref{ss:corr}). This allows to generalize and  simplify results in \cite{GW} that tell how to obtain more Hadamard states out of an already given one.

We also derive a number of results for the classical Yang-Mills theory linearized around a non-vanishing background, for instance our formula for the phase space of $P$ in terms of Cauchy data appears to be new (see  \ref{ss:phhypsurf}).

\subsection*{Outlook} An evident limitation of our method is that we have to assume that the Cauchy surface $\Sigma$ is either compact or equal $\rr^d$, as the construction is based on standard pseudodifferential operator classes. We also use Hardy's inequality in the case $\Sigma=\rr^d$. We expect, however, that it would be possible to extend our results to other Cauchy surfaces by considering extensions of the standard pseudodifferential calculus on classes of non-compact manifolds on which a generalized form of Hardy's inequality still holds true.

Let us also stress that all our results are formulated in the subsidiary condition framework to gauge theories. Especially for applications in perturbative Quantum Field Theory, a different approach --- the {\em BRST framework}, is commonly believed to be more efficient \cite{hollands2}. We do not consider it here, although it seems plausible that one can transport Hadamard states from one framework to the other, as illustrated in \cite[Appendix B]{FS}. Another assumption that we implicitly make is that $\bA$ is a connection on a \emph{trivial} principal bundle and one can ask whether the methods of this paper can be applied to the non-trivial case. We plan to address these issues in a future work. 

\subsection*{Structure of the paper} The paper is structured as follows.

Sect. \ref{sec:classical} concerns the classical theory. We first recall well-known facts on ordinary field theories, then in Subsect. \ref{ss:subsidiary} review gauge theories on curved spacetime in the (simplified) subsidiary condition framework. We introduce the corresponding quantities on a Cauchy surface in \ref{ss:phhypsurf} and then in Subsect. \ref{lnym} we show how the linearized Yang-Mills equation fits into this framework.

Sect. \ref{secmain} discusses Hadamard states  for both ordinary field theories and  for gauge theories in the subsidiary framework in general terms. We introduce in Subsect. \ref{ss:haddef} the definition of Hadamard states that we use for ordinary field theories.  We then set up in Subsect. \ref{ss:corr} a correspondence between Hadamard states and parametrices subject to special conditions. Next, we discuss in Subsect. \ref{ss:psecauchy}  two-point functions  in gauge theory, and formulate the conditions $\musc$, $\gi$, $\pos$ and the Cauchy surface analogues of the latter two. In the same subsection we outline our method to cope with $\gi$ and $\pos$, and discuss the main technical obstructions. The section ends with an extended version of the Fulling, Narcowich \& Wald argument in Subsect. \ref{ss:fnw}, which allows us to reduce the construction of Hadamard states for the Yang-Mills equation to a situation where the spacetime is static, but the equations of motions still depend on the time coordinate.

Sect. \ref{sec1} reviews  the vector and scalar Klein-Gordon equations on ultra-static spacetimes.

In Sect. \ref{sec2} we give a detailed construction of the parametrix for the vector Klein-Gordon equations considered in Sect. \ref{sec1}, generalizing  results from \cite{GW}.

In Sect. \ref{sec3}, using the results of Sect. \ref{sec2} we obtain two-point functions for the vector and scalar Klein-Gordon  equations on an ultra-static spacetime and study their properties. At this point, the properties $\gi$ and $\pos$ are not satisfied and only their weaker versions are available.   

As a byproduct of our constructions, we prove that for vector Klein-Gordon equations, where the natural hermitian product is not positive-definite on the fibers, there {\em does not exist Hadamard states}, but only Hadamard {\em pseudo-states}.

In Sect. \ref{sec5}, we study the relationship between the two-point functions constructed in Sect. \ref{sec3} in the vector and scalar case. 
In particular  Thm. \ref{th:diff-cov} will be important later on. 

In Sect. \ref{sec4} we prove Thm. \ref{maintheo2} by the method described  in Subsect. \ref{ss:psecauchy}. This is the most technical part of the paper.

In Appendix \ref{sec1opd} we introduce the necessary background on pseudodifferential calculus. It includes amongst other a version of Egorov's theorem adapted to the case of matrices of pseudodifferential operators.

Appendix \ref{appB} gathers independent results, used in several parts of the main text. In \ref{ss:hardy} we prove a version of Hardy's inequality adapted to our applications for the Yang-Mills equation. 
In \ref{ss:tempgauge} we recall the transition to the temporal gauge for the non-linear Yang-Mills equation. In \ref{ss:constraints} we discuss the constraint equations on Cauchy data for the non-linear Yang-Mills equation and show how to construct examples of solutions satisfying our hypotheses. In \ref{ss:global-existence} we sketch the proof of Prop. \ref{prop:idiotic}.
\section{Classical gauge field theory}\label{sec:classical}\init

\subsection{Notation}
Let $V$ be a finite rank vector bundle  over a smooth manifold $M$. We denote by $\Gamma(M; V)$, resp. $\Gamma_{\rm c}(M; V)$, $\Gamma_{\rm sc}(M; V)$ the space of smooth, resp. smooth with compact, space-compact support sections of $V$, the later notation requiring that $M$ is equipped with some causal structure.

If $V_{1}$, $V_{2}$ are two vector bundles, the set of differential operators (of order $m$) $\Gamma(M;V_{1})\to\Gamma(M;V_{2})$ is denoted  $\Diff(M;V_{1},V_{2})$ ($\Diff^m(M;V_{1},V_{2})$), we also use the notation $\Diff(M;V)=\Diff(M;V,V)$. 

By a \emph{bundle with hermitian structure} we will mean a vector bundle $V$ equipped with a fiber wise non-degenerate hermitian form (in the literature the name `hermitian bundle' is usually  reserved for positive definite hermitian structures). 

Suppose now that $(M,g)$ is a pseudo-Riemannian oriented manifold. If $V$ is a bundle on $M$ with hermitian structure, we denote $V^*$ the anti-dual bundle. The hermitian structure on $V$ and the volume form on $M$ allow to embed $\Gamma(M; V)$ into $\Gamma_{\rm c}'(M; V)$, using the non-degenerate hermitian form on $\Gamma_{\rm c}(M; V)$
\begin{equation}
\label{defdepscalM}
(u|v)_{V}\defeq  \int_{M}(u(x)| v(x))_{V}d{\rm Vol}_{g}, \ u, v\in \Gamma_{\rm c}(M; V).
\end{equation}Therefore, we have a well-defined notion of the formal adjoint $A^*:\Gamma_{\rm c}(M;W)\to\Gamma(M;V)$ of an operator $A:\Gamma_{\rm c}(M;V)\to\Gamma(M;W)$.

If $E,F$ are vector spaces, the space of linear operators is denoted $L(E,F)$. If $E,F$ are additionally endowed with some topology, we write $A:E\to F$ if $A\in L(E,F)$ is continuous. 

To distinguish between the same operator $A$ acting on different spaces of functions and distributions, for instance $A:\Gamma_{\rm c}(M;V)\to\Gamma_{\rm c}'(M;W)$ and $A:\Gamma(M;V)\to\Gamma(M;W)$, we use the notation $A|_{\Gamma_{\rm c}}$ and $A|_{\Gamma}$.

\subsection{Quotient spaces}

In the sequel we will frequently encounter operators and sesquilinear forms on quotients of linear spaces, we recall thus the relevant basic facts.

\subsubsection{Operators on quotient spaces}

Let $F_i\subset E_i$, $i=1,2$ be vector spaces and let $A\in L(E_1, E_2)$. Then the induced map
\[
[A]\in L( E_1/F_1, E_2/F_2),
\]
defined in the obvious way, is
\begin{itemize}
\item well-defined if $A E_1\subset E_2$ and $A F_1\subset F_2$;
\item injective iff $A^{-1}F_2=F_1$;
\item surjective iff $E_2=A E_1+F_2$. 
\end{itemize} 

\subsubsection{Sesquilinear forms on quotients} Let now $E\subset F$ be vector spaces and let $C\in L( E, E^*)$, where $E^*$ is the anti-dual space of $E$. Then the induced map
\[
[C]\in L( E/F,(E/F)^*),
\]
defined as before, is
\begin{itemize}
\item well-defined if $CE\subset F^{\circ}$ (where $F^\circ\subset E^*$ denotes the annihilator of $F$) and $F\subset\Ker\, C$;
\item non-degenerate iff $F=\Ker\,C$. 
\end{itemize}

If $C$ is hermitian or anti-hermitian (which will usually be the case in our examples) then the condition $F\subset\Ker\, C$ implies the other one $CE\subset F^{\circ}$ (and vice versa).

\subsection{Ordinary classical field theory}\label{ssec:classical}
We recall now some standard results, see eg \cite{BGP, HS}.
Let $(M,g)$ be a globally hyperbolic spacetime (we use the convention $(-,+,\dots,+)$ for the Lorentzian signature). If $V$ is a vector bundle over $M$, we denote $\Gamma_{\rm sc}(M;V)$ the space of space-compact sections, i.e. sections in $\Gamma(M;V)$ such that their restriction to a Cauchy surface has compact support.

One says that $D\in\Diff(M;V)$ is \emph{Green hyperbolic} if $D$ and $D^*$ possess retarded and advanced propagators --- the ones for $D$ will be denoted respectively $G^{+}_\sDo$ and $G^-_\sDo$ (for the definition, see \cite{BGP}). The \emph{causal propagator} (or Pauli-Jordan commutator function) of $D$ is then by definition $G_\sDo\defeq G^+_\sDo-G^-_\sDo$. Normally hyperbolic and prenormally hyperbolic operators (defined below) are  Green hyperbolic.

\begin{definition}\label{def:prenorm}
\ben
\item An operator $D\in\Diff(M;V)$  is {\em normally hyperbolic} if its principal symbol equals $-\xi_\mu \xi^\mu \one_V$.
\item An operator $D\in\Diff(M;V)$ is {\em prenormally hyperbolic} if there exists $\widetilde D\in\Diff(M;V)$ s.t. $D\widetilde D$ is normally hyperbolic.
\een
\end{definition}
This terminology is slightly more general than the one used in e.g. \cite{muehlhoff}, cf. \cite{static,wrothesis} for examples.

\begin{proposition}\label{prop:prenormal}If $D,\widetilde D\in\Diff(M;V)$ are such that $D\widetilde D$ is Green hyperbolic then $D$ is Green hyperbolic and their retarded/advanced propagators $G^{\pm}$ and $G^\pm_{{\scriptscriptstyle D} {\scriptscriptstyle\widetilde{D}}}$ are related by
\[
G^{\pm}_\sDo = \widetilde D G^\pm_{{\scriptscriptstyle D} {\scriptscriptstyle\widetilde{D}}}.
\]
\end{proposition}

The proof of Prop. \ref{prop:prenormal} is a straightforward generalization of the arguments of Dimock \cite{dimock,muehlhoff}.

Before discussing gauge theories, let us recall the basic data that define an ordinary classical field theory (i.e., with no gauge freedom built in) on a globally hyperbolic manifold $(M,g)$.

\begin{assumption}\label{as:ord}Suppose that we are given:
\begin{enumerate}
\item a bundle $V$ over $M$ with hermitian structure;
\item a Green hyperbolic operator $D\in\Diff(M;V)$ s.t. $D^*=D$.
\end{enumerate}
\end{assumption}

\begin{proposition}\label{prop:basic}As a consequence of Hypothesis \ref{as:ord},
\begin{enumerate}
\item\label{prop:basic2} the induced map 
\[
[G_\sDo]:\,\frac{\Gamma_{\rm c}(M;V)}{\Ran\,D|_{\Gamma_{\rm c}}}\longrightarrow\Ker\,D|_{\Gamma_{\rm sc}}
\]
is well defined and bijective.
\item $(G^{\pm}_{\sDo})^*=G^{\mp}_{\sDo}$ and consequently $G^*_{\sDo}=-G_{\sDo}$;
\end{enumerate}
\end{proposition}

To fix some terminology, by a phase space we mean a pair $(\cV,q)$ consisting of a complex vector space $\cV$ and a sesquilinear form $q$ on $\cV$. Actual physical meaning can be associated to $(\cV,q)$ if $q$ is hermitian.
The classical phase space associated to $D$ is $(\cV,q)$, where
\beq\label{defo}
\cV\defeq \frac{\Gamma_{\rm c}(M;V)}{\Ran\,D|_{\Gamma_{\rm c}}}, \quad \bar u \,q v\defeq \i(u| [G_\sDo] v)_{\sV}.
\eeq
By (2) of Prop. \ref{prop:basic} the sesquilinear form $q$ is hermitian, and it is not difficult to show that it is non-degenerate. As a rule, we will work with hermitian forms rather than with real symplectic ones, but it should be kept in mind that the two approaches are equivalent.

\subsubsection{Phase space on Cauchy surface}\label{ssec:cauchy}

Let us fix a Cauchy surface $\Sigma$ of $(M,g)$. Consider a Green hyperbolic operator $D\in\Diff^m(M;V)$. Let $V_\rho$ be a vector bundle over $\Sigma$ and $\rho_\sDo :\Gamma_{\rm sc}(M;V)\to\Gamma_{\rm c}(\Sigma;V_\rho)$ an operator which is the composition of a differential operator of order $\leq m$ with the pullback $\imath^\pback$ of the embedding $\imath:\Sigma\hookrightarrow M$.

We equip $V_{\rho}$ with a hermitian structure $(\cdot | \cdot)_{V_{\rho}}$, which extends to $\Gamma_{\rm c}(\Sigma; V_{\rho})$  as in (\ref{defdepscalM}), using the volume form on $\Sigma$ induced by $g$. It is convenient to assume that this hermitian structure is {\em positive definite}.  The adjoint map:
\[
\rho^{*}: \Gamma_{\rm c}(\Sigma; V_{\rho})\to \Gamma'(M; V)
\]
is defined using the two hermitian structures $(\cdot | \cdot)_{V}$ and $(\cdot| \cdot)_{V_{\rho}}$.  

\begin{assumption}\label{cauchyhyp}Assume that for each initial datum $\varphi\in\Gamma_{\rm c}(\Sigma;V_\rho)$, the Cauchy problem
\beq\label{eq:cauchyD}
\begin{cases}Df=0, \quad f\in\Gamma_{\rm sc}(M;V)\\
\rho_\sDo f= \varphi,
\end{cases}
\eeq
has a unique solution.
\end{assumption}

In other words, the map $\rho_\sDo:\Ker\,D|_{\Gamma_{\rm sc}}\to\Gamma_{\rm c}(\Sigma;V_\rho)$ is a bijection. If $D$ satisfies Hypothesis \ref{cauchyhyp}, we will say that it is \emph{Cauchy hyperbolic} (for the map $\rho_\sDo$). It can be proved that if $D$ is Green hyperbolic then there exists $\rho$ s.t. $D$ is Cauchy hyperbolic\footnote{Of course one has to choose $\rho$ sensibly, cf. the example in \cite[Sec. 2.7]{BG}.}, cf. the reasoning in \cite[Sec. 4.3]{khavkine}.

By Hypothesis \ref{cauchyhyp}, assuming additionally $D=D^*$ and using (\ref{prop:basic2}) of Prop. \ref{prop:basic} we deduce that the phase space $(\cV_\sDo,q_\sDo)$ is isomorphic to $(\cV_{\sDo\Sig},q_{\sDo\Sig})$, defined in the following way:
\beq\label{eq:defdeqsigma}
\cV_{\sDo\Sig}\defeq\Gamma_{\rm c}(\Sigma;V_\rho), \quad \bar u \,q_{\sDo\Sig} v\defeq \i(u| G_{\sDo\Sig} v)_{V_\rho},
\eeq
where $G_{\sDo\Sig}$ is uniquely defined by 
\[
G_\sDo\eqdef (\rho_\sDo  G_\sDo )^* G_{\sDo\Sig} (\rho_\sDo  G_\sDo ).
\]
(Let us stress again that the stars refer to formal adjoints using the hermitian structures of $V$ and $V_\rho$, the latter can be chosen quite arbitrarily.) As a consequence of this definition,
\beq
\label{eq:idU}
\one=G_\sDo^*\rho_\sDo^* G_{\sDo\Sig}\rho_\sDo \mbox{\ \ on \ } \Ker\,D|_{\Gamma_{\rm sc}}.
\eeq
%As a consequence of this definition and $G^*_\sDo=-G_\sDo$, we get the identity $\one=-\rho_\sDo^* G_{\sDo\Sig} \rho_\sDo G_\sDo$. Taking the adjoint, this yields
%\beq\label{eq:idU}
%\one=-G_{\sDo}\rho_\sDo^* G_{\sDo\Sig} \rho_\sDo.
%\eeq 
This also implies $\rho_\sDo=\rho_\sDo G_\sDo^*\rho_\sDo^* G_{\sDo\Sig}\rho_\sDo$ on $\Ker\,D|_{\Gamma_{\rm sc}}$, hence
\beq\label{eq:idU2}
\one=\rho_\sDo G_{\sDo}^*\rho_\sDo^* G_{\sDo\Sig} \mbox{\ \ on \ } \Gamma_{\rm c}(\Sigma;V_\rho).
\eeq

It is useful to introduce the {\em Cauchy evolution operator}:
\beq
U_{\sDo}\defeq G_\sDo^* \rho_\sDo^* G_{\sDo\Sig}.
\eeq
By (\ref{eq:idU}) and (\ref{eq:idU2}), it satisfies $\rho_\sDo U_\sDo =\one$ and $U_\sDo \rho_\sDo = \one$ (on space-compact solutions of $D$). Moreover, since $G_\sDo^*=-G_\sDo$ we get $D U_\sDo=0$. Applying both sides of (\ref{eq:idU}) to $f$ we obtain a formula for the solution of the Cauchy problem (\ref{eq:cauchyD}).
\begin{proposition}Assume $D$ is Cauchy hyperbolic for $\rho_\sDo$ and $D=D^*$. Then the unique solution of the Cauchy problem (\ref{eq:cauchyD}) equals 
\[
f=U_\sDo \varphi = G_\sDo^* \rho_\sDo^* G_{\sDo\Sig}\varphi=-G_\sDo \rho_\sDo^* G_{\sDo\Sig}\varphi.
\] 
\end{proposition}

\subsection{Gauge theory in subsidiary condition formalism}\label{ss:subsidiary}

The following data is used to define a classical linearized gauge field theory on a globally hyperbolic manifold $(M,g)$. This is a special case of the setting proposed by Hack and Schenkel in \cite{HS}, well suited for the case of  Yang-Mills fields.

\begin{assumption}\label{as:subsidiary}Suppose that we are given:
\begin{enumerate}
\item bundles with hermitian structures $V_{0},V_{1}$ over $M$;
\item a formally self-adjoint operator $P\in\Diff(M;V_{1})$;
\item an operator $K\in\Diff(M;V_{0},V_{1})$, such that $K\neq0$ and 
\begin{enumerate}
\item $PK=0$,
\item $D_0\defeq K^* K\in\Diff(M;V_0)$ is Green hyperbolic,
\item $D_1\defeq P+K K^*\in\Diff(M;V_1)$ is Green hyperbolic.
\end{enumerate}
\end{enumerate}
\end{assumption}

The operator $P$ accounts for the equations of motion, linearized around a background solution. The operator $K$ defines the linear gauge transformation $f\mapsto f+K g$, and the condition $PK=0$ states that $P$ is invariant under this transformation, which entails that $P$ is not hyperbolic. Making use of the assumption on $D_0$, the non-hyperbolic equation $Pf=0$ can  be reduced by gauge transformations to the subspace $K^* f=0$ of solutions of  the hyperbolic problem $D_1 f=0$. The equation $K^*f=0$ is traditionally called \emph{subsidiary condition} and can be thought as a covariant fixing of gauge. 

{  The canonical example is the Maxwell theory, in which case $K$ is the differential $d$ acting on $0$-forms on $M$ and $P=\delta d$, where $\delta$ is the codifferential. The subsidiary condition $K^* f =0$ is then simply the Lorenz gauge. This example will be further discussed in Subsect. \ref{lnym} as a special case of Yang-Mills theory.}

 %The relation b) is used to prove that the subsidiary condition is compatible with the evolution.

Let us first observe that the differential operators from Hypothesis \ref{as:subsidiary} satisfy the algebraic relations
\[
K^* D_1=D_0 K^*, \quad D_1 K=K D_0. 
\]
These have the following consequences on the level of propagators and spaces of solutions, proved in \cite{HS}.

% (statements (\ref{prop:notbasic1})--(\ref{longpropi4}) are ).

\begin{proposition}\label{prop:notbasic}As a consequence of Hypothesis \ref{as:subsidiary},
\begin{enumerate}
\item\label{prop:notbasic1} $K^* G^{\pm}_{\sD}=G^{\pm}_\sQ K^*$ on $\Gamma_{\rm c}(M;V_1)$ and $K G^{\pm}_{\sQ}=G^{\pm}_\sD K$ on $\Gamma_{\rm c}(M;V_0)$;
\item For all $\psi\in\Gamma_{\rm sc}(M;V_1)$ there exists $h\in\Gamma_{\rm sc}(M;V_0)$ s.t. $\psi-Kh\in\Ker\,K^*|_{\Gamma_{\rm sc}}$. If moreover $\psi\in\Ker\,P|_{\Gamma_{\rm sc}}$ then $\psi-K h\in \Ker\,P|_{\Gamma_{\rm sc}} \cap \Ker\,K^*|_{\Gamma_{\rm sc}}$;
\item\label{prop:notbasic3}We have 
\[
\Ker\,P|_{\Gamma_{\rm sc}} \cap \Ker\,K^*|_{\Gamma_{\rm sc}}\subset G_\sD \Ker\,K^*|_{\Gamma_{\rm c}} + G_\sD \Ran\,K|_{\Gamma_{\rm c}};
\]
\item\label{longpropi4} $\Ran\,P|_{\Gamma_{\rm c}}=\Ker\,K^*|_{\Gamma_{\rm c}}\cap G_\sD^{-1}\Ran\,K|_{\Gamma_{\rm sc}}$.
%\item\label{longpropi5} $\Ran\,K|_{\cD}\cap\Ker\,K^*|_{}\subset\Ran\,P|_{\cD}$.
\end{enumerate}
\end{proposition}
%\proof (\ref{longpropi5}): By (\ref{longpropi4}) it suffices to prove that if $u=Kf$ and $K^*u=0$ then $G_\sD u\in\Ran\,K|_{\Gamma_{\rm sc}}$. We have $Qf=K^*K f=0$, hence $f=G_{\sQ} h$ for some $h\in\cD$. It follows that $G_\sD u = G_\sD K G_{\sQ} h= K G_{\sQ} G_{\sQ}h$.
%\qeds

Since the auxiliary operators $D_1,D_0$ are Green hyperbolic, we can associate to them phase spaces $(\cV_\sD,q_\sD)$, $(\cV_\sR,q_\sR)$ as in the previous subsection. 

In the `subsidiary condition' framework, the physical phase space associated to $P$, denoted $(\cV_\sP,q_\sP)$, is defined by
\[
\cV_\sP\defeq \frac{\Ker\,K^*|_{\Gamma_{\rm c}}}{\Ran\,P|_{\Gamma_{\rm c}}}, \quad \bar u \,q_\sP v\defeq \i(u| [G_\sD] v)_{V_1}.
\]
The first thing to check is that the propagator $G_\sD$ of $D_1$ induces a well-defined linear map on the quotient space above.

\begin{proposition}The sesquilinear form $q_\sP$ is well defined on $\cV_\sP$.
\end{proposition}
\proof We need to show that $(u| G_\sD v)_{\sV_1}=0$ if $u\in\Ker\,K^*|_{\Gamma_{\rm c}}$ and $v=Pf$ for some $f\in\Gamma_{\rm c}(M;V_1)$. We have in such case
\[
G_\sD P f= - G_\sD K K^* f=-K G_\sR K^* f, 
\]
hence $(u|G_\sD P f)_{\sV_1}=-(K^* u | G_\sR K^* f)_{\sV_0}=0.$\qeds

The definition of the phase space $\cV_\sP$ agrees with the one considered in \cite{dimock2,FP,pfenning,HS} and is arguably the most natural one. Other possible definitions are discussed in \cite{SDH,HS,benini}. Let us also mention that the form $q_\sP$ needs not be non-degenerate on $\cV_\sP$, cf. examples and further discussion in \cite{SDH,HS,benini}.

It is possible to give different generalizations of Prop. \ref{prop:basic}, (\ref{prop:basic2}) (claim a) below is proved in \cite{HS}).

\begin{proposition}\label{prop:Gpasses}The induced maps
\[
\begin{aligned} 
{\rm a)}\quad &[G_\sD]:\,\frac{\Ker\,K^*|_{\Gamma_{\rm c}}}{\Ran\,P|_{\Gamma_{\rm c}}}\longrightarrow \frac{\Ker\,P|_{\Gamma_{\rm sc}}}{\Ran\,K|_{\Gamma_{\rm sc}}},\\
{\rm b)}\quad &[G_\sD]:\,\frac{\Ker\,K^*|_{\Gamma_{\rm c}}}{\Ran\,P|_{\Gamma_{\rm c}}}\longrightarrow\frac{\Ker\,D_1|_{\Gamma_{\rm sc}}\cap\Ker\,K^*|_{\Gamma_{\rm sc}}}{\Ran\,G_\sD K |_{\Gamma_{\rm c}}},
\end{aligned}
\]
are  well defined and bijective. 
\end{proposition}
\proof b):  For well-definiteness we  check that $G_{1}\Ker\,K^*|_{\Gamma_{\rm c}}\subset \Ker D_{1}$ which is easy,  and $G_{1}\Ker\,K^*|_{\Gamma_{\rm c}}\subset \Ker K^*$, which follows from $K^*G_\sD=G_\sQ K^*$. We need also to check that $G_{1}\Ran P\subset \Ran G_{1}K$ which follows from Hypothesis \ref{as:subsidiary} (c).

For injectivity we see that if $K^{*}u=0$ and $G_{1}u = G_{1}Kv$, then $u-Kv=D_{1}f$ for $f\in \Gamma_{\rm c}(M; V_{1})$, hence $D_{0}(v+ K^{*}f)=0$, which implies that $v+ K^{*}f=0$ and hence $u= Pf$.

Surjectivity amounts to showing
\[
\Ker\,D_1|_{\Gamma_{\rm sc}}\cap\Ker\,K^*|_{\Gamma_{\rm sc}}=G_\sD\Ker K^*|_{\Gamma_{\rm c}}+G_\sD\Ran\,K|_{\Gamma_{\rm c}}.
\]
The inclusion `$\supset$' is easy, the other one follows from Prop. \ref{prop:notbasic}, (\ref{prop:notbasic3}).\qeds

%\begin{remark}It is possible to construct directly a bijection
%\beq\label{eq:defI}
%I : \, \frac{\Ker\,P|_{\Gamma_{\rm sc}}}{\Ran\,K|_{\Gamma_{\rm sc}}}\longrightarrow \frac{\Ker\,D_1|_{\Gamma_{\rm sc}}\cap\Ker\,K^*|_{\Gamma_{\rm sc}}}{\Ran\,G_\sD K |_{\Gamma_{\rm c}}}
%\eeq
%by setting for $\psi\in\Ker\,P|_{\Gamma_{\rm sc}}$
%\[
%I\psi\defeq \{ \psi-Kh : \ h \in \Gamma_{\rm sc}(M;V_0), \ D_0 h=K^*\psi \}.
%\]
%Using similar arguments as before, one can show that $I\psi$ is not empty and 
%\begin{itemize}
%\item $I{\psi}+\Ran\,G_\sD K|_{\Gamma_{\rm c}}\subset I{\psi}$,
%\item $\phi_1,\phi_2\in I\psi$ implies $\phi_1-\phi_2\in \Ran\,G_\sD K|_{\Gamma_{\rm c}}$, 
%\item $I(\psi+Kf)=I\psi$ for all $\psi\in\Ker\,P|_{\Gamma_{\rm sc}}$, $f\in\Gamma_{\rm sc}(M;V_0)$.
%\end{itemize}
%These properties ensure that (\ref{eq:defI}) is well defined.
%\end{remark}

Finally, let us quote another useful result, shown in the present context in \cite{HS}, and often called the \emph{time-slice property} (or time-slice axiom). Below, $J^+(O)$ (resp. $J^-(O)$) denotes the causal future (resp. causal past) of $O\subset M$.

\begin{proposition}\label{prop:timeslice}
Let $\Sigma_+$, $\Sigma_-$ be two Cauchy surfaces s.t. $J^-(\Sigma_+)\cap J^+(\Sigma_-)$ contains properly a Cauchy surface. Then for all $[f]\in \Ker\,K^*|_{\Gamma_{\rm c}} / \Ran\,P|_{\Gamma_{\rm c}} $ there exists $\tilde f\in\Ker\,K^*|_{\Gamma_{\rm c}}$ s.t. 
\[
[f]=[\tilde f], \quad \supp\,\tilde f\subset  J^-(\Sigma_+)\cap J^+(\Sigma_-).
\] 
\end{proposition}

\subsubsection{Phase spaces on a Cauchy surface}\label{ss:phhypsurf}

Let us now discuss the corresponding phase spaces on a fixed Cauchy surface $\Sigma\subset M$. Recall that in Hypothesis \ref{as:subsidiary} we have required  that the operators $D_1$ and $D_0$ are Green hyperbolic, and thus Cauchy hyperbolic. The corresponding maps will be denoted 
\[
\begin{aligned}
\rho_\sD: \ \Gamma(M;V_1)\to\Gamma_{\rm c}(\Sigma;V_{\rho_1}),\\
\rho_\sR: \ \Gamma(M;V_0)\to\Gamma_{\rm c}(\Sigma;V_{\rho_0}).
\end{aligned}
\]
We also recall  that we have defined operators $G_{i\Sig}$ such that  $G_i= (\rho_i  G_i )^* G_{i\Sig} (\rho_i  G_i )$ and Cauchy evolution operators $U_{i}$   for $i=0, 1$. 

To the operator $K$ we associate an operator $K_\Sig\in\Diff(\Sigma;V_{\rho_0},V_{\rho_1})$:
\beq\label{eq:relc}
K_\Sig\defeq \rho_\sD K U_\sQ.
\eeq

It is useful to introduce the adjoint $K^{\dagger}_{\Sig}\in\Diff(\Sigma;V_{\rho_1},V_{\rho_0})$ w.r.t. the hermitian forms $q_{\sD\Sig}$ and $q_{\sQ\Sig}$ (the so-called \emph{symplectic adjoint}), i.e.
\beq\label{eq:defkdag}
G_{\sR\Sig} K^{\dagger}_{\Sig}\defeq K^{*}_{\Sig} G_{\sD\Sig}.
\eeq
The notation $^\dagger$ is used to avoid confusion with the formal adjoint $^*$ w.r.t. the hermitian structures on the bundles $V_{\rho_0}$, $V_{\rho_1}$, appearing for instance in the LHS of the above equation.    

\begin{lemma}\label{lem:cauchyrel}As a consequence of Hypothesis \ref{as:subsidiary},
\begin{enumerate}
\item\label{cauchyrelit1} $K U_\sR = U_\sD K_\Sig$ and $K^*U_\sD=U_\sR K^{\dagger}_{\Sig}$;
\item\label{cauchyrelit0} $\rho_\sD K =K_\Sig \rho_\sR$ on $\Ker\,D_0|_{\Gamma_{\rm sc}}$ and $\rho_\sR K^*=K^{\dagger}_{\Sig}\rho_\sD$ on $\Ker\,D_1|_{\Gamma_{\rm sc}}$;
\item\label{cauchyrelit2} $\Ker\,K^{\dagger}_{\Sig}|_{\Gamma_{\rm c}}=\rho_\sD G^{*}_\sD \Ker\,K^*|_{\Gamma_{\rm c}}$;
\item\label{cauchyrelit3} $\Ran\,K_{\Sig}|_{\Gamma_{\rm c}}=\rho_\sD G_\sD^{*} \Ran\,K|_{\Gamma_{\rm c}}$;
\item\label{cauchyrelit4} $K^{\dagger}_{\Sig} K_\Sig =0$.
\end{enumerate}
\end{lemma}
\proof (\ref{cauchyrelit1}): Let us prove the second assertion (the first one is trivial). By (\ref{eq:defkdag}) and Prop. \ref{prop:notbasic}, (\ref{prop:notbasic1}), 
\[
\begin{aligned}
U_\sR K^{\dagger}_{\Sig} &= G_\sR^* \rho_\sR^* G_{\sR\Sig} K_{\Sig}^{\dag} = G_\sQ^* \rho_\sQ^* K^{*}_{\Sig} G_{\sD\Sig}=G_\sQ^* \rho_\sQ^* U_\sQ^* K^* \rho_\sD^* G_{\sD\Sig} \\
&= G_\sQ^* K^{*} \rho_\sD^* G_{\sD\Sig} =   K^{*} G_\sD^* \rho_\sD^* G_{\sD\Sig} = K^*U_\sD.
\end{aligned}
\] 

(\ref{cauchyrelit0}): By (\ref{cauchyrelit1}) we have $\rho_\sR K^*=\rho_\sR K^* U_\sD \rho_\sD=\rho_\sR U_\sR K^{\dagger}_{\Sig}\rho_\sD=K^{\dagger}_{\Sig}\rho_\sD$. The other assertion is trivial.

(\ref{cauchyrelit2}): If $u=\rho_\sD G_\sD^* f$ with $f\in\Ker\,K^*|_{\Gamma_{\rm c}}$ then $K^{\dagger}_{\Sig}u=\rho_\sR K^* G_\sD^* f=\rho_\sR G^{*}_\sR K^* f=0$. Conversely, if $u\in\Ker\,K^{\dagger}_{\Sig}|_{\Gamma_{\rm c}}$ then using that $\one=\rho_\sD G_\sD^* \rho_\sD^* G_{\sD\Sig}$ we get $u=\rho_\sD G_\sD^* f$ with $f=\rho_\sD^* G_{\sD\Sig} u$ and 
\[
K^*f =K^* \rho_\sD^* G_{\sD\Sig} u =\rho_\sR^* K^{*}_{\Sig} G_{\sD\Sig} u = \rho_\sR^*  G_{\sR\Sig}  K^{\dagger}_{\Sig} u=0.
\]

(\ref{cauchyrelit3}): If $u=\rho_\sD G_\sD^* K f$ then $u=\rho_1 K G_\sD f=K_\Sig\rho_0 G_0 f$. Conversely, if $u=K_\Sig h$ then using that $\one=\rho_\sD G_\sD^* \rho_\sD^* G_{\sD\Sig}$ we get 
\[
u=\rho_\sD G_\sD^* \rho_\sD^* G_{\sD\Sig} K_\Sig h = \rho_\sD G_\sD^* K \rho_\sR^* G_{\sR\Sig} h.
\]

(\ref{cauchyrelit4}): By (\ref{cauchyrelit1}), $K^{\dagger}_{\Sig} K_\Sig= \rho_\sR U_\sR K^{\dagger}_{\Sig} K_\Sig=\rho_\sR K^* U_\sD K_\Sig=\rho_\sR K^*K U_\sR=0$.\qeds

\begin{proposition}\label{prop:rhopasses} The induced map
\[
[\rho_\sD]: \ \frac{\Ker\,D_1|_{\Gamma_{\rm sc}}\cap\Ker\,K^*|_{\Gamma_{\rm sc}}}{\Ran\,G_\sD K |_{\Gamma_{\rm c}}}\longrightarrow\frac{\Ker\,K^{\dagger}_{\Sig}|_{\Gamma_{\rm c}}}{\Ran\,K_\Sig|_{\Gamma_{\rm c}}}
\]
is well defined and bijective.
\end{proposition}
\proof Recall that we proved $\Ker\,D_1|_{\Gamma_{\rm sc}}\cap\Ker\,K^*|_{\Gamma_{\rm sc}}=G_\sD\Ker K^*|_{\Gamma_{\rm c}}+G_\sD\Ran\,K|_{\Gamma_{\rm c}}$.

For well-definiteness and surjectivity of $[\rho_\sD]$ it is thus sufficient to check that 
\[
\rho_\sD(G_\sD\Ker K^*|_{\Gamma_{\rm c}}+G_\sD\Ran\,K|_{\Gamma_{\rm c}})=\Ker\,K^{\dagger}_{\Sig}|_{\Gamma_{\rm c}},
\]
which  follows directly from (2) and (3) of Lemma \ref{lem:cauchyrel} (using  $G_\sD^*=-G_\sD$).

For injectivity we need to show that if $u\in G_\sD\Ker K^*|_{\Gamma_{\rm c}}+G_\sD\Ran\,K|_{\Gamma_{\rm c}}$ and $\rho_\sD u\in \Ran\,K_\Sig|_{\Gamma_{\rm c}}$ then $u\in\Ran\,G_\sD K|_{\Gamma_{\rm c}}$. This follows from (4) of Lemma \ref{lem:cauchyrel}.
\qeds

We deduce from Prop. \ref{prop:Gpasses} and Prop. \ref{prop:rhopasses} that the map $\rho_\sD G_\sD$ induces an isomorphism between the phase space $(\cV_\sP,q_\sP)$ and the phase space $(\cV_{\sP\Sig},q_{\sP\Sig})$, defined in the following way:
\[
\cV_{\sP\Sig}\defeq\frac{\Ker\,K^{\dagger}_{\Sig}|_{\Gamma_{\rm c}}}{\Ran\,K_\Sig|_{\Gamma_{\rm c}}}, \quad \bar u \,q_{\sP\Sig} v\defeq \i(u| [G_{\sD\Sig}] v)_{V_{\rho_1}}.
\]
%Moreover, the assumption $K=T$ implies that and $\cV_{\sP\Sig}$ simplifies to 
%\[
%\cV_{\sP\Sig}=\frac{\Ker\,K^{\dagger}_{\Sig}|_{\Gamma_{\rm c}}}{\Ran\,K_\Sig|_{\Gamma_{\rm c}}}.
%\]
\subsection{Linearized Yang-Mills}\label{lnym} We now recall  how  the formalism of Subsect. \ref{ss:subsidiary} applies to Yang-Mills equations linearized around a background solution $\bA$. We follow \cite{MM,HS}.

Let $\mathfrak{g}$ be a real compact Lie algebra as in Hypothesis \ref{as:group}. We still denote by $\mathfrak{g}$ its complexification. The complexification of the Killing form yields a sesquilinear form 
\[
\killing\in L(\fg,\fg^*), \quad \killing>0.
\]

For simplicity we will work in a geometrically trivial situation\footnote{Otherwise one has to use the language of principal bundles, some indications can be found in \cite{MM,zahn}.}.

As in \cite{HS} we take $V_0$ to be the trivial bundle
\[
V_0\defeq M\times \fg,
\] 
equipped with the hermitian structure induced by  $\killing$, and $V_1$ the corresponding $1$-form bundle
\[
V_1\defeq  T^*M\times\fg.
\]
We equip $V_{1}$ with the hermitian structure given by the tensor product of the canonical hermitian structure on $T^{*}M$ with $\killing$.

 Note that under Hypothesis \ref{as:spacetime} this bundle is trivial since  $\Sigma$ and hence $M$ is   then parallelizable.

Let us denote   by $\cE^p(M)$  the space of smooth $p$-forms on $M$ and by $\cE^{\oplus}(M)=\bigoplus_p\cE^p(M)$ the space of smooth forms on $M$.  As explained in \ref{as:group}, the spaces of sections $\Gamma(M; V_{i})$ $i=0, 1$ can be identified respectively with $\cE^{0}(M)\otimes\fg$ and $\cE^{1}(M)\otimes \fg$.  
The exterior product on $\cE^{\oplus}(M)\otimes\fg$ is defined  by
\[
 (\alpha\otimes a) \wedge (\beta\otimes b) \defeq  (\alpha \wedge \beta)\otimes[a, b] \quad a,b\in\fg, \ \alpha, \beta\in \cE^{\oplus}(M),
\]
(note that in the physics literature a bracket notation is sometimes used instead). The interior product is defined by
\[
( \alpha\otimes a) \Int (\beta\otimes b) \defeq (\alpha \Int \beta)\otimes [b, a] , \quad a,b\in\fg, \ \alpha, \beta\in \cE^{\oplus}(M).
\]
We also define
\[
A\tnI\,\cdot : \ \cE^{\oplus}(M)\otimes\fg  \ni B \mapsto B \Int A\in \cE^{\oplus}(M)\otimes\fg.
\]
It holds that 
\[
(B\wedge\,\cdot\,)^*=\overline{B}\Int\,\cdot\, , \quad B\in\cE^p(M)\otimes\fg
\]
where the bar stands for ordinary complex conjugation. Note also that for $0$-forms the interior product reduces to
\beq\label{eq:reduwedge}
f\Int\,\cdot\, = -f\wedge\,\cdot\, , \quad f\in\cE^0(M)\otimes\fg. 
\eeq

Let $d:\cE^p(M) \to \cE^{p+1}(M)$ be the ordinary differential and let $\bA\in\cE^1(M)\otimes\fg $ (the thick bar is designed to distinguish $\bA$ from dynamical variables $A$, it should not to be confused with complex conjugation $\bar{A}$). The \emph{covariant differential} $\bardel: \cE^p(M)\otimes\fg \to \cE^{p+1}(M)\otimes\fg$ respective to $\bA$ is defined by
\[
 \bardel f \defeq d f + \bA\wedge f, \quad f\in\cE^p(M)\otimes\fg.
\]
Despite its name, it is in general not a differential in the sense that $\bardel\bardel$ would vanish, instead it holds that
\beq\label{eq:notdif}
\bardel\bardel= \bF\wedge\,\cdot\,,
\eeq
where $\bF\defeq d\bA + \bA\wedge \bA\in\cE^2(M)\otimes\fg$ is the \emph{curvature} of $\bA$.  The covariant co-differential $\bdelta:\cE^{p+1}(M)\otimes\fg \to \cE^p(M)\otimes\fg$ is by definition the formal adjoint $\bardel^*$ of $\bardel$.  
The covariant differential satisfies
\[
\bardel(A\wedge B)= (\bardel A) \wedge B+(-1)^p A \wedge (\bardel B), \quad A\in\cE^p(M)\otimes\fg, \ B\in\cE^q(M)\otimes\fg.
\]
This can be written as an identity for operators and by taking their adjoints, one gets
\beq\label{eq:grbdelta}
A \Int \bdelta B =  (\bardel A)\Int B +(-1)^p\bdelta (A\Int B), \quad A\in\cE^p(M)\otimes\fg, \ B\in\cE^q(M)\otimes\fg.
\eeq

A consequence of the definition $\bF=\bardel \bA$ is the {\em Bianchi identity}
\beq
\bardel \bF =0.
\eeq
The {\em non-linear Yang-Mills equation} for $\bA$ reads
\beq\label{eq:nlYM}
 \bdelta\bardel\bA\ (= \bdelta\bF)=0.
\eeq

This system can be linearized as follows. We fix a real-valued section $\bA\in\cE^1(M)\otimes\fg$ and assume it is \emph{on-shell}, i.e. satisfies the Yang-Mills equation (\ref{eq:nlYM}). The {\em linearized Yang-Mills operator}  is \beq
P\defeq\bdelta\bardel + \bF\tnI\, \in\Diff^2(M;V_1),
\eeq
where  $\bardel$, $\bdelta$ and $\bF$  refer to the background solution $\bA$. 
The linearized Yang-Mills equation is
\begin{equation}
\label{eq:linYM}
PA=0.
\end{equation}

Gauge transformations are  described in this linearized setting by the differential operator 
\[
K\defeq \bardel\in\Diff^1(M;V_0,V_1).
\]
It is not difficult to see that Hypothesis \ref{as:subsidiary} is satisfied by $P$ and $K$. More precisely, the operators $D_0=K^* K$ and $D_1=P+KK^*$ equal
\[
\begin{aligned}
D_0&=\bdelta \bardel \in \Diff^2(M;V_0),\\
D_1&=\bardel\bdelta + \bdelta \bardel + \bF\tnI\, \in \Diff^2(M;V_1).
\end{aligned}
\] 
To show $PK=0$, we compute using (\ref{eq:notdif}), (\ref{eq:grbdelta}) and (\ref{eq:reduwedge})
\[
\begin{aligned}
PK f &= \bdelta\bardel\bardel f + \bF\tnI(\bardel f) =   \bdelta(\bF\wedge f) + (\bardel f)\Int \bF\\
 &=\bdelta(f\Int \bF ) + (\bardel f)\Int \bF = f\Int(\bdelta\bF) \quad \forall f\in\cE^0(M)\otimes\fg.
 \end{aligned}
\]
By the assumption that $\bA$ is on-shell (\ref{eq:nlYM}) this vanishes.

\subsubsection{Adapted Cauchy data}\label{ss:adapted} Let us denote by $n$ the future directed unit normal vector field to a Cauchy surface $\Sigma$.

Since $D_1,D_0$ are normally hyperbolic, they are Cauchy hyperbolic for the maps $\rho_1,\rho_0$ defined by taking the restriction to $\Sigma$ of a given section and of its first derivative along $n$. 

For many purposes it will however be more convenient to consider different maps $\rho_1^{\rm F}$, $\rho_0^{\rm F}$, which appear to be due to Furlani \cite{furlani} (cf. also \cite{pfenning}), and which are defined as follows\footnote{To be precise, reference \cite{furlani} uses Cauchy data which are denoted $(A_{(n)},A_{(0)},A_{(\delta)},A_{(d)})$ therein and are related to ours by $g^{0}_{t}= A_{(n)}$, $g^{0}_{\Sig}= A_{(0)}$, $g^{1}_{t}= \i^{-1}A_{(\delta)}$, $g^{1}_{\Sig}= \i^{-1}A_{(d)}$.}.

 We equip $\cE^{p}_{\rm c}(\Sigma)\otimes \fg$ with their standard (positive) hermitian scalar products, obtained from $\killing$ and the Riemannian metric $h$ induced by $g$ on $\Sigma$.  We also recall that $\imath^\pback: \cE_{\rm sc}^{p}(M)\otimes\fg\to \cE^{p}_{\rm c}(\Sigma)\otimes\fg$ is the pullback map induced by the embedding $\imath:\Sigma\to M$.

\begin{definition}
 If $\zeta\in \cE^{1}_{\rm sc}(M)\otimes\fg$, we set:
 \[
\begin{aligned}
g^{0}_{t}&\defeq  \imath^\pback n\lrcorner \zeta\in \kE^{0}_{\rm c}(\Sigma)\otimes\fg,\\[2mm]
g^{0}_{\Sig}&\defeq  \imath^\pback \zeta\in \kE^{1}_{\rm c}(\Sigma)\otimes\fg,\\[2mm]
g^{1}_{t}&\defeq  \i^{-1}\imath^\pback \wbar{\delta}\zeta\in \kE^{0}_{\rm c}(\Sigma)\otimes\fg,\\[2mm]
g^{1}_{\Sig}&\defeq  \i^{-1}\imath^\pback n\lrcorner \wbar{d} \zeta\in \kE^{1}_{\rm c}(\Sigma)\otimes\fg.
\end{aligned}
\]
For $g^{i}\defeq  (g^{i}_{t}, g^{i}_{\Sig})\in \cE^{0}_{\rm c}(\Sigma)\otimes\fg\oplus \cE^{1}_{\rm c}(\Sigma)\otimes\fg$ we set:
\[
g\defeq  (g^{0}, g^{1})\eqdef  \rho^{\rm F}_1 \zeta.
\]
Analogously, if $\zeta\in \cE^{0}_{\rm sc}(M)\otimes\fg$, we set
\[
\begin{aligned}
g^{0}&\defeq  \imath^\pback \zeta\in \kE^{0}_{\rm c}(\Sigma)\otimes\fg,\\[2mm]
g^{1}&\defeq  \i^{-1}\imath^\pback n\lrcorner \wbar{d} \zeta\in \kE^{0}_{\rm c}(\Sigma)\otimes\fg,
\end{aligned}
\]
and
\[
g\defeq  (g^{0}, g^{1})\eqdef  \rho^{\rm F}_0 \zeta.
\]
\end{definition}
In the terminology of Sect. \ref{ss:phhypsurf}, $\rho_i^{\rm F}:\Gamma_{\rm c}(M;V_i)\to\Gamma_{\rm c}(\Sigma;V_{\rho_i^{\rm F}})$ where the bundles
\[
V_{\rho_1^{\rm F}}=(T^{*}\Sigma\oplus T^{*}\Sigma)\times\fg, \quad V_{\rho_0^{\rm F}}=(\Sigma\oplus \Sigma)\times\fg
\]
are equipped with their canonical hermitian structures inherited from the inverse  Riemannian metric on $\Sigma$ and the Killing form $\killing$. 

As in \cite{furlani,pfenning}, it can be checked that the corresponding Cauchy problems are well-posed and that the operators $G_{i\Sig}$ (defined using the $\rho^{\rm F}_i$ data) can be written as
\beq\label{eq:Gsig}
G_{1\Sig}=\i^{-1}\begin{pmatrix} 0 & 0 & -\one & 0 \\ 0 & 0 & 0 & \one \\ -\one & 0 & 0 & 0 \\ 0 & \one & 0 & 0\end{pmatrix}, \quad G_{0\Sig}=\i^{-1}\begin{pmatrix} 0 & \one \\ \one & 0 \end{pmatrix}.
\eeq

We denote by $\bds,\bdeltas$ the covariant differential and co-differential on $\Sigma$ respective to $\bA_\Sig\defeq \imath^\pback \bA$, i.e.
\[
\begin{array}{rl}
&{\bds}\defeq  d_{\Sig} + \wbar{A}_{\Sig}\wedge \,\cdot\,: \ \kE^{p}_{\rm c}(\Sigma)\otimes\fg\to \kE^{p+1}_{\rm c}(\Sigma)\otimes\fg,\\[2mm]
&{\bdeltas}\defeq  {\bds}^{*}:  \ \kE^{p}_{\rm c}(\Sigma)\otimes\fg\to \kE^{p-1}_{\rm c}(\Sigma)\otimes\fg,
\end{array}
\]
where now the adjoint is computed using the inverse metric on $\Sigma$ and the Killing form $\killing$. 

The $\rho_i^{\rm F}$ Cauchy data are particularly useful to express the operators $K_{\Sig}$ and $K_{\Sig}^{\dag}$. In the lemma below we still denote by $K_{\Sigma}^{(\dag)}$ the maps $R_{1{\rm F}}\circ K_{\Sigma}^{(\dag)}\circ R_{0{\rm F}}^{-1}$, where $R_{i{\rm F}}= \rho_{i}^{\rm F}\circ \rho_{i}^{-1}$ for $i=0,1$ are the transition maps from standard to adapted Cauchy data.  The map $R_{1{\rm F}}$ will be simply denoted by $R_{\rm F}$ later on. Its concrete expression is given  in Lemma \ref{n1.1} below .
\begin{lemma}\label{ln.1}
We have:
\[
K_{\Sig}= \left(\begin{array}{cc}
0&\i\\
{\bds}&0\\
0&0\\
\i^{-1}\vara&0
\end{array}\right), \quad K_{\Sig}^{\dag}=\left(
\begin{array}{cccc}
0&0&\i&0\\
0&\i\,\vara^{*}&0&{\bdeltas}
\end{array}\right),
\]
where $\vara\defeq\imath^\pback( n\lrcorner \bF) \wedge \cdot $.\end{lemma}
\proof The formula for $K_{\Sig}$ is a routine computation. To obtain the formula for $K_{\Sig}^{\dag}$ we use (\ref{eq:Gsig}) and (\ref{eq:defkdag}).\qeds

Using Lemma \ref{ln.1} and the identity $K_{\Sig}^{\dag} K_{\Sig}=0$ (Lemma \ref{lem:cauchyrel}, (\ref{cauchyrelit4})), we obtain the following important identity:
\begin{equation}
\label{en.4}
{\bdeltas}\circ \vara= \vara^{*}\circ {\bds}\hbox{ in }L(\kE^{0}(\Sigma)\otimes\fg).
\end{equation}

\section{Hadamard states}\label{secmain}\init 
In this section we discuss Hadamard states both in ordinary field theory and the subsidiary condition framework. 
In Subsect. \ref{ss:quasi-free} we recall basic facts on quasi-free states on complex symplectic spaces.  The Hadamard condition in ordinary field theory is recalled in Subsect. \ref{ss:haddef}. Subsect. \ref{ss:corr} contains a streamlined version of the arguments in \cite{GW},
dealing with the correspondence between Hadamard states and parametrices for the Cauchy problem in the ordinary framework.
In Subsect. \ref{ss:psecauchy} we consider the subsidiary gauge framework. We explain there in detail the strategy we will follow in later sections to construct Hadamard states in this case, thereby proving Thm. \ref{maintheo2}. 

Finally in Subsect. \ref{ss:fnw} we explain the version of the Fulling-Narcowich-Wald deformation argument adapted to the Yang-Mills case, which we use to deduce Thm. \ref{maintheo1} from Thm. \ref{maintheo2}.
\subsection{Quasi-free states}\label{ss:quasi-free}
Let $\cV$ a complex vector space, $\cV^{*}$ its anti-dual and $L_{\rm h}(\cV, \cV^{*})$ the space of hermitian sesquilinear forms on $\cV$. If $q\in L_{\rm h}(\cV, \cV^{*})$ then we can define the 
{\em polynomial  CCR $*$-algebra} ${\rm CCR}^{\rm pol}(\cV,q)$ (see eg \cite[Sect. 8.3.1]{derger}) \footnote{See \cite{GW,wrothesis} for remarks on the transition between real and complex vector space terminology.}. The  (complex) field operators $\cV\ni v\mapsto \psi(v), \psi^{*}(v)$,   which generate ${\rm CCR}^{\rm pol}(\cV,q)$ are anti-linear, resp. linear in $v$ and  satisfy the canonical commutation relations
\[
[\psi(v), \psi(w)]= [\psi^{*}(v), \psi^{*}(w)]=0,  \ \ [\psi(v), \psi^{*}(w)]=  \bar{v} q w \one, \ \ v, w\in \cV.
\]
The \emph{complex covariances}  $\Lambda^\pm\in L(\cV,\cV^*)$ of a (gauge-invariant\footnote{Here by gauge invariance we mean invariance w.r.t. transformations generated by the complex structure. We always consider states that are gauge-invariant in this sense and not mention it anymore in order to avoid confusion with other possible meanings of gauge invariance.}) state $\omega$ on ${\rm CCR}^{\rm pol}(\cV,q)$ are defined in terms of the abstract field operators  by
\[
\bar{v}\Lambda^+ w \defeq \omega\big(\psi(v)\psi^*(w)\big), \quad \bar{v}\Lambda^- w \defeq \omega\big(\psi^*(w)\psi(v)\big), \quad v,w\in \cV
\]
By the canonical commutation relations, one has $\Lambda^+ - \Lambda^- = q$.

In what follows we will consider only \emph{quasi-free states}, which means that they are uniquely determined by their covariances  $\Lambda^\pm$ (since $\Lambda^+ - \Lambda^- = q$ it suffices to know one of them). 

\begin{definition}
 A pair $\Lambda^{\pm}$ of hermitian forms on $\cV$ such that $\Lambda^{+}- \Lambda^{-}=q$ will be called a pair of {\em pseudo-covariances}.
\end{definition}

Let us recall the following characterization of covariances  of  quasi-free states on ${\rm CCR}^{\rm pol}(\cV,q)$ (cf. \cite{araki,GW}).

\begin{proposition}Pseudo-covariances $\Lambda^\pm\in L_{\rm h}(\cV,\cV^*)$ are covariances  of a (bosonic, gauge-invariant) quasi-free state on ${\rm CCR}^{\rm pol}(\cV,q)$ iff
\beq\label{eq:condpositiv}
\Lambda^\pm\geq 0.
\eeq
If $q$ is non-degenerate then this is equivalent to $\pm q c^{\pm}\geq 0$, where $c^{\pm}\defeq \pm q^{-1}\Lambda^\pm$. If moreover, $( c^+)^2= c^+$ on the completion of $\cV$ w.r.t. $\Lambda^++\Lambda^-$, then the associated state is pure.
\end{proposition}
Hence a pair of (pseudo-)covariances $\Lambda^{\pm}\in L_{\rm h}(\cV, \cV^{*})$ uniquely define a (pseudo-)state on ${\rm CCR}^{\rm pol}(\cV,q)$, where by pseudo-state we mean a $*-$invariant linear functional on ${\rm CCR}^{\rm pol}(\cV,q)$.

\begin{definition}A (bosonic) charge reversal on $(\cV,q)$ is an anti-linear operator $\kappa$ on $\cV$ such that $\kappa^2=\pm \one$ and $\kappa^* q \kappa=-\bar q$, where the bar stands for ordinary complex conjugation. A quasi-free state on $\CCR^{\rm pol}(\cV,q)$ with two-point function $\Lambda^+$ is said to be invariant under  charge reversal if $\Lambda^-=-\kappa^* \overline{\Lambda^+} \kappa$. If $q$ is non-degenerate then this is equivalent to $ c^{-}=-\kappa  c^+\kappa$.
\end{definition}

Clearly, if $\Lambda^+$ is  a covariance  of a quasi-free state invariant under charge conjugation then one of the two conditions in (\ref{eq:condpositiv}) implies the other. Note that one can always obtain a state invariant under charge reversal by taking $\12(\Lambda^+ - \kappa^* \overline{\Lambda^-} \kappa)$ instead of $\Lambda^+$. For this reason, we will disregard this issue and consider states that need not be invariant under a charge reversal (contrarily to most of the literature on Hadamard states).

%The two conditions (\ref{hadamard}) will be altogether named the \emph{Hadamard condition}.
%\proof We have to check that $ c_\Sig^+|_{\Gamma_{\rm c}}$ extends to a projection as an operator on $\Gamma_{\rm c}(\Sigma;V_\rho)^{\rm cpl}$ --- the completion of $\Gamma_{\rm c}(\Sigma;V_\rho)$ w.r.t. $\lambda_\Sig^++\lambda_\Sig^-$. But $\lambda_\Sig^+=q_\Sig c_\Sig^+$ and $q_\Sig$ is non-degenerate, therefore $\Gamma_{\rm c}(\Sigma;V_\rho)^{\rm cpl}\subset\Dom\, c_\Sig^+$ and consequently $(c_\Sig^+)^2= c_\Sig^+$ on this space.\qed

\subsection{Hadamard two-point functions}\label{ss:haddef}
\subsubsection{Two-point functions}
Let $D\in {\rm Diff}^{m}(M; V)$ be prenormally hyperbolic and formally selfadjoint for $(\cdot | \cdot)_{V}$.  Let us  introduce the assumptions:
\begin{equation}
\label{eq:titu}
\begin{array}{rl}
i)&\lambda^{\pm}: \Gamma_{\rm c}(M; V)\to \Gamma(M; V)\\[2mm]
ii)& \lambda^{\pm}= \lambda^{\pm*} \hbox{ for }(\cdot| \cdot)_{V} \hbox{ on }\Gamma_{\rm c}(M; V),\\[2mm]
iii)& \lambda^{+}-\lambda^{-}= \i G,\\[2mm]
iv)& D\lambda^{\pm} = \lambda^{\pm}D=0,\\[2mm]
\end{array}
\end{equation}
\begin{equation}
\label{eq:titupos}
\lambda^{\pm}\geq 0 \hbox{ for }(\cdot| \cdot)_{V} \hbox{ on }\Gamma_{\rm c}(M; V).
\end{equation}
Note that (\ref{eq:titu})  implies that $\lambda^{\pm}: \Gamma'(M; V)\to \Gamma_{\rm c}'(M; V)$.
Let us set 
\[
\overline{u}\Lambda^{\pm}v\defeq  (u|\lambda^{\pm}v)_{V}, \ u, v\in \Gamma_{\rm c}(M; V).
\]
If  (\ref{eq:titu}) hold,   then $\Lambda^{\pm}$  define a pair of complex pseudo-covariances on the phase space $(\cV, q)$ defined in (\ref{defo}), hence define a unique quasi-free pseudo-state  on $\CCR^{\rm pol}(\cV, q)$. If additionally (\ref{eq:titupos}) holds, they are (true) covariances, and define  
 a unique quasi-free state on $\CCR^{\rm pol}(\cV, q)$.
\begin{definition}
 A pair  of maps  $\lambda^{\pm}: \Gamma_{\rm c}(M; V)\to \Gamma(M; V)$  satisfying (\ref{eq:titu})   will be called a pair of {\em  spacetime two-point functions}. 
\end{definition}
%Later on we will introduce the corresponding two-point functions $\lambda^{\pm}_{\Sig}$ on a given Cauchy surface $\Sigma$, which will be called {\em Cauchy surface two-point functions}. Both objects will often simply be called {\em two-point functions}.
 \subsubsection{Hadamard condition}

The (primed) wave front set of $\lambda^\pm$ is by definition the (primed) wave front set of its Schwartz kernel. For $x\in M$, we denote $V_x^{\pm*}$ the {\em positive/negative energy cones}, dual future/past light cones and set
\[
\cN^\pm\defeq \{(x,\xi)\in T_x^* M\setminus\{0\} : \ g^{\mu\nu}(x)\xi_\mu \xi_\nu =0, \ \xi\in V_x^{\pm*}\}, \ \cN\defeq \cN^{+}\cup \cN^{-}.
\]

\begin{definition}\label{def:hadamard}A pair of  two-point functions $\lambda^\pm$  satisfying (\ref{eq:titu}) is  \emph{Hadamard} if \beq\label{hadamard}\tag{Had}
\wf'(\lambda^\pm)\subset \cN^\pm\times\cN^\pm.
\eeq
\end{definition}
{  This form of the Hadamard condition is taken from \cite{SV,hollands}, see also \cite{wrothesis} for a review on the equivalent formulations.}

\begin{remark}
 Assume that there exists an anti-linear operator $\kappa: \Gamma(M; V)\to \Gamma(M; V)$ with $\kappa^{2}= \pm \one$ and $D\kappa= \kappa D$. It follows that $\kappa$ induces a charge reversal on  $(\cV, q)$ defined in (\ref{defo}). If moreover $\kappa$ has the property that 
 \[
\kappa(f u)= \overline{f}\kappa u, \ f\in \Gamma(M), \ u\in \Gamma(M; V)
\]
then it is easy to see that 
\[
 \WF(\kappa u)= \overline{\WF(u)}, \ u\in \Gamma_{\rm c}'(M; V)
\]
where 
\[
\overline{\Gamma}\defeq \{(x, -\xi) : \ (x, \xi)\in \Gamma\}, \hbox{ for }\Gamma\subset T^{*}M.
\]
If $\lambda^{\pm}$ are the two-point functions of a (pseudo-)state $\omega$ invariant under the charge reversal $\kappa$, then the relation between $\lambda^{+}$ and $\lambda^{-}$ shows that the two conditions in (\ref{hadamard}) are equivalent.  Most of the literature on Hadamard states deals only with the charge-reversal invariant case, see however \cite{hollands,wrothesis}.
\end{remark}
\subsection{Correspondence between Hadamard states and parametrices}\label{ss:corr}

One of the  methods to impose  $\musc$ is to construct a sufficiently explicit parametrix for the Cauchy problem on a given Cauchy surface $\Sigma$, as was done in \cite{GW} for the scalar Klein-Gordon equation.
In the present subsection, we  will derive the precise relation between two-point functions of Hadamard states in ordinary field theory and parametrices.  

\subsubsection{Two-point functions  on a Cauchy surface}\label{sss:covcauchy}
Let $D\in {\rm Diff}^{m}(M; V)$ be  prenormally hyperbolic, formally selfadjoint on $\Gamma_{\rm c}(M; V)$ and Cauchy hyperbolic for some map $\rho$ as in \ref{ssec:cauchy}.
\begin{lemma}\label{lem:sobolev}
 The operator $\rho G$ extends continuously to a surjection
 \[
\rho G: \Gamma'(M; V)\to \Gamma'(\Sigma; V_{\rho})
\]
with $\Ker \rho G|_{\Gamma'}= \Ran D|_{\Gamma'}$.
\end{lemma}
\proof To show that $\rho G: \Gamma'(M; V)\to \Gamma_{\rm c}'(\Sigma; V_{\rho})$ is well-defined and continuous, it suffices to use the well-known fact that 
\beq\label{eq:WFofG}
{\rm WF}'(G)\subset \cN\times \cN
\eeq
and the rules for composition of distributional kernels in terms of the wavefront set (see \cite{H1}).
The fact that $\rho G: \Gamma'(M; V)\to \Gamma'(\Sigma; V_{\rho})$ follows then from the support properties of $G$. To prove the surjectivity it suffices to show that the identity
\[
\one = - \rho G \rho^{*} G_{\Sig}\hbox{ valid on }\Gamma_{\rm c}(\Sigma; V_{\rho})
\]
extends to $\Gamma'(\Sigma; V_{\rho})$. This is indeed the case because $G_{\sDo\Sig}$ is a differential operator (this is usually shown using Green's formula) and consequently acts continuously from $\Gamma'$ to $\Gamma'$, hence $\rho_\sDo^* G_{\sDo\Sig}:\Gamma'(\Sigma;V_\rho)\to\Gamma'(M;V)$.

The fact that $\Ker \rho G|_{ \Gamma'}= \Ker G|_{ \Gamma'}= \Ran D|_{\Gamma'}$ follows by the same proof as before. \qeds

Let us introduce the assumptions:
\begin{equation}
\label{eq:titucauchy}
\begin{array}{rl}
i)&\lambda^{\pm}_{\Sig}: \Gamma_{\rm c}(\Sigma; V_{\rho})\to\Gamma(\Sigma; V_{\rho}), \\[2mm]
ii)&\lambda^{\pm}_{\Sig}= (\lambda^{\pm}_{\Sig})^{*} \hbox{ for }(\cdot| \cdot)_{V_{\rho}},\\[2mm]
iii)& \lambda_{\Sig}^{+}- \lambda_{\Sig}^{-}= \i G_{\Sig}.
\end{array}
\end{equation}
\begin{definition}
 A pair of maps $\lambda_{\Sig}^{\pm}$ satisfying (\ref{eq:titucauchy}) will be called a pair of {\em Cauchy surface two-point functions}.
\end{definition}
In the proposition below we recall a well known bijection between spacetime and Cauchy surface two-point functions.
\begin{proposition}\label{minusu}
The maps:
 \beq\label{eq:nimnim}\lambda_{\Sig}^{\pm}\mapsto \lambda^{\pm}\defeq  (\rho G)^{*}\lambda^{\pm}_{\Sig}(\rho G),
 \eeq
 and 
 \begin{equation}
\label{eq:nim}
\lambda^{\pm}\mapsto\lambda_{\Sig}^{\pm}\defeq (\rho^{*}G_{\Sig})^{*} \lambda^{\pm} (\rho^{*}G_{\Sig})
\end{equation}
are bijective and inverse from one another.
Moreover, $\lambda^{\pm}$ are the two-point functions of a quasi-free state iff
\[
\lambda_{\Sig}^{\pm}\geq 0 \hbox{ for }(\cdot| \cdot)_{V_{\rho}}.
\]
 \end{proposition}

 \proof  
(1): let $\lambda^{\pm}_{\Sig}$ satisfy (\ref{eq:titucauchy}). Clearly $\lambda^{\pm}$ is well defined as a map from $\Gamma_{\rm c}(M; V)$ to $\Gamma_{\rm c}'(M; V)$. If $u\in \Gamma_{\rm c}(M; V)$, then $f^{\pm}\defeq\lambda^{\pm}_{\Sig}\rho G u\in \Gamma(\Sigma; V_{\rho})$, hence ${\rm WF} (\rho^{*}f^{\pm})\subset N^{*}_{\Sig}M$, the {\em conormal bundle} to $\Sigma$ in $M$. We use now (\ref{eq:WFofG}), the fact that $\Sigma$ is non-characteristic i.e. $N^{*}_{\Sig}M\cap \cN= \emptyset$ and standard arguments with wave front sets (see \cite{H1}) to obtain that $\lambda^{\pm} u= -G \rho^{*}f^{\pm}\in \Gamma(M; V)$.  The other conditions in (\ref{eq:titu}) are straightforward.

(2): let $\lambda^{\pm}$ satisfies (\ref{eq:titu}).  Since $\lambda^{\pm} D=0$, we  have ${\rm WF}'(\lambda^{\pm})\subset T^{*}M\times \cN$ which implies that $\lambda^{\pm} (\rho^{*}G_{\Sig}): \Gamma_{\rm c}(\Sigma; V_{\rho})\to \Gamma(M; V)$. Next we use that $G_{\Sig}$ is a differential operator hence $G_{\Sig}: \Gamma(\Sigma; V_{\rho})\to \Gamma(\Sigma; V_{\rho})$ to obtain that $\lambda^{\pm}_{\Sig}: \Gamma_{\rm c}(\Sigma; V_{\rho})\to \Gamma(\Sigma; V_{\rho})$. The other conditions in (\ref{eq:titucauchy}) are straightforward.
 
The fact that the two maps are inverse from each other  follows from $\rho U= \rho G^{*}\rho^{*}G_{\Sig}=\one$.  The last statement about positivity is obvious. \qeds
 
Prop. \ref{minusu} leads to the following definition:
\begin{definition}
 A pair $\lambda^{\pm}_{\Sig}$ of Cauchy surface two-point functions is {\em Hadamard} if the associated spacetime  two-point functions $\lambda^{\pm}$ are Hadamard.
\end{definition}

\subsubsection{ Hadamard two-point functions and parametrices}
Let us now discuss the link between Hadamard two-point functions and parametrices for the Cauchy problem. 
Let  $\lambda^{\pm}$ be   the two-point functions of a state. We set\footnote{For instance, if $\lambda^\pm$ are the two-point functions of the vacuum for the scalar Klein-Gordon equation on Minkowski space then $H^0(\Sigma;V_\rho)=H^{\12}(\rr^d)\oplus H^{-\12}(\rr^d)$, where $H^m(\rr^d)$ are the usual Sobolev spaces.}
\beq\label{sp:HS}
H^0(\Sigma;V_\rho)\defeq \big(\Gamma_{\rm c}(\Sigma;V_\rho))^{\rm cpl}
\eeq
where the completion is taken w.r.t. $(\cdot|(\lambda^{+}_{\Sig}+ \lambda^{-}_{\Sig}) \cdot)_{V_\rho}$.
\begin{theoreme}\label{thm:corres} Let $D\in\Diff^m(M;V)$ be prenormally hyperbolic, formally self-adjoint and Cauchy hyperbolic. Let $\lambda^{\pm}$ be  the two-point functions of a quasi-free Hadamard state and define
\[
U^{\pm}\defeq U  c^\pm:\, \Gamma'(\Sigma;V_\rho)\to \Gamma_{\rm c}'(M;V),
\]
where $c^\pm=\pm(\i G_{\Sig})^{-1}\lambda^\pm_\Sig$. Then 
\begin{enumerate}
\item[(1)] $U^+ + U^- = U_\sDo$.
\item[(2$a$)] The spaces $\Ker\,U^+|_{H^0}$ and $\Ker\,U^-|_{H^0}$ are orthogonal for $q_\Sig$. 

\item [(2$b$)] if the state is pure then
\[
\quad H^0(\Sigma;V_\rho)=\Ker\,U^+|_{H^0} \oplus \Ker\,U^-|_{H^0}.
\] 
\item[(3)] $\pm\i G_{\Sig}$ is positive on $\Ker\,U^\pm|_{H^0}$ for $(\cdot| \cdot)_{V_{\rho}}$.
\item[(4)] $\wf(U^\pm f)\subset\cN^\pm$ for all $f\in \Gamma'(\Sigma;V_\rho)$. 
\end{enumerate}
\end{theoreme}
\proof  (1) follows from $c^{+}+ c^{-}= \one$.  To prove (2a) we note that for $u^{\pm}\in \Ker c^{\mp}$ and $q_{\Sig}$ defined in (\ref{eq:defdeqsigma}) one has:
\[
 \overline{(u^{+}+ z u^{-})}q_{\Sig}u^{+}= \overline{(u^{+}+ z u^{-})}q_{\Sig}c^{+}(u^{+}+ z u^{-})\in \rr, \ \forall \ z\in \cc, 
\]
which implies that $\overline{u^{-}}q_{\Sig}u^{+}=0$.  (2b) follows from the fact that $c^{\pm}$ are bounded projections on $H^{0}$ if the state $\omega$ is pure, (3) follows from the conditions $\lambda_{\Sig}^{\pm}\geq 0$.
 To show (4), observe that for all $u\in \Gamma'(M;V)$
\[
\lambda^+ u = (\rho_\sDo G_\sDo)^*\lambda_\Sig^+ \rho_\sDo G_\sDo u= U^+ \rho_\sDo G_\sDo u.
\] 
Thus, the Hadamard condition entails that $\wf(U^+ \rho_\sDo G_\sDo u)\subset\cN^+$. Since $\rho_\sDo G_\sDo$ is surjective this means $\wf(U^+ f)\subset\cN^+$ for all $f\in \Gamma'(\Sigma;V_\rho)$. The proof for $U^-$ is analogous.\qeds

To obtain a converse statement, we need spaces that can replace the space $H^{0}(\Sigma; V_{\rho})$, and that will allow to compose operators.

To this end, suppose $\cH(\Sigma;V_\rho)$ is a topological vector space s.t.
\[
\Gamma_{\rm c}(\Sigma;V_\rho)\subset \cH (\Sigma;V_\rho) \subset \Gamma(\Sigma;V_\rho),
\]
with continuous and dense embedings.
Examples of such spaces are (intersections of) scales of Sobolev spaces associated to a positive, elliptic pseudodifferential operator.
The dual space of $\cH(\Sigma;V_\rho)$, denoted $\cH'(\Sigma;V_\rho)$, satisfies
\[
\Gamma'(\Sigma;V_\rho)\subset \cH' (\Sigma;V_\rho) \subset \Gamma_{\rm c}'(\Sigma;V_\rho).
\]
We will denote $B^{-\infty}(\Sigma;V_\rho)$ the class of operators that map $\cH' (\Sigma;V_\rho)$ into $\Gamma (\Sigma;V_\rho)$.

We assume that 
\beq\label{eq:gsig1}
G_{\Sig}, \ G_{\Sig}^{-1}: \cH (\Sigma;V_\rho)\to \cH (\Sigma;V_\rho),
\eeq
which  since $\i G_{\Sig}$ is selfadjoint for $(\cdot| \cdot)_{V_{\rho}}$ implies of course 
\[
G_{\Sig}, \ G_{\Sig}^{-1}: \cH' (\Sigma;V_\rho)\to \cH' (\Sigma;V_\rho),
\]
The corresponding natural assumption for a pair of   Cauchy surface two-point functions $\lambda_{\Sig}^{\pm}$ is
\begin{equation}
\label{eq:blaireau}
\begin{array}{rl}
&\lambda^{\pm}_{\Sig}:\cH (\Sigma;V_\rho)\to \cH (\Sigma;V_\rho),\\[2mm]
&\lambda^{\pm}_{\Sig}:\cH'(\Sigma;V_\rho)\to \cH' (\Sigma;V_\rho),
\end{array}
\end{equation}
where as before one of the above conditions implies the other.
\begin{theoreme}\label{thm:corres2} Assume  that there exist operators $U^{\pm}:\cH'(\Sigma;V_\rho)\to \Gamma_{\rm c}'(M;V)$ such that $U^\pm:\cH(\Sigma;V_\rho)\to \Gamma(M;V)$ and 
\[
D U^\pm=0, \ \ U^+ + U^- = U_\sDo,
\]
up to remainders that map  $\cH' (\Sigma;V_\rho)\to\Gamma (M;V)$.

Assume moreover that\begin{enumerate}
\item[(1)] the spaces $\Ker\,U^+|_{\cH}$ and $\Ker\,U^-|_{\cH}$ are orthogonal for $q_\Sig$ and
\[
\cH(\Sigma;V_\rho)=\Ker\,U^+|_{\cH} \oplus \Ker\,U^-|_{\cH}.
\] 
\item[(2)] $\wf(U^\pm f)\subset\cN^\pm$ for all $f\in \Gamma'(\Sigma;V_\rho)$. 
\end{enumerate}

Let $c^{\pm}:\cH(\Sigma;V_\rho)\to \cH(\Sigma;V_\rho)$ be the projection s.t. 
\[
\Ran\, c^{\pm}=\Ker\,U^\mp|_{\cH}, \quad \Ker\, c^{\pm}=\Ker\,U^\pm|_{\cH}.
\]
Then  $\lambda_\Sig^\pm\defeq  \pm \i G_\Sig c^\pm$ are  Hadamard  Cauchy surface two-point functions. If moreover 
\begin{enumerate}
\item[(3)] $\pm\i G_{\Sig}c^{\pm}\geq 0$ for $(\cdot| \cdot)_{V_{\rho}}$,
\end{enumerate} then $\lambda_\Sig^\pm$ are the Cauchy surface two-point functions of a Hadamard state.

\end{theoreme}\proof (1) implies   $ c^+ +  c^-=\one$. By duality, $c^\pm:\cH'(\Sigma,V_\rho)\to\cH'(\Sigma,V_\rho)$. Next, for all $f\in \Gamma'(\Sigma;V_\rho)$ we have:
\[
U_\sDo  c^\pm f= (U^+ + U^-) c^\pm f = U^\pm  c^\pm f = U^\pm (\one- c^\mp) f = U^\pm f \ \ \mod \ \cinf.
\]
Therefore,
\[
\lambda^{\pm} u =  \pm \i U c^{\pm}\rho G u= \pm\i U^{\pm}\rho G u \ \ \mod \ C^\infty, \quad u\in \Gamma'(M;V).
\]
Let $a^{\pm}$ be a properly supported pseudodifferential operator, non-characteristic on $\cN^{\pm}$ and with essential support disjoint from  $\cN^{\mp}$. From {  (2)} and the relation above it follows that $a^{\pm}\lambda^{\pm}$ is smoothing, hence $\wf'(\lambda^{\pm})\subset \cN^{\pm}\times \cN$. Since $\lambda^{\pm}=(\lambda^{\pm})^*$ this implies $\wf'(\lambda^\pm)\subset \cN^\pm\times \cN^\pm$.  This proves the first statement of the proposition. The second statement is obvious. \qeds

Thm. \ref{thm:corres2} allows to simplify the construction of Hadamard states for the scalar Klein-Gordon equation given in \cite{GW} --- it is in fact not difficult to check properties (1)-(3) directly from the construction of the parametrix therein. The space $\cH(\Sigma;V_\rho)$ is taken there to be the intersection of usual Sobolev spaces on $\rr^{d}$. The next proposition is an abstract version of a result from \cite{GW}.

\begin{proposition}Assume that  $\lambda^\pm_\Sig,\tilde\lambda^\pm_\Sig$ satisfy (\ref{eq:blaireau})  and are the   Cauchy surface two-point functions of two quasi-free states, and suppose the first of them is pure and Hadamard. Then the other one is Hadamard iff 
\beq\label{eq:ppm}
 c^- \tilde c^+  c^-,\,  c^+ \tilde c^+  c^-,\,  c^+ \tilde c^-  c^+  \in B^{-\infty}(\Sigma;V_\rho)
\eeq   
or, equivalently, iff
\beq\label{eq:dif}
\tilde c^{\pm} - c^{\pm} \in B^{-\infty}(\Sigma;V_\rho)
\eeq
\end{proposition}
\proof 
$\Leftarrow$: \ if (\ref{eq:ppm}) or (\ref{eq:dif}) holds then
\[
U\tilde c^\pm - U  c^\pm \tilde c^\pm  c^\pm : \  \cH'(\Sigma;V_\rho)\to \Gamma(M;V).
\]
By Thm. \ref{thm:corres}, it follows that $\wf(U\tilde c^\pm f)\subset\cN^\pm$ for all $f\in \Gamma'(\Sigma;V_\rho)$ and consequently $\tilde\lambda$ is Hadamard by Thm. \ref{thm:corres2}.

$\Rightarrow$: \ for all $f\in \Gamma'(\Sigma;V_\rho)$, 
\[
U_\sDo  c^- \tilde c^+  c^\pm f = U_\sDo  \tilde c^+  c^\pm f - U  c^+  \tilde c^+  c^\pm f.
\]
By Thm. \ref{thm:corres}, the wave front set of the LHS is contained in $\cN^-$, and the wave front set of the RHS is contained in $\cN^+$. This shows that the operators $U_\sDo  c^- \tilde c^+  c^\pm$ are smoothing, therefore $ c^- \tilde c^+  c^\pm=\rho_\sDo U_\sDo  c^- \tilde c^+  c^\pm$ are smoothing. The assertion $ c^+ \tilde c^-  c^+  \in B^{-\infty}(\Sigma;V_\rho)$ is shown similarly.

Moreover, (\ref{eq:ppm}) entails that
\[
\begin{aligned}
\tilde c^+ -  c^+&=( c^+ +  c^- ) \tilde c^+ ( c^+ +  c^- ) - c^+ =   c^+ \tilde c^+  c^+ - c^+ \\ &= c^+ (\tilde c^+-\one)  c^+ =-  c^+ \tilde c^- c^+ \ \ \mod \ B^{-\infty}(\Sigma;V_\rho),
\end{aligned}
\]
where the last term belongs to $B^{-\infty}(\Sigma;V_\rho)$. This proves (\ref{eq:dif}).\qeds

\begin{corollary}If $\lambda^\pm_\Sig$ satisfying (\ref{eq:blaireau}) are  Hadamard Cauchy surface two-point functions then so are $v^* \lambda^\pm_\Sig v$ for any $v\in\one+B^{-\infty}(\Sigma;V_\rho)$ s.t. $v^*G_\Sig v = G_\Sig$.
\end{corollary}

%It is possible to generalize the results of this section to gauge theories and give a correspondence between parametrices for $D_1$ and pseudo-covariances that satisfy $\musc$, $\gi$ and $\pos$. Unfortunately, one would obtain a list of conditions on the parametrices that are difficult to obtain using standard pseudo-differential calculus methods. We will therefore focus our attention on pseudo-covariances on a Cauchy surface.

\subsection{Hadamard states in the subsidiary condition formalism}  \label{ss:psecauchy}
\subsubsection{Hadamard states in the subsidiary condition formalism}

Definition \ref{def:hadamard} can be generalized to gauge theories in the `subsidiary condition' framework. Recall that to a given non-hyperbolic operator $P$ we have assigned a hyperbolic operator $D_1$ and introduced phase spaces $\cV_\sP=\Ker K^*/\Ran P$, $\cV_\sD=\Gamma_{\rm c}/\Ran D$. We consider the following definition, which generalizes the one used by \cite{FP,FS}.

\begin{definition}A quasi-free state $\omega$ on $\CCR^{\rm pol}(\cV_\sP,q_\sP)$ is \emph{Hadamard}   if  there exists Hadamard two-point functions $\lambda_{1}^{\pm}$ on $\Gamma_{\rm c}(M; V_{1})$ such that the complex covariances of $\omega$ are given by:
\[
\overline{[u]}\Lambda^{\pm}[v]=  (u| \lambda_{1}^{\pm}v)_{V}, \ u, v\in\Ker K^{*}|_{\Gamma_{\rm c}},
\]
 where $\Ker K^{*}|_{\Gamma_{\rm c}}\ni u\mapsto [u]\in \Ker K^*/\Ran P$ is the canonical map.
\end{definition}
We say that $\lambda_{1}^{\pm}$ are the two-point functions of    the Hadamard state $\omega$ on $\CCR^{\rm pol}(\cV_\sP,q_\sP)$.
The following lemma is straightforward.
\begin{lemma}
 $\lambda_{1}^{\pm}: \Gamma_{\rm c}(M; V_{1})\to \Gamma(M; V_{1})$ are the two-point functions of  a Hadamard state on $\CCR^{\rm pol}(\cV_\sP,q_\sP)$ if:
\beq\label{defodefo}
\begin{aligned}
\musc &\quad  D_1\lambda^\pm_1=\lambda^\pm_1 D_1 =0, \quad \wf'(\lambda^\pm_1)\subset\cN^\pm\times\cN^\pm,\\
\gi & \quad(\lambda^\pm_1)^*=\lambda^\pm_1 \mbox{ \ and \ } \lambda^\pm_1: \ \Ran\,K|_{\Gamma_{\rm c}}\to\Ran\,K|_{\Gamma_{\rm c}'},\\
\pos & \quad \lambda^\pm_1  \geq 0 \mbox{ \ on \ } \Ker\,K^*|_{\Gamma_{\rm c}}.
\end{aligned}
\eeq
\end{lemma}

{  It is worth mentioning that in perturbative interacting Quantum Field Theory, some constructions  seem to survive if one replaces gauge-invariance $\gi$ by a condition `modulo smooth terms' \cite{rejzner}. Nevertheless, $\musc$ and positivity $\pos$ are still essential (cf. \cite{DF} and \cite[Sec. 4.1.2]{hollands2} for discussion on the latter), and gauge-invariance $\gi$ is needed to have a reasonable non-interacting theory, we will thus aim at solving all of them when possible.}

We  now discuss gauge-invariance and positivity on the level of Cauchy surface two-point functions $\lambda^\pm_{1\Sig}$.   We explain the main steps of the construction of Hadamard states for the linearized Yang-Mills equations, leading to a proof of Thm. \ref{maintheo2}, which will be completed in Sect. \ref{sec4}.

The construction is  somewhat complicated by the need to justify that various operators can be composed. These technical points can be bypassed on the first reading.

We fix spaces $\cH(\Sigma; V_{\rho_{i}})$, $i=0,1$ as in  Subsect. \ref{ss:corr} and assume that $G_{i\Sig}$ satisfy (\ref{eq:gsig1}). 
The corresponding assumption on $K_{\Sig}$ is:
\begin{equation}
\label{eq:terrier}
\begin{array}{rl}
&K_{\Sig}: \cH(\Sigma; V_{\rho_{0}})\to \cH(\Sigma; V_{\rho_{1}}), \\[2mm]
&K_{\Sig}: \cH'(\Sigma; V_{\rho_{0}})\to \cH'(\Sigma; V_{\rho_{1}}).
\end{array}
\end{equation}
 The operator $K_\Sig^\dag$ has then the same properties as $K_{\Sig}$.
 \subsubsection{Cauchy surface two-point functions}
 Assume that we are given  Cauchy surface two-point functions $\lambda_{i\Sig}^{\pm}$  for $i=0,1$ satisfying (\ref{eq:titucauchy}) and (\ref{eq:blaireau})  for $V= V_{i}$.  
 
 To $\lambda^\pm_{i\Sig}$ we associate as before operators $c_i^\pm\defeq \pm \i G_{i\Sig}^{-1} \lambda^\pm_{i\Sig}$ which by the above assumptions satisfy: \begin{equation}
\label{eq:azert}
\begin{array}{rl}
i)&c_{i}^{\pm}:\cH(\Sigma; V_{\rho_{i}})\to \cH(\Sigma; V_{\rho_{i}}),\\[2mm]
ii)&c_{i}^{\pm}:\cH'(\Sigma; V_{\rho_{i}})\to \cH'(\Sigma; V_{\rho_{i}}),\\[2mm]
iii)&c^+_i + c^-_i=\one.
\end{array}
\end{equation}
Conditions $\pos$, $\gi$  on $\lambda_{1}^{\pm}$  in (\ref{defodefo}) can be rewritten as
\[
\begin{array}{rl}
 \pos& \lambda_{1\Sig}^{\pm}= \pm \i G_{1\Sig} c_1^\pm \geq 0 \mbox{ for }(\cdot| \cdot)_{V_{\rho_{1}}}\mbox{\ \ on\ \ } \Ker K_\Sig^\dag,\\[2mm]
 \gi& (c_1^{\pm})^\dag=c_1^{\pm}, \quad c_1^{\pm}:\Ran K_\Sig\to\Ran K_\Sig.
\end{array}
\]
Note that the last condition can be rewritten as:
\[
\begin{aligned}
 \gi &\quad(c_1^{\pm})^\dag=c_1^{\pm}, \quad c_1^{\pm}:\Ker K_\Sig^\dag \to\Ker K_\Sig^\dag.
\end{aligned}
\]
Let us now set:
\begin{equation}
\label{eq:defderinfty}
 c_1^{\pm} K_\Sig- K_\Sig c_0^{\pm}\eqdef\pm R_{-\infty}.
\end{equation}
Condition $\gi$ is clearly satisfied if $R_{-\infty}=0$.

The operators $c_{i}^{\pm}$ are obtained from parametrices $U_{i}^{\pm}$ for the Cauchy problems for $D_{i}$ as in Thm. \ref{thm:corres2}, in order  to enforce the Hadamard condition for $\lambda_{1}^{\pm}$.  The construction of parametrices done in Sect. \ref{sec2}  relies on pseudodifferential calculus, from which we will only be able to obtain that 
$R_{-\infty}$  is smoothing. 

Nevertheless, it is possible to ensure $\gi$ by subtracting  to $c_{1}^{\pm}$ a term $c_{1\,{\rm reg}}^{\pm}$, which is expected to be smoothing, and hence will not invalidate  the Hadamard property. 

The method works as follows.

\subsubsection{Construction of a projection}\label{sss:proj} Let $\Proj$ be a projection s.t.
\beq\label{nmis.0}
\begin{array}{rl}
&\Ker\,\Proj=\Ran\,K_\Sig,\\[2mm]
&\Proj: \cH(\Sigma; V_{\rho_{1}})\to \cH(\Sigma; V_{\rho_{1}}),\\[2mm]
&\Proj:\cH'(\Sigma; V_{\rho_{1}})\to \cH'(\Sigma; V_{\rho_{1}}).
\end{array}
\eeq
Clearly $\Proj^{\dag}$ has the same mapping properties as $\Proj$. Moreover one has:
  \begin{equation}
\label{nmis.1}
\Ran \Proj^{\dag}= \Ker K_{\Sig}^{\dag}, \quad \Ran (\one - \Proj)= \Ran K_{\Sig},\quad \Ker (\one -\Proj^{\dag})= \Ker K_{\Sig}^{\dag}.
\end{equation}
Since $\Ran K_{\Sig}\subset \Ker K_{\Sig}^{\dag}$ we also have:
 \begin{equation}
\label{nmis.2}
\Proj^{\dag}K_{\Sig}= K_{\Sig}, \ K_{\Sig}^{\dag}\Proj= K_{\Sig}^{\dag}.
\end{equation}
\subsubsection{Construction of a right inverse to $K_{\Sig}$}\label{sss:rightinverse}
Let also $B:\Gamma_{\rm c}(\Sigma; V_{\rho_{1}})\to \Gamma(\Sigma; V_{\rho_{0}})$ an operator such that
\begin{equation}
\label{crim}
K_{\Sig}B = \one - \Proj, \hbox{ and hence } B^{\dag}K_{\Sig}^{\dag}= \one - \Proj^{\dag}.
\end{equation}
The operator $B$ is typically {\em unbounded} from $\cH(\Sigma; V_{\rho_{1}})$ to $\cH(\Sigma; V_{\rho_{0}})$, because of infrared problems. 
To control its unboundedness, we introduce a smooth positive function $\x: \Sigma\to \rr$ and still denote by $\x$ the operator of multiplication by $\x$, acting on $\Gamma(\Sigma; V_{\rho_{i}})$. If  $\Sigma$  is compact the weight is unnecessary and one can take $\x= \one$.

We assume that:
\beq\label{eq:fouine}
\begin{array}{rl}
i)&\x G_{i\Sig}\x^{-1}: \cH(\Sigma; V_{\rho_{i}})\to \cH(\Sigma; V_{\rho_{i}}),\ i=0,1,\\[2mm]
ii)&\x^{-1}K_{\Sig}\x: \cH(\Sigma; V_{\rho_{0}})\to \cH(\Sigma; V_{\rho_{1}}),\\[2mm]
iii)&\x^{-1}c_{0}^{\pm}\x : \cH(\Sigma; V_{\rho_{0}})\to \cH(\Sigma; V_{\rho_{0}}),\\[2mm]
\end{array}
\eeq
Concerning the operator $B$ we assume that:
 \begin{equation}
\label{eq:moufette}
\begin{array}{rl}
&B: \cH(\Sigma; V_{\rho_{1}})\to \x \cH(\Sigma; V_{\rho_{0}}), \\[2mm]
&B: \cH'(\Sigma; V_{\rho_{1}})\to \x\cH'(\Sigma; V_{\rho_{0}}),
\end{array}
\end{equation}

\begin{theoreme}
 \label{thm:fincov}
 Let $c_{i}^{\pm}$, $\Proj$, $B$ be as above. Let us set:
 \[
\begin{aligned}
\tilde{c}_{1}^{\pm}&\defeq  \Proj^{\dag} c_{1}^{\pm}\Proj+ B^{\dag}c_{0}^{\pm}K_{\Sig}^{\dag}+ K_{\Sig}c_{0}^{\pm}B,\\[2mm]
c_{1\, {\rm reg}}^{\pm}&\defeq\pm( B^{\dag} R^{\dag}_{-\infty}+ \Proj^{\dag}R_{-\infty}B),\\[2mm]
\tilde{\lambda}_{1\Sig}^{\pm}&\defeq\pm \i G_{1\Sig}\tilde{c}_{1}^{\pm}.
\end{aligned}
\]
Then: 
\ben
\item $\tilde{c}_{1}^{\pm}:\x^{-1}\cH(\Sigma; V_{\rho_{1}})\to \x\cH(\Sigma; V_{\rho_{1}})$, hence   $\tilde{c}_{1}^{\pm}:\Gamma_{\rm c}(\Sigma; V_{\rho_{1}})\to \Gamma(\Sigma; V_{\rho_{1}})$.
%Assume that
%\[
%c_{1\, {\rm reg}}^{\pm}: \Gamma_{\rm c}(\Sigma; V_{\rho_{1}})\to \Gamma(\Sigma; V_{\rho_{1}}). 
%\]
\item  One has:
\[
\begin{array}{rl}
i)& (\tilde{c}_{1}^{\pm})^{\dag}= \tilde{c}_{1}^{\pm},\\[2mm]
ii)&\tilde{c}_{1}^{+}+ \tilde{c}_{1}^{-}= \one,\\[2mm]
iii)& \tilde{c}_{1}^{\pm}: \Ker K_{\Sig}^{\dag}\to \Ker K_{\Sig}^{\dag},\\[2mm]
%iv)& \tilde{\lambda}_{1\Sig}^{\pm}= \Proj^{*}\circ \lambda_{1\Sig}^{\pm} \circ \Proj\hbox{ on }\Ker K_{\Sig}^{\dag},\\[2mm]
{ {iv)}}&   {  (f|\tilde{\lambda}_{1\Sig}^{\pm}g)_{V_{\rho_{1}}}= (\Proj f|\lambda_{1\Sig}^{\pm}\Proj g)_{V_{\rho_{1}}} , \ f, g\in \Ker K_{\Sig}^{\dag} ,         }\\[2mm]
v)&c_{1}^{\pm}=  \tilde{c}_{1}^{\pm}+ c_{1\, {\rm reg}}^{\pm},
\end{array}
\]
in particular   $\tilde{\lambda}_{1\Sig}^{\pm}$ satisfy $\gi$. 
\item If  the projection $\Proj$ is  such that 
\beq\label{D520}
\lambda_{1\Sig}^{\pm}\geq 0 \hbox{ on }\Proj\Ker K^{\dag}_{\Sig},
\eeq
then $\tilde{\lambda}_{1\Sig}^{\pm}$ satisfy also $\pos$. 
\item If moreover
\[
c_{1\,{\rm reg}}^{\pm}: \Gamma_{\rm c}'(\Sigma; V_{\rho_{1}})\to \Gamma(\Sigma; V_{\rho_{1}})
\]
 and ${\lambda}_{1\Sig}^{\pm}$ are Hadamard, then  $\tilde{\lambda}_{1\Sig}^{\pm}$ are Hadamard.
 \een
\end{theoreme}

\proof Let us first prove (1).  Clearly $\Proj^{\dag} c_{1}^{\pm}\Proj: \cH(\Sigma; V_{\rho_{1}})\to \cH(\Sigma; V_{\rho_{1}})$, by (\ref{eq:azert}), (\ref{nmis.0}). Next we obtain that $K_{\Sig}c_{0}^{\pm}B: \cH(\Sigma; V_{\rho_{1}})\to \x \cH(\Sigma; V_{\rho_{1}})$, by (\ref{eq:moufette}), (\ref{eq:fouine}). Using the same assumptions and duality we obtain that $B^{\dag}c_{0}^{\pm}K_{\Sig}^{\dag}: \x^{-1}\cH(\Sigma; V_{\rho_{1}})\to \cH(\Sigma; V_{\rho_{1}})$.

Let  us now prove (2). $i)$ is easy. To prove $ii)$ we write
\[
\begin{aligned}
\tilde{c}_{1}^{+}+ \tilde{c}_{1}^{-}&=\Proj^{\dag}\Proj + B^{\dag}K_{\Sig}^{\dag}+ K_{\Sig}B\\[2mm]
&=\Proj^{\dag}\Proj + B^{\dag}K_{\Sig}^{\dag}\Proj+ K_{\Sig}B\\[2mm]
&=\Proj^{\dag}\Proj + (\one - \Proj^{\dag})\Proj + (\one - \Proj)= \one,
\end{aligned}
\]
using successively  $c_{i}^{+}+ c_{i}^{-}= \one$,  (\ref{nmis.2}),  and (\ref{crim}).
$iii)$  follows from $\Ran \Proj^{\dag}= \Ker K_{\Sig}^{\dag}$ (see (\ref{nmis.1})), and  $\Ran K_{\Sig}\subset \Ker K_{\Sig}^{\dag}$.
$iv)$ follows from the definition of $\tilde{\lambda}_{1\Sig}^{\pm}$.  To prove $v)$ we write:
\[
\begin{aligned}
\tilde{c}_{1}^{\pm}&= \Proj^{\dag} c_{1}^{\pm}\Proj+ B^{\dag}c_{0}^{\pm}K_{\Sig}^{\dag}+ K_{\Sig}c_{0}^{\pm}B\\[2mm]
&= \Proj^{\dag} c_{1}^{\pm}\Proj+ B^{\dag}c_{0}^{\pm}K_{\Sig}^{\dag}+ \Proj^{\dag}K_{\Sig}c_{0}^{\pm}B\\[2mm]
&= \Proj^{\dag} c_{1}^{\pm}\Proj+ B^{\dag}K_{\Sig}^{\dag}c_{1}^{\pm}+ \Proj^{\dag}c_{1}^{\pm}K_{\Sig}B\mp B^{\dag}R_{-\infty}^{\dag}\mp \Proj^{\dag}R_{-\infty}B\\[2mm]
&= \Proj^{\dag} c_{1}^{\pm}\Proj+ (\one - \Proj)^{\dag}c_{1}^{\pm}+ \Proj^{\dag}c_{1}^{\pm}(\one -\Proj)\mp B^{\dag}R_{-\infty}^{\dag}\mp\Proj^{\dag}R_{-\infty}B\\[2mm]
&=c_{1}^{\pm}- c_{1\, {\rm reg}}^{\pm}.
\end{aligned}
\]
(3) follows from the fact that  $(\cdot| \tilde{\lambda}_{1\Sig}^{\pm}\cdot)_{V_{\rho_{1}}}= (\cdot| \lambda_{1\Sig}^{\pm}\cdot)_{V_{\rho_{1}}}$ on $\Ker K^{\dag}_{\Sig}$. 

Under the hypotheses of (4) $\lambda_{1\Sig}^{\pm}- \tilde{\lambda}_{1\Sig}$ is smoothing, hence  so is  $\lambda_{1}^{\pm}- \tilde{\lambda}_{1}^{\pm}$.  This completes the proof of the theorem.\qed

{ 
\begin{remark}If $B$ satisfies additionally $B K_\Sig=\one$ (as will be the case in Sect. \ref{sec4}), then $\tilde{c}^\pm_1$ satisfies a stronger version of gauge-invariance, namely
\beq
\tilde c_1^{\pm} K_\Sig= K_\Sig c_0^{\pm}.
\eeq
Such property is needed to construct two-point functions in the BRST framework, cf. \cite{hollands2} for discussion in the case of Yang-Mills fields with flat background connection and \cite{WZ} for generalization and computations on the Cauchy surface.   
\end{remark}
}

\subsection{Reduction to ultra-static spacetimes by deformation}\label{ss:fnw} A well-known argument due to Fulling, Narcowich and Wald \cite{FNW} allows one to reduce the construction of Hadamard states for the Klein-Gordon equation to the special case of an ultra-static spacetime, and an extension of this method can be used for the Maxwell equations \cite{FP} and Yang-Mills linearized around $\bA=0$ \cite{hollands2}.

Let us first recall the FNW deformation argument for ordinary field theory: let $g,g'$ be Lorentzian metrics on $M$ such that $(M, g)$ and $(M, g')$ are globally hyperbolic and $\Sigma\subset M$ a Cauchy surface for $(M, g)$ and $(M, g')$.  Assume that $g= g'$ on a causal neighborhood $O(\Sigma)$ of $\Sigma$. Assume also that $D, D'\in {\rm Diff}^{m}(M; V)$ are normally hyperbolic operators  satisfying the assumptions in Subsect. \ref{ssec:classical} such that $D= D'$ on $O(\Sigma)$.  Then by the time-slice property and H\"{o}rmander's propagation of singularities theorem, the restriction of a Hadamard state for $D'$ to $O(\Sigma)$ yields a Hadamard state for $D$.

In the subsidiary condition formalism, one has to assume the existence of operators $P, K$, $P', K'$ as in Hypothesis \ref{as:subsidiary} such that $P= P'$, $K= K'$ on $O(\Sigma)$. The same argument using the gauge invariant version of the time slice property, i.e. Prop. \ref{prop:timeslice}, shows that the restriction of a Hadamard state for $(P,' K')$ to $O(\Sigma)$ yields a Hadamard state for $(P, K)$. 

In the ordinary case one fixes an ultra-static metric $g_{\rm us}$, a normally hyperbolic operator $D_{\rm us}$, an  interpolating metric $g'$ sharing a Cauchy surface $\Sigma$ with $g$ and a Cauchy surface $\Sigma_{\rm us}$ with $g_{\rm us}$, and finally a normally hyperbolic operator $D'$ with $D'= D$ near $O(\Sigma)$ and $D'= D_{\rm us}$ near $O(\Sigma_{\rm us})$. Applying twice the above argument, one obtains a one-to-one correspondence between Hadamard states for $D$ and Hadamard states for $D_{\rm us}$. The construction of Hadamard states for $D_{\rm us}$ is easier since {  $D_{\rm us}$ can be chosen in such way that its coefficients are independent on the time coordinate and then it admits a natural vacuum state which can be shown to be Hadamard.}

\subsubsection{Deformation argument for Yang-Mills}

In the subsidiary condition formalism, it is not obvious how to find interpolating operators $P', K'$ equal to $P, K$ near $O(\Sigma)$ and satisfying Hypothesis \ref{as:subsidiary} {\em globally} on $M$. 
Moreover even  if $(M, g')$ is ultra-static on some $O(\Sigma_{\rm us})$, this does not imply in general that $P', K'$ will be independent on the time coordinate  on $O(\Sigma_{\rm us})$. 

 For linearized Yang-Mills equations, it is possible to find interpolating operators $P', K'$ if we can find a $1-$form $\wbar{A}'$ on $(M, g')$ such that $\wbar{\delta}'\wbar{F}'=0$ and $\wbar{A}'= \wbar{A}$ near $O(\Sigma)$.  This will follow in turn from a  result of {\em global existence} of {\em smooth} solutions of the non-linear Yang-Mills equation, on the spacetime $(M, g')$, with smooth Cauchy data on $\Sigma$. 
 
Assuming this problem is solved,   there is another issue that we need to consider:
 
 by the deformation argument explained above, to prove the existence of Hadamard states for the linearized Yang-Mills equations on $(M, g)$,  we may assume that $(M, g)$ is ultra-static, i.e. $g= g_{\rm us }= -dt^{2}+ h_{ij}(x) dx^{i}dx^{j}$ on $M = \rr_{t}\times \Sigma_{x}$. 

Recall that we assume that $\Sigma$ is either a compact manifold or $\Sigma= \rr^{d}$.  The Riemannian metric $h_{ij}(x)dx^{i}dx^{j}$ on $\Sigma$ can be chosen as we wish, in particular if $\Sigma= \rr^{d}$ is not compact, we may assume that it satisfies Hypothesis \ref{as:metric}.
However if $\Sigma= \rr^{d}$, we need also to ensure Hypothesis \ref{as:background} on the {  (non necessarily time-independent)} background solution $\wbar{A}_{\rm us}$ (recall that this is a decay condition at spatial infinity).  Moreover we have to assume that $\wbar{A}_{\rm us}$ is in the {\em temporal gauge}, i.e. that $\wbar{A}_{{\rm us}, t}\equiv 0$. 

 If our model problem is obtained from the above deformation argument, $\wbar{A}_{\rm us}$ is obtained by solving  two Cauchy problems for  non-linear Yang-Mills equations:
 
  in the first step one has to solve  it on $(M, g')$, from a Cauchy surface $\Sigma$ in the future (where $g'= g$) to a Cauchy surface $\Sigma_{\rm us}$ in the past (where $g'= g_{\rm us}$).  In a second step one has to solve it globally on $(M, g_{\rm us})$ with the Cauchy data on $\Sigma_{\rm us}$ obtained in the first step. 
  
  Clearly if the Cauchy problem for the Yang-Mills equation (\ref{eq:nlYM})  on a globally hyperbolic spacetime $(M,g)$ can be globally solved in the space of smooth {\em space-compact} solutions, then all the intermediate background fields $\wbar{A}'$ and $\wbar{A}_{\rm us}$ will be space compact, and hence $\wbar{A}_{\rm us}$ will satisfy the decay condition (\ref{as:background}). As a consequence the FNW deformation argument can be applied, giving the existence of Hadamard states if the background field $\wbar{A}$ is space-compact.

 Fortunately it is not very difficult to deduce the result we need in dimensions lower than $4$, from the existing literature, in particular from the work  by 
 Chru\'sciel \& Shatah \cite[Thm. 1.1]{CS}.  
 The proof of the following proposition will be sketched in Appendix \ref{ss:global-existence}.
\begin{proposition}\label{prop:idiotic}
 Assume that $\dim M \leq 4$ and $(M, g)$ is globally hyperbolic. Let $\wbar{A}\in \cE^{1}_{\rm sc}(M; \mathfrak{g})$  a local solution of the Yang-Mills equation (\ref{eq:nlYM}) near some Cauchy surface $\Sigma$. Then there exists  $\wbar{A}'\in \cE^{1}_{\rm sc}(M; \mathfrak{g})$ such that:
 \ben
 \item $\wbar{A}'\sim \wbar{A}$ near $\Sigma$, where $\sim$ denotes gauge equivalence,
 \item $\wbar{A}'_{t}\equiv 0$, ie $\bar{A}'$ is in the temporal gauge,
 \item $\wbar{A}'$ is a global solution of (\ref{eq:nlYM}).
 \een
\end{proposition}

Combining Prop. \ref{prop:idiotic} with the above discussion, we see that Thm. \ref{maintheo1} follows from Thm. \ref{maintheo2}. 
\section{Vector and scalar Klein-Gordon equations on ultra-static spacetimes}\init\label{sec1}\init 

In this section we consider a general framework containing the operators $D_{0}=  \bdelta \bardel$  and $D_1=\bardel\bdelta + \bdelta \bardel + \bF\tnI\,$ associated to the Yang-Mills equation (defined in Subsect. \ref{lnym}) on ultra-static spacetimes. 
This will provide a basis for the construction of the parametrix in Sect. \ref{sec2}. 

\subsection{Preparations}\label{sec1.1}
The operator $D_{1}$, (resp. $D_{0}$) acts on  $\cE^{1}(M)\otimes \fg$ (resp. $\cE^{0}(M)\otimes\fg$). Since by Hypothesis \ref{as:spacetime} $M= \rr_{t}\times \Sigma$ is parallelizable, we fix  a global trivialization of $T^{*}M$ and identify $\cE^{1}(M)\otimes \fg$ (resp. $\cE^{0}(M)\otimes \fg$) with $C^{\infty}(M; W)$
for\begin{equation}
\label{en.-1}
W\defeq  V\otimes \fg \quad \mbox{and \ } V= \cc^{1+d} \ (\mbox{resp. \ } V= \cc).
\end{equation}
We refer to the two  cases as the {\em vector case} (resp. {\em scalar case}).
 
The background metric is  ultra-static:
\[
g= -dt^{2}+ h_{ij}(x)dx^{i}dx^{j},
\]
on $M= \rr\times\Sigma$, with either $\Sigma=\rr^{d}$ or $\Sigma$ a compact manifold.
We obtain a splitting 
\begin{equation}
\label{en.4c}
V= V_{t}\oplus V_{\Sig}, \ W_{t, \Sig}\defeq  V_{t, \Sig}\otimes\fg, \ W= W_{t}\oplus W_{\Sig},
\end{equation} by writing a $1-$form as $A= A_{t}dt+ A_{\Sig}dx$, and we identify $V_{t}$ with $\cc$. 
In the scalar case we take $V_{t}= \{0\}$, $V_{\Sig}= \cc$. Defining $J\in L(V)$ by  \begin{equation}
\label{e1.1.00}
J\defeq  \mat{-1}{0}{0}{\one}  \mbox{ if }V= \cc^{1+d}, \ J\defeq 1 \mbox{ if }V= \cc,
\end{equation}
we see that $V_{t}= \Ker(J+\one), \ V_{\Sig}=  \Ker(J-\one)$.

We denote by $(\cdot| \cdot)$ the canonical positive definite scalar product on $\coinf(M; W)$.  In the scalar case we set:
\[
(u|v)\defeq\int_{M}\overline{u}(t, x)\killing v(t, x)|h|^{\12}dtdx,
\]
in the vector case we set:
\[
{ (}u|v)\defeq  \int_{M}\overline{u}(t,x)Jg^{-1}(x)\otimes \killing v(t,x)|h|^{\12}dtdx,
\]
To avoid introducing too much notation, we also denote by $(\cdot| \cdot)$ the analogous scalar product on $\coinf(\Sigma; W)$, i.e.:
\beq\label{e-1.1.0}
\begin{aligned}
(u|v)\defeq & \int_{\Sigma}\overline{u}(x)\otimes \killing v(x)|h|^{\12}dx,\hbox{ resp.}\\[2mm]
(u|v)\defeq&  \int_{\Sigma}\overline{u}(x)Jg^{-1}(x)\otimes \killing v(x)|h|^{\12}dx,
\end{aligned}
\eeq
which is also positive definite.

%\subsubsection{Connection coefficients.}
We denote by $\Gamma_{a}\in \cinf(\Sigma; L(V))$ the coefficients of the Levi-Civita connection for $(M,g)$. 
Since this connection  is metric  for $g^{-1}$, we have:
\beq\label{e-2.1}
\p_{a}g^{-1}= \Gamma_{a}^{*} g^{-1}+ g^{-1}\Gamma_{a}.
\eeq
Since the metric is ultra-static we have moreover $\Gamma_{0}=0$,
and $\Gamma_{i}$ are  the  Levi-Civita connection  coefficients for $(\Sigma; h_{ij}dx^{i}dx^{j})$. 

We denote by $M_{a}= {\rm ad}_{\bar{A}_{a}}\in \cinf(\rr\times \Sigma; L(\fg))$
the connection coefficients for the algebra degrees of freedom. They  can also depend  on $x^{0}$ because the background Yang-Mills solution is in general time-dependent. We have of course $M_{a}^{*}\killing + \killing M_{a}=0$.

In the vector case we set
\[
T_{a}\defeq  \Gamma_{a}\otimes\one_{\fg}+ \one_{V}\otimes M_{a}\in\cinf(\rr\times \Sigma;  L(W)),
\]
and $T_{a}:= M_{a}$ in the scalar case. 

  In the vector case we also fix a map $\rho\in C^{\infty}(\rr\times\Sigma; L(W))$ representing  the term $F\llcorner$ such that
\[
\rho^{*}(g^{-1}\otimes \killing)= (g^{-1}\otimes \killing)\rho,
\]
in the scalar case we take $\rho= 0$. We set:
%\subsubsection{Operators.}
\beq\label{e-2.2biso}
\nabla_{a}^{T}\defeq  \p_{a}+ T_{a}, \ D\defeq - |g|^{-\12}\nabla_{a}^{T} |g|^{\12}g^{ab}\nabla_{b}^{T}+ \rho.
\eeq
%\subsubsection{Conserved charge.}

The {\em charge} $q$ defined in (\ref{defo}) equals:
\beq\label{def-de-q}
\overline{\zeta} q\zeta\defeq  \int_{\{t\}\times\Sigma}\overline{\i^{-1}\nabla_{0}^{T}\zeta}\cdot g^{-1}\otimes \killing\  \zeta+ \overline{\zeta}\cdot g^{-1}\otimes \killing\ \i^{-1}\nabla_{0}^{T}\zeta|h|^{\12}dx,
\eeq
in the vector case and 
\beq\label{def-de-qzero}
\overline{\zeta} q\zeta\defeq  \int_{\{t\}\times\Sigma}\overline{\i^{-1}\nabla_{0}^{T}\zeta}\cdot\killing\  \zeta+ \overline{\zeta}\cdot  \killing\ \i^{-1}\nabla_{0}^{T}\zeta|h|^{\12}dx,
\eeq
in the scalar case.

\subsection{Temporal gauge}

The temporal gauge is $\wbar{A}_{0}(t, x)\equiv 0$, which since $M_{a}= {\rm ad}_{\wbar{A}_{a}}$ implies that $T_{0}=0$, i.e. $\nabla^{T}_{0}= \p_{t}$. It is well known that one can always assume that one is in the temporal gauge, cf. Appendix \ref{ss:tempgauge}.

In this case the operator $D$ takes the form:
\beq\label{en.0}
D= \p_{t}^{2}+ a(t, x, D_{x}), \quad a(t, x, D_{x})= - |h|^{-\12}\nabla^{T}_{i}h^{ij}(x)|h|^{\12}\nabla^{T}_{j}+ \rho(t, x).
\eeq
Denoting by $a^{*}$ the formal adjoint of $a$ for the positive  scalar product $(\cdot|\cdot)$ , we deduce from the fact that $q$ defined in (\ref{def-de-q}), (\ref{def-de-qzero}) is independent on $t$ that:
\beq\label{en.1}
a^{*}J= Ja,
\eeq
for $J$ defined in (\ref{e1.1.00}). In other terms, $D$ is self-adjoint for $(\cdot|\cdot)_V\defeq(\cdot|J\cdot)$. In the next sections we will use primarily the product $(\cdot|\cdot)$.
\subsection{Cauchy problem}
The standard Cauchy problem for  the operator $D$ is
\begin{equation}
\label{e3.9}
\left\{
\begin{array}{l}
D\zeta=0, \\[2mm]
\rho \zeta=f,
\end{array}
\right.
\end{equation}
for $\rho \zeta(x)= (\zeta(0,x), \i^{-1}\p_{t}\zeta(0,x))$, $f= (f^{0}, f^{1})$.  We denote by $\zeta= U f$ the solution of (\ref{e3.9}). 
We will denote by $f^{i}_{t}$, $f^{i}_{\Sig}$, $i=0, 1$ the time and space components of $f^{i}$, according to the decomposition
$ W= W_{t}\oplus W_{\Sig}$.

Denoting still by $q$ the charge expressed in terms of Cauchy data we obtain that in the vector case:
\begin{equation}
\label{en.4b}
\begin{aligned}
\bar{f}q f&= (f^{1}| J f^{0})+ (f^{0}| J f^{1})\\[2mm]
&=(f^{1}_{\Sig}| f^{0}_{\Sig})+ (f^{0}_{\Sig}| f^{1}_{\Sig})- (f^{1}_{t}| f^{0}_{t})- (f^{0}_{t}| f^{1}_{t}).
\end{aligned}
\end{equation}
In the first line above the positive scalar product $(\cdot | \cdot)$  is defined in (\ref{e-1.1.0}),  the positive scalar products in the second line are equal to
\begin{equation}
\label{en.1.2}
(f_{\Sig}| f_{\Sig})\defeq  \int_{\Sigma}\overline{f_{\Sig}} h^{-1}\otimes \killing f_{\Sig}|h|^{\12}dx,\
(f_{t}|f_{t})\defeq  \int_{\Sigma}\overline{f_{t}}\cdot\killing f_{t}|h|^{\12}dx.
\end{equation}
In the scalar case  we have instead
\[
\bar{f} q f= (f^{1}| f^{0})+ (f^{0}| f^{1}), \mbox{ for } (u|v)= \int_{\Sigma}\overline{u}\cdot\killing v |h|^{\12}dx.
\]

\subsection{Adapted Cauchy data}
The above choice of Cauchy data is the usual one for an operator obtained from a metric connection.  In the vector  case, however, it will often be more convenient to work with the adapted Cauchy data $\rho^{\rm F}_i$ defined in Sect. \ref{ss:adapted}. In this subsection we discuss the transition from one choice of Cauchy data to the other.

\subsubsection{Identifications}\label{sss:ident}
The space  $\kE^{1}_{\rm sc}(M)\otimes\fg$ equals $\cinf_{\sc}(M; W)$.

For $A\in \kE^{1}_{\rm sc}(M)\otimes\fg$ we set:
\beq\label{en.3}
A\eqdef  A_{t}dt+ A_{\Sig},
\eeq
for $ A_{t}\in \cinf(\rr, \kE^{0}_{\rm c}(\Sigma)\otimes\fg)$, $ A_{\Sig}\in  \cinf(\rr, \kE^{1}_{\rm c}(\Sigma)\otimes\fg)$, which corresponds to the decomposition
$ \zeta= \zeta_{t}\oplus \zeta_{\Sig}$, using (\ref{en.4c}). 
We will use the corresponding identifications for restrictions to $\Sigma$, i.e.:
\begin{equation}
\label{en.7}
\coinf(\Sigma; W)\sim \coinf(\Sigma; W_{t})\oplus  \coinf(\Sigma; W_{\Sig}) \sim(\kE^{0}_{\rm c}(\Sigma)\otimes\fg)\oplus (\kE^{1}_{\rm c}(\Sigma)\otimes\fg).
\end{equation}

We have also corresponding decompositions for $2-$forms. Namely, if  $F\in \kE^{2}_{\rm sc}(M)\otimes\fg$ we set:
\beq\label{en.4d}
 F\eqdef   dt\wedge F_{t}+ F_{\Sig},
\eeq
for $ F_{t}\in \cinf(\rr, \kE^{1}_{\rm c}(\Sigma)\otimes\fg)$, $ F_{\Sig}\in  \cinf(\rr, \kE^{2}_{\rm c}(\Sigma)\otimes\fg)$.

We recall that $\wbar{A}\in \kE^{1}_{\rm sc}(M)\otimes\fg$ is the background connection, which we assume to be in the temporal gauge. We introduce the derivative and co-derivative on $\Sigma$:
\[
\begin{array}{rl}
&{\bds}\defeq  d_{\Sig} + \wbar{A}_{\Sig}\wedge\,\cdot\,: \ \kE^{p}_{\rm c}(\Sigma)\otimes\fg\to \kE^{p+1}_{\rm c}(\Sigma)\otimes\fg,\\[2mm]
&{\bdeltas}\defeq  {\bds}^{*}:  \ \kE^{p}_{\rm c}(\Sigma)\otimes\fg\to \kE^{p-1}_{\rm c}(\Sigma)\otimes\fg,
\end{array}
\]
and one has  ${\bds}{\bds}= \wbar{F}_{\Sig}\wedge\cdot\,$ using the notation in (\ref{en.4d}).
An easy computation using that  $\wbar{A}_{t}\equiv 0$  shows that:
\begin{equation}
\label{en.2}
\begin{aligned}
\wbar{d}u&=\p_{t}u dt + {\bds}u, \ u\in \kE^{0}_{\sc}(M)\otimes\fg,\\[2mm]
\wbar{d}A&=dt\wedge (\p_{t}A_{\Sig}- {\bds}A_{t})+ {\bds}A_{\Sig},\ \ A\in \kE^{1}_{\sc}(M)\otimes\fg,\\[2mm]
\wbar{\delta}A&=\p_{t}A_{t}+ {\bdeltas}A_{\Sig},\ A\in \kE^{1}_{\sc}(M)\otimes\fg,\\[2mm]
\wbar{\delta}F&=- ({\bdeltas}F_{t})dt+ \p_{t}F_{t}+\bar\delta_{\Sig}F_{\Sig}, \ \ F\in \kE^{2}_{\rm sc}(M)\otimes\fg.
\end{aligned}
\end{equation}
Using (\ref{en.2}), we see that 
\[
 \wbar{F}_{t}= \p_{t}\wbar{A}_{\Sig}, \ \ \wbar{F}_{\Sig}= {\bds}\wbar{A}_{\Sig},
\]
 and that the Yang-Mills equation $\wbar{\delta}\wbar{F}=0$ is equivalent to:
\begin{equation}
\label{en.3bidi}
{\bdeltas}\wbar{F}_{t}=0, \ \ \p_{t}\wbar{F}_{t}+ {\bdeltas}\wbar{F}_{\Sig}=0, 
\end{equation}
where of course (\ref{en.3bidi}) holds for all $t\in \rr$.

\subsubsection{Transition to adapted Cauchy data} The adapted Cauchy data were defined in Sect. \ref{ss:adapted}. Using (\ref{en.2}) we obtain the following relation between the standard Cauchy data $\rho_1$ and the adapted ones $\rho_1^{\rm F}$.

%We recall that if $i: \Sigma\to M$ is the standard embedding, then $i^{*}: \kE^{i}_{\rm sc}(M)\otimes\fg\to \kE^{i}_{\rm sc}(\Sig)\otimes\fg$. 

\begin{lemma}\label{n1.1}
 Let $R_{\rm F}\defeq  \rho^{\rm F}_1\circ \rho_{1}^{-1}$. Then:
 \ben
 \item
  \[
R_{\rm F}=\left(\begin{array}{cccc}
\one&0&0&0\\
0&\one&0&0\\
0&-\i {\bdeltas}&\one&0\\
\i {\bds}&0&0&\one
\end{array}\right), \ R_{\rm F}^{-1}=\left(\begin{array}{cccc}
\one&0&0&0\\
0&\one&0&0\\
0&\i {\bdeltas}&\one&0\\
-\i {\bds}&0&0&\one
\end{array}\right).
\]
\item
We have:
\[
 R_{\rm F}^{*}qR_{\rm F}= q,
 \]
 i.e. $R_{\rm F}$ is symplectic.
\een
\end{lemma}

Note that the precise form of $R_{\rm F}$ relies on the assumption that the spacetime is ultra-static. It enjoys some good properties particular to that case, like for instance $JR_{\rm F}=R_{\rm F}J$, which is used implicitly in some computations in Sect. \ref{sec4}. 

%\subsection{The operators $K_{\Sig}$ and $K^{\dag}_{\Sig}$}

% We will denote by $a$ the operator: 
% \[
%a: \kE^{0}_{\rm c}(\Sigma)\otimes\fg\ni u\mapsto \wbar{F}_{t}\wedge u= [\wbar{F}_{t}, u]\in \kE^{1}_{\rm %c}(\Sigma)\otimes\fg.
%\]
%This identity can also be deduced from ${\bdeltas}\wbar{F}_{t}=0$, which holds  by (\ref{en.3bidi}).

\section{Parametrices for the Cauchy problem}\init\label{sec2}\init
In this section we give a construction of the parametrix for the Cauchy problem (\ref{e3.9}), by adapting arguments in \cite{GW} to vector-valued Klein-Gordon equations.  
In the rest of the paper,  the principal part of the operator $a(t, x, D_{x})$ below is time-independent, since the background metric is ultra-static. In this section however we treat the more general case where the principal part is time-dependent, which corresponds to the case when  the riemannian metric $h_{ij}(t, x)dx^{i}dx^{j}$ is time-dependent.
 The completely general situation of a metric $-\beta(t, x)dt^{2}+ h_{ij}(t, x)dx^{i}dx^{j}$ could be treated as well by  our methods.

The construction of a parametrix for the Cauchy problem given later on will rely heavily on pseudodifferential calculus. For the necessary basic facts and definitions we refer the reader to Appendix \ref{sec1opd}.
%\subsection{Notations}
%We will assume that $\Sigma$ is either compact or equal $\rr^d$.
%
%Let $V$, $V_{1}$, $V_{2}$ be finite dimensional hermitian spaces. Then the standard (H\"ormander) classes of pseudodifferential operators of order $p\in\rr$ are denoted $\Psi^{p}(\Sigma; V_{1}, V_{2})$, $\Psi^{p}(\Sigma; V)\defeq\Psi^{p}(\rr^{d}; V,V)$ (sometimes they will be abbreviated $\Psi^p$). Furthermore,
%\[\textstyle
%\Psi^{\infty}(\Sigma; V_{1}, V_{2})\defeq \bigcup_{p\in\rr}\Psi^{p}(\Sigma; V_{1}, V_{2}), \quad \Psi^{-\infty}(\Sigma; V_{1}, V_{2})\defeq \bigcap_{p\in\rr}\Psi^{p}(\Sigma; V_{1}, V_{2}).
%\]
%The space of {\em scalar }pseudodifferential operators, i.e. of the form $a\otimes \one_{V}$ for $a\in \Psi^{p}(\Sigma)$, will be denoted by $\Psi^{p}_{\rm scal}(\Sigma; V)\subset \Psi^{p}(\Sigma; V)$. We will often identify $a\in \Psi^{p}(\Sigma)$ with $a\otimes \one_{V}\in \Psi^{p}_{\rm scal}(\Sigma; V)$ and denote the later operator also by $a$ if there is no risk of confusion.

% We denote  $D_{x_j}\defeq \i^{-1}\p_{x_{j}}$.

\subsection{Setup and notation}
We consider an operator 
 \[
D= \p_{t}^{2}+ a(t, x, D_{x}), \ a(t, x, D_{x})= - |h|^{-\12}\nabla^{T}_{i}h^{ij}(t, x)|h|^{\12}\nabla^{T}_{j}+\rho(t, x),
\]
where $T$, $\rho$ etc. are as in Sect. \ref{sec1}. 

We assume that  the metric $h_{ij}(t, x)dx^{i}dx^{j}$ satisfies Hypothesis \ref{as:metric}, locally uniformly in $t$, and that the background Yang-Mills solution $\wbar{A}$ satisfies Hypothesis \ref{as:background} {\it ii)}.

In the sequel we denote  $a(t, x, D_{x})$ simply by $a(t)\in \cinf(\rr, \Psi^{2}(\Sigma; W))$ (see Appendix \ref{sec1opd} for the definition of pseudodifferential operators classes $\Psi^m$, $\Psi^m_{\rm scal}$).  One has:
\beq\label{e1.0}
 \sigma_{\rm pr}(a(t))=k_{i}h^{ij}(t, x)k_{j}\otimes\one_{W}, 
 \eeq
 hence $a(t)$ has a scalar principal part.
For $V$ a finite dimensional vector space, we set
 \beq\label{eq:def-de-sob}
\cH(\Sigma; V)\defeq \bigcap_{m\in \zz}H^{m}(\Sigma; V), \ \cH'(\Sigma; V)\defeq \bigcup_{m\in \zz}H^{m}(\Sigma; V),
\eeq
equipped with their natural topologies,  where $H^{m}(\Sigma; V)$ are the Sobolev spaces, which are canonically defined since $\Sigma$
is equal either to $\rr^{d}$ or to a compact manifold. We set also
\[
L^{2}(\Sigma; W)= H^{0}(\Sigma; W),
\]
where in the situation considered in Sect. \ref{sec1}, $L^{2}(\Sigma; W)$ is equipped with the scalar product (\ref{e-1.1.0}).
\subsection{Some classes of pseudodifferential operators}\label{sec1.1b}
  In this subsection we introduce some special classes of pseudodifferential operators which will play an  important role later on.
\subsubsection{High momenta localization}
 A first problem that we have to face is the need to construct {\em exact} inverses to some elliptic operators, not only inverses modulo smoothing errors.   Let us explain the well-known way to solve this problem on a simple scalar example:

if $r\in\Psi^{-1}(\rr^{d})$, the operator $\one + r$ is not necessarily invertible on $L^{2}(\rr^{d})$. However if we fix some cutoff function 
$\chi\in\cinf(\rr)$, with $\chi(s)\equiv 0$ for $|s|< 1$, $\chi(s)\equiv 1$ for $|s|\geq 2$ and set  
\begin{equation}
\label{e-1.1}
r_{R}(x, k)\defeq  \chi(R^{-1}|k|)r(x, k), \ r_{R}\defeq  r_{R}(x, D_{x}),
\end{equation}
 then $r-r_{R}\in \Psi^{-\infty}(\rr^{d})$ and 
$r_{R}\to 0$ in $\Psi^{0}(\rr^{d})$ as $R\to+\infty$. It follows that  \beq\label{e-1.3}
\one + r_{R}\hbox{  is invertible on }L^{2}(\rr^{d})\hbox{ for }R\gg 1, \ (\one +r_{R})^{-1}\in \one + \Psi^{-1}(\rr^{d}).
\eeq 
We formalize this method by introducing the following definition.
\begin{definition}\label{defo1}
 Let $V_{1}, V_{2}$ be finite dimensional hermitian spaces. We denote by $\Psi^{p}_{\rm as}(\Sigma; V_{1}, V_{2})$ the space of $R-$dependent pseudodifferential operators $c_{R}$ such that:
 
i) $c_{R}$ is uniformly bounded in $\Psi^{p}(\Sigma; V_{1}, V_{2})$,

ii)    $c_{R}\to 0$ in $\Psi^{p+\varepsilon}(\Sigma;V_{1}, V_{2})$ when $R\to +\infty$ for some (and hence for all) $\varepsilon>0$.

The space $\Psi^{p}_{\rm as}(\Sigma; V,V)$ will be simply denoted by $\Psi^{p}_{\rm as}(\Sigma; V)$.
\end{definition}
We now collect some easy properties of the above classes (the meaning of statement (2) below is explained in the proof). 
\begin{lemma}\label{lem1}
 \ben
 \item $\big(\Psi^{p}_{\rm as}(\Sigma; V_{1}, V_{2})\big)^{*}=\Psi^{p}_{\rm as}(\Sigma; V_{2}, V_{1})$,
 \item  $\Psi^{p}(\Sigma; V_{1}, V_{2})\subset \Psi^{p}_{\rm as}(\Sigma; V_{1}, V_{2})+ \Psi^{-\infty}(\Sigma; V_{1}, V_{2})$,
 \item let $c_{R}\in \Psi^{-\varepsilon}_{\rm as}(\Sigma; V)$ for $\varepsilon>0$ and let $\alpha\in \rr$. Then for $R\geq R_{0}$ we have:
 \[
(\one + c_{R})^{\alpha}\in \one + \Psi^{-\varepsilon}_{\rm as}(\Sigma; V).
\]
 \een
\end{lemma}
\proof (1) follows from the definition. If $c\in S^{p}(\Sigma; V_{1}, V_{2})$ we set $c_{R}(x, k)= \chi(R^{-1}|k|)c(x, k)$, for $\chi$ as in (\ref{e-1.1}), and obtain that $c_{R}(x, D_{x})\in\Psi^{p}_{\rm as}(\Sigma; V_{1}, V_{2})$, $c(x, D_{x})- c_{R}(x, D_{x})\in\Psi^{-\infty}(\Sigma; V_{1}, V_{2})$, which proves (2). Let us now prove (3).  We obtain that $c_{R}\to 0$ in $\Psi^{0}(\Sigma; V)$, hence in $B(L^{2}(\Sigma; V))$. It follows that for $R\geq R_{0}$ $(\one + c_{R})^{\alpha}$ is well defined by the holomorphic functional calculus of bounded operators.  The map $c_{R}\mapsto (\one + c_{R})^{\alpha}- \one$ is then continuous on $\Psi^{-\varepsilon}(\Sigma; V)$ for all $\varepsilon>0$, from which we deduce that $(\one+ c_{R})^{\alpha}\in \one + \Psi^{-\varepsilon}_{\rm as}(\Sigma; V)$. \qeds

\subsubsection{Infrared  cutoffs}
Some operators will need to contain additional low energy (infrared) cutoffs, defined using some selfadjoint operators.
  These cutoffs will play an important role in Sect. \ref{sec4}.

 In the rest of the paper we denote by $\chi_{\ls}\, , \chi_{\gs}\in \cinf(\rr)$ two cutoff functions with 
\begin{equation}
\label{cutoffs}
\chi_{\ls}+ \chi_{\gs}=1, \quad \supp \chi_{\gs}\subset ]-\infty,-1]\cup[1, +\infty[, \quad \supp\chi_{\ls}\subset [-2, 2]. 
\eeq

\begin{definition}
Let $V_{1}, V_{2}$ be finite dimensional hermitian spaces and $h_{i}\in \Diff^{2}(\Sigma; V_{i})$ be  elliptic, selfadjoint and bounded from below.  We denote by $\Psi^{p}_{\rm reg}(\Sigma; V_{1}, V_{2})$ 
 the space of $R-$dependent pseudodifferential operators $c_{R}$ such that:

i) $c_{R}\in \Psi^{p}_{\rm as}(\Sigma; V_{1}, V_{2})$,

ii) $c_{R}= \chi_{\gs}(h_{2})c_{R}\chi_{\gs}(h_{1})$ for  some $\chi_{\gs}$ as in (\ref{cutoffs}).

The space $\Psi^{p}_{\rm reg}(\Sigma; V, V)$ will be simply denoted by $\Psi^{p}_{\rm reg}(\Sigma;V)$.
\end{definition}
\begin{lemma}\label{lem2}
 \ben
 \item $\big(\Psi^{p}_{\rm reg}(\Sigma; V_{1}, V_{2})\big)^{*}=\Psi^{p}_{\rm reg}(\Sigma;V_{2}, V_{1})$,
 \item  $\Psi^{p}(\Sigma;V_{1}, V_{2})\subset \Psi^{p}_{\rm reg}(\Sigma;V_{1}, V_{2})+ \Psi^{-\infty}(\Sigma;V_{1}, V_{2})$,
 \item let $c_{R}\in \Psi^{-\varepsilon}_{\rm reg}(\Sigma; V)$ for $\varepsilon>0$ and let $\alpha\in \rr$. Then for $R\geq R_{0}$ we have: 
 \[
(\one + c_{R})^{\alpha}\in \one + \Psi^{-\varepsilon}_{\rm reg}(\Sigma; V).
\]
 \een
\end{lemma}

\proof
(1) follows from the definition.   (2) follows from Lemma \ref{lem1} (2) and the fact that $\chi_{\ls}(h_{i})\in \Psi^{-\infty}(\Sigma; V_{i})$, since $h_{i}$ is elliptic and bounded below.  Next  $(1+ c_{R})^{\alpha}$ is well defined for $R$ large enough by Lemma \ref{lem1}.   For $f(\lambda)= (1+\lambda)^{\alpha}$ we have (denoting $\chi_{\gs}(h)$ simply by $ \chi_{\gs}$):
\[
f(c_{R})= f(\chi_{\gs} c_{R}\chi_{\gs})= \one + f'(0)\chi_{\gs} c_{R} \chi_{\gs} +  \chi_{\gs} c_{R} \chi_{\gs} g(\chi_{\gs} c_{R} \chi_{\gs})\chi_{\gs} c_{R} \chi_{\gs}, 
\]
for $g(\lambda)= \lambda^{-2}(f(\lambda)-1- f'(0)\lambda)$. Since $g$ is analytic near $0$, we obtain that $g(\chi_{\gs} c_{R} \chi_{\gs})\in \Psi^{0}(\Sigma; V)$ and moreover that $g(\chi_{\gs} c_{R}\chi_{\gs})$ is uniformly bounded in $\Psi^{0}(\Sigma; V)$.  This implies (3). \qeds

We will use the above operators classes for $V= W_{t}$, $W_{\Sig}$, $W$ or $W\oplus W$.
We start by defining the operators $h$ that will be used in our case. 

\begin{definition}\label{n.def1}
 We set:
 \[
 \begin{aligned}
h_{t}&\defeq  {\bdeltas}{\bds}: \ \kE^{0}_{\rm c}(\Sigma)\otimes\fg\to \kE^{0}_{\rm c}(\Sigma)\otimes\fg,\\[2mm]
h_{\Sig}&\defeq {\bdeltas}{\bds}+ {\bds}{\bdeltas}+ \wbar{F}_{\Sig}\tnI \cdot: \ \kE^{1}_{\rm c}(\Sigma)\otimes\fg\to \kE^{1}_{\rm c}(\Sigma)\otimes\fg,
\end{aligned}
\]
and denote still by $h_{t}$, $h_{\Sig}$ their selfadjoint extensions, with domains $H^{2}(\Sigma; W_{t})$, $H^{2}(\Sigma; W_{\Sig})$. We set:
\[
h\defeq  h_{t}\oplus h_{\Sig}\hbox{ acting on }\  L^{2}(\Sigma; W).
\]
\end{definition}
{  Note that from Hypothesis \ref{as:metric} we obtain that $h_{t}$, resp. $h_{\Sig}$  belong to $\Psi^{2}(\Sigma; V_{t})$, resp. $\Psi^{2}(\Sigma;V_{\Sig})$ with principal symbol equal to $h^{ij}(0, x)k_{i}k_{j}$. It is well-known that this implies that their closures  are selfadjoint with domains   equal to $H^{2}(\Sigma; W_{t})$,  resp. $H^{2}(\Sigma; W_{\Sig})$.  }

 We equip then the spaces $W_{t}$, $W_{\Sig}$, $W$ and $W\oplus W$ with the elliptic operators $h_{t}$, $h_{\Sig}$, $h$ and $h\oplus h$ and define the various spaces $\Psi^{p}_{\rm reg}$ using the above operators.

Finally we choose a number $C\gg 1$ such that $h+C\one\geq \one$ and set:
\beq\label{defdeepsi}
\epsilon\defeq  (h+C\one)^{\12}= \epsilon_{t}\oplus \epsilon_{\Sig},
\eeq
where $\epsilon_{t}\defeq (h_{t}+C\one)^{\12}$, $\epsilon_{\Sig}\defeq (h_{\Sig}+C\one)^{\12}$. Let us collect some useful properties of  the above operators. 
\begin{lemma}\label{ln.2}
\ben
\item  $h\in {\rm Diff}^{2}(\Sigma; W)$  is an elliptic differential operator with principal symbol
\[
\sigma_{\rm pr}(h)(x, k)= k_{i}h^{ij}(0, x)k_{j}\otimes \one_{W}. 
\]
\item $\epsilon\in \Psi^{1}(\Sigma; W)$ is an elliptic pseudodifferential operator with principal symbol:
\[
\sigma_{\rm pr}(\epsilon)(x, k)= (k_{i}h^{ij}(0, x)k_{j})^{\12}\otimes \one_{W}. 
\]
\item  \[
\begin{array}{rl}
i)& h=h^{*}, \ \epsilon= \epsilon^{*}, \ [h, J]=[\epsilon, J]=0,\\[2mm]
ii)&h_{\Sig}\bards=\bards h_{t}+\bardels \wbar{F}_{\Sig}\wedge\cdot\, , \ \bardels h_{\Sig}= h_{t}\bardels+ \bardels \wbar{F}_{\Sig}\lrcorner \cdot\,.
\end{array}
\]
\een
\end{lemma}
\proof
(1)  and (3) {\it i)} are straightfoward.  (2) follows from Prop. \ref{1.1opd}.  (3) {\it ii)} follows from the Riemannian version of the computations at the end of Subsect. \ref{lnym}.\qed

\subsection{Construction of generators}
In this subsection we construct the two generators for the parametrix of the Cauchy problem, by modifying arguments from \cite{GW}.

We first introduce a convenient family $\rr\ni t\mapsto \epsilon(t)\in \cinf(\rr,\Psi^{1}(\Sigma; W))$ with the properties below. The  operators $\epsilon(t)$ will serve as elliptic `weight' operators. 
\begin{equation}
\label{toto.1}
\begin{array}{rl}
i)&\sigma_{\rm pr}(\epsilon(t))= (k_{i}h^{ij}(t, x)k_{j})^{\12}\otimes \one_{W},\\[2mm]
ii)&\epsilon(t)\hbox{ is selfadjoint on } L^{2}(\Sigma; W)\hbox{ with domain }H^{1}(\Sigma; W),\\[2mm]
iii)&\epsilon(t)\geq \one,  \ \epsilon(t)J = J \epsilon(t), \ \epsilon(0)= \epsilon,
\end{array}
\end{equation}
where $\epsilon$ is defined in \eqref{defdeepsi}. It is easy to construct such a family $\epsilon(t)$, one way being to  introduce the operator $h(t)$ as in Def. \ref{n.def1} using the metric $h^{ij}(t, x)dx^{i}dx^{j}$ and connection coefficients $T_{a}(t,x)$ at time $t$ instead of at time $0$, and to set $\epsilon(t)= (h(t)+ C(t))^{\12}$ for some $C(t)\gg 1$.
\begin{proposition}\label{p1.1}
 There exists  for $R\geq 1$ a family  $\rr\ni t\mapsto b_{R}(t)$ such that:
 \[
\begin{array}{rl}
ia)&b_{R}(t)= \epsilon(t)+ \tpdo{0}{W},\\[2mm]
ib)&\i \p_{t}b_{R}(t)- b_{R}^{2}(t)+ a(t)=r_{-\infty}(t)\in \tpdo{-\infty}{W},\\[2mm]
ic)& \begin{cases}b_{R}(t)+ J b^{*}_{R}(t)J: H^{1}(\Sigma; W)\to L^{2}(\Sigma; W)\hbox{ is invertible},\\[2mm]
 b_{R}(t)+ J b^{*}_{R}(t)J= \epsilon(t)^{\12}(2\one + \tpdo{-1}{W})\epsilon(t)^{\12},
 \end{cases}\\[5mm]
id)&(b_{R}(t)+ J b^{*}_{R}(t)J)^{-\12}= \epsilon(t)^{-\12}(\12 \one + \tpdo{-1}{W})\epsilon(t)^{-\12}.
\end{array}
\]
Moreover we have:
\[
\begin{array}{rl}
ii)&b_{R}(0)+ J b^{*}_{R}(0)J= 
\epsilon^{\12}(2\one + \Psi^{-1}_{\rm reg}(\Sigma; W))\epsilon^{\12},\\[2mm]
iii)& (b_{R}(0)+ Jb^{*}_{R}(0)J)^{-1}= \epsilon^{-\12}(\12 \one +  \Psi^{-1}_{\rm reg}(\Sigma; W))\epsilon^{-\12}\\[2mm]
iv)&b_{R}(0)= \epsilon^{\12}(\one + r_{-1, R})\epsilon^{\12},\ r_{-1, R}\in \Psi^{-1}_{\rm reg}(\Sigma; W).
\end{array}
\]
\end{proposition}
\begin{remark}\label{r1.1}
 It is easy to see that  the equation
 \[
\i \p_{t}b(t)- b^{2}(t)+ a(t)=r_{-\infty}(t)
\]
is equivalent to
\beq\label{eq:likejunker}
(\p_{t}+ \i b(t))\circ (\p_{t}-\i b(t))= \p_{t}^{2}+ a(t)-r_{-\infty}(t)
\eeq
{  The idea of factorizing the Klein-Gordon operator modulo a smoothing error term   was already used in \cite{junker} to construct Hadamard states in the scalar case. However, in contrast to \cite{junker}, instead of solving (\ref{eq:likejunker}) on the level of symbols we work with the operators and supplement arguments from microlocal analysis by Hilbert space techniques (cf. \cite{GW} for the scalar case).}
\end{remark}
\proof  
{\it Step 1}:  in Step 1  the parameter $R$ will be absent, so we suppress the subscript $R$ to simplify notation.   We  look for $b(t)$ of the form:
\beq\label{e1.3a}
 b(t)\eqdef \epsilon(t)+ b_{0}(t), \ b_{0}(t)\in \tpdo{0}{W}.
\eeq
Using that $a(t)= \epsilon^{2}(t)+ r_{1}(t)$  by  \eqref{toto.1} {\it i)},  we obtain that $b_{0}(t)$ should solve:
\begin{equation}
\label{e1.4}
\begin{aligned}
b_{0}&= (2\epsilon)^{-1}\i \p_{t}\epsilon+ (2\epsilon)^{-1}(r_{1}(t)-\one) + (2\epsilon)^{-1}(\i \p_{t}b_{0}- b_{0}^{2}+ [\epsilon, b_{0}])\\[2mm]
&=(2\epsilon)^{-1}(\i \p_{t}\epsilon+ r_{1}-\one)+ F(b_{0}),
\end{aligned}
\end{equation}
Since $\epsilon(t)$ as a scalar principal symbol, we have $\epsilon(t)\in \cinf(\rr, \pdoscal{1}{W})+ \tpdo{0}{W}$. Therefore   we obtain that 
$[\epsilon, c]\in \cinf(\rr, \Psi^{m}(\Sigma; W))$ for any operator $c\in \cinf(\rr, \Psi^{m}(\Sigma; W))$. It follows that  we can apply \cite[Lemma A.1]{GW} and find  $b(t)= \epsilon(t)+ b_{0}(t)$, unique modulo $\tpdo{-\infty}{W}$ such that
\begin{equation}
\label{e1.5}
\i \p_{t}b(t) -b^{2}(t)+ a(t)\in \tpdo{-\infty}{W},
\end{equation}
hence we have satisfied conditions {\it ia)}, {\it ib)}.

{\it Step 2}:
in Step 2 we modify $b(t)$ by subtracting an $R-$dependent term in $\Psi^{-\infty}(W)$ to ensure the remaining conditions.
 We first write $b(t)$ as
 \[
b(t)= \epsilon(t)^{\12}(\one + r_{-1}(t))\epsilon(t)^{\12}, \ r_{-1}(t)\in \tpdo{-1}{W}.
\]
We fix a cutoff function $\chi\in\cinf(\rr)$, with $\chi(\lambda)=0$ for $|\lambda|\leq 1$, $\chi(\lambda)= 1$ for $|\lambda|\geq 2$, and set for $R\geq 1$ and a function $\lambda\in \cinf(\rr)$ to be determined later:
\[
 r_{-1, R}(t)= \chi\left(\frac{\epsilon(t)}{R\lambda(t)}\right)r_{-1}(t) \chi\left(\frac{\epsilon(t)}{R\lambda(t)}\right)
\]
We know that for fixed $t$ we have $\chi(\frac{\epsilon(t)}{\lambda})r_{-1}(t)\chi(\frac{\epsilon(t)}{\lambda})\to 0\hbox{ in }\pdo{0}{W}$, 
when $\lambda\to +\infty$. Therefore we can find a smooth function $\rr\ni t\mapsto \lambda(t)\in \rr$ such that 
\begin{equation}
\label{toto2}
\|r_{-1, R}(t)\|_{B(L^{2}(\Sigma; W))}\leq \12, \ \forall \ t\in \rr, \ R\geq 1.
\end{equation}
 Moreover we have 
 \[
r_{-1}(t)- r_{-1, R}(t)= r_{-\infty, R}(t) \in \tpdo{-\infty}{W}. 
\]
Finally we set
\[
b_{R}(t)= \epsilon(t)^{\12}(\one + r_{-1, R}(t))\epsilon(t)^{\12},
\] 
so that $b_{R}(t)= b(t)+ \tpdo{-\infty}{W}$, hence $b_{R}(t)$ still satisfies {\it ia)}, {\it ib)}.

To verify the remaining conditions, we write:
\[
b_{R}(t)+ J b^{*}_{R}(t)J= \epsilon(t)^{\12}( 2\one + r_{-1, R}(t)+ J r_{-1,R}^{*}(t)J)\epsilon(t)^{\12},
\]
since $\epsilon(t)$ is selfadjoint and $[J, \epsilon(t)]=0$, by \eqref{toto.1}. Since $\| r_{-1, R}(t)\|+ \| J r_{-1, R}^{*}(t)J\|\leq 1$ by (\ref{toto2}), we have
\[
\begin{aligned}
(b_{R}(t)+ J b^{*}_{R}(t)J)^{-\12}&= \epsilon(t)^{-\12}( 2\one + r_{-1, R}(t)+ J r_{-1,R}^{*}(t)J)^{-1}\epsilon(t)^{-\12}\\[2mm]
&=\epsilon(t)^{-\12}(\12 \one + \tpdo{-1}{W})\epsilon(t)^{-\12},
\end{aligned}
\]
 by Prop. \ref{1.1opd}. This proves conditions {\it ic)}, {\it id)}. 
 
 It remains to check {\it ii)}, {\it iii)}, {\it iv)}. This follows from the fact that $\epsilon(0)= \epsilon$, hence $r_{-1, R}(0)\in \Psi^{-1}_{\rm reg}(\Sigma; W)$. It suffices then to apply  the properties of the space $\Psi^{-1}_{\rm reg}(\Sigma; W)$ recalled in Lemma \ref{lem2}.
  \qeds

\subsection{Parametrices for the Cauchy problem}\label{sec1.3}
It is well known that if $f\in \cH(\Sigma; W\oplus W)$, then the Cauchy problem (\ref{e3.9}) has a unique solution $\zeta= U(t)f\in \cinf(\rr, \cH(\Sigma; W))$.
In this subsection we give 
 a representation of $U(t)$  by generalizing to vector-valued wave equations  the constructions in \cite[Sect. 5]{GW} for the scalar case. 
\begin{theoreme}\label{1.1}
Let $b(t)=b_{R}(t)\in \tpdo{1}{W}$ be the operator constructed in Prop. \ref{p1.1} and let us set:
\[
\begin{aligned}
b^{+}(t)&\defeq  b(t), \ b^{-}(t)\defeq  - J b^{*}(t)J, \\[2mm]
 u^{\pm}(t)&\defeq  \Texp(\i \textstyle \int_{0}^{t}b^{\pm}(\sigma)d\sigma)\\[3mm]
r^{0\pm}&\defeq \mp(b^{+}(0)- b^{-}(0))^{-1}b^{\mp}(0)\in \Psi^{0}(\Sigma; W),\\[2mm]
r^{1\pm}&\defeq \pm(b^{+}(0)- b^{-}(0))^{-1}\in \Psi^{-1}(\Sigma; W),
\end{aligned} 
\]
 and 
 \beq\label{e1.9bc}
r^{\pm}f\defeq  r^{0\pm}f^{0}+ r^{1\pm}f^{1}, \ f\in \cH(\Sigma; W\oplus W).
\eeq
Then 
\[
 U(t)= u^{+}(t)r^{+}+ u^{-}(t)r^{-}+ r_{-\infty}(t), \ r_{-\infty}(t)\in \tpdos{-\infty}{W\oplus W}{W}.
\]
\end{theoreme}
\proof
It is convenient to generalize slightly the situation and to denote by $U(t,s)$ the Cauchy evolution operator for initial data  at time $s$, so that $U(t)= U(t, 0)$. We set also
\[
T(t,s)\defeq\lin{U(t,s)}{\i^{-1}\p_{t}U(t,s)}: \cH(\Sigma, W\oplus W)\to \cH(\Sigma; W\oplus W),
\]
so that
\begin{equation}
\label{toto1b}
\i^{-1}\p_{t}T(t,s)= A(t)T(t,s), \ A(t)= \mat{0}{\one}{a(t)}{0}.
\end{equation}
Note that the operators $r^{\pm}(t)$, defined as in (\ref{e1.9bc}) with  $b^{\pm}(0)$ replaced by $b^{\pm}(t)$ are well defined, by Prop. \ref{p1.1}.  Similarly we set $u^{\pm}(t,s)= \Texp(\i \textstyle \int_{s}^{t}b^{\pm}(\sigma)d\sigma)$, and
\[
 \tilde{U}(t,s)\defeq u^{+}(t,s)r^{+}(s)+ u^{-}(t,s)r^{-}(s), 
\]
\[
\tilde{T}(t,s)\defeq \lin{\tilde{U}(t,s)}{\i^{-1}\p_{t}\tilde{U}(t,s)}= \lin{u^{+}(t,s)r^{+}(s)+ u^{-}(t,s)r^{-}(s)}{b^{+}(t)u^{+}(t,s)r^{+}(s)+ b^{-}(t)u^{-}(t,s)r^{-}(s)}.
\]
An easy computation shows that
\beq\label{toto3}
\tilde{T}(s,s)= \one, \ u^{\pm}(t,s)r^{\pm}(s)= r^{\pm}(t)\tilde{T}(t,s),
\eeq
which implies that $(t,s)\mapsto \tilde{T}(t,s)$ is a two-parameter group.  From Prop. \ref{p1.1}
\[
\i^{-1}\p_{t}\left(b^{\pm}(t)u^{\pm}(t)\right)= \i^{-1}\p_{t}b^{\pm}(t)+b^{\pm 2}(t)= a(t)-r_{-\infty}^{\pm}(t),
\]
for $r_{-\infty}^{\pm}(t)\in \tpdo{-\infty}{W}$. Using then \eqref{toto3} we obtain that
\[
\i^{-1}\p_{t}\tilde{T}(t,s)= \tilde{A}(t)\tilde{T}(t,s), 
\]
for
\[
\tilde{A}(t)= A(t)-R_{-\infty}(t), 
\]
\[
R_{-\infty}(t)=\lin{0}{r_{-\infty}^{+}(t)}\circ r^{+}(t)+ \lin{0}{r_{-\infty}^{-}(t)}\circ r^{-}(t)\in \cinf(\rr^{2}, \Psi^{-\infty}(\Sigma; W\oplus W)).
\]
We can then express $T(t,s)$ in terms of $\tilde{T}(t,s)$ by setting
\begin{equation}
\label{toto5b}
T(t,s)\eqdef \tilde{T}(t,s)\circ R(t,s),
\end{equation}
where $R(t,s)$ solves the equation
\begin{equation}
\label{toto4}
\begin{cases}\i^{-1}\p_{t}R(t,s)- \tilde{T}(s,t) R_{-\infty}(t)\tilde{T}(t,s)\circ R(t,s)=0, \\[2mm]
R(s,s)= \one.\end{cases}
\end{equation}
By Lemma \ref{budaopd} we first obtain that
\[
\tilde{R}_{-\infty}(t,s)\defeq \tilde{T}(s,t) R_{-\infty}(t)\tilde{T}(t,s)\in \cinf(\rr^{2}, \Psi^{-\infty}(\Sigma; W\oplus W)).
\]
The solution of (\ref{toto4}) is then given by
\beq\label{toto5}
R(t,s)= \Texp(\i {\textstyle\int_{s}^{t}}\tilde{R}_{-\infty}(\sigma, s)d\sigma)
= \one +\i \int_{s}^{t}\tilde{R}_{-\infty}(\sigma, s)R(\sigma, s)d\sigma.
\eeq
By the argument in the proof of Prop. \ref{1.2opd} (see the properties of $m(t,s)$ in the proof), we first obtain that $R(t,s)\in  \cinf(\rr^{2}, \Psi^{0}(\Sigma; W\oplus W))$. \eqref{toto5} then implies that
\[
R(t,s)= \one +  \cinf(\rr^{2}, \Psi^{-\infty}(\Sigma; W\oplus W)).
\]
By Lemma \ref{budaopd} we obtain finally that
\[
T(t,s)= \tilde{T}(t,s)+  \cinf(\rr^{2}, \Psi^{-\infty}(\Sigma; W\oplus W)),
\]
hence
\[
U(t,s)= \tilde{U}(t,s)+  \cinf(\rr^{2}, \Psi^{-\infty}(\Sigma; W\oplus W, W)).
\]
Setting $s=0$ completes the proof of the theorem. \qeds

{  At this point, we could set 
\[
U^\pm\defeq U(t) r^\pm= u^{\pm}(t)r^{\pm}+ \tpdo{-\infty}{W},
\]
and prove directly that these are parametrices that satisfy the properties listed in Thm. \ref{thm:corres2}, with the exception of positivity (positivity w.r.t. the product $(\cdot|\cdot)_V$ does not hold if $J\neq \one$, instead one gets positivity w.r.t. the `non-physical' inner product $(\cdot|\cdot)$). Thus, we could associate to them (non-positive) pseudo-covariances $\lambda^\pm$ in an abstract manner as in Thm. \ref{thm:corres2}. However, we prefer to construct them in a more systematic way in Sect. \ref{sec3} in order to derive additional information needed to cope later on with the conditions $\gi$ and $\pos$ in gauge theory.}

 \section{Hadamard two-point functions}\init\label{sec3}
\init\subsection{Preparations}

In the present section, we continue with the setup of Sect. \ref{sec2} and deduce expressions for Hadamard two-point functions from the construction of the parametrix. This is done in a similar way as in \cite{GW}, i.e. we construct an operator $T_R$ that diagonalizes the symplectic form and separates Cauchy data that propagate with positive and negative energies in the wave front set.
We also show in Subsect. \ref{ss:non-existence} that Hadamard states do not exist for vector Klein-Gordon equations if the scalar product is not positive-definite on the fibers.

In the sequel, if $b_{R}(t)$ is the operator constructed in Prop. \ref{p1.1} we denote  $b_{R}(0)$ simply by $b_{R}$.
\begin{lemma}\label{1.2}
There exists $Z_{R}\in \Psi^{\12}(\Sigma; W)$ such that:
\beq\label{e1.98}
 b_{R}J+Jb_{R}^{*}= Z_{R}^{*}J Z_{R},
\eeq
and additionally:
\[
 Z_{R}= (\one +\Psi^{-1}_{\rm reg}(\Sigma; W))(2\epsilon)^{\12}, \ Z_{R}^{-1}= (2\epsilon)^{-\12}(\one + \Psi^{-1}_{\rm reg}(\Sigma; W)).
\]
\end{lemma}
\proof 
By Prop. \ref{p1.1} we have 
\[
Jb_{R}+ b_{R}^{*}J= (2\epsilon)^{\12}(J + J c_{R}+ c_{R}^{*}J)(2\epsilon)^{\12}, \ c_{R}\in \Psi^{-1}_{\rm reg}(\Sigma; W).
\]
We look for $Z_{R}$  in the lemma under the form $Z_{R}= S_{R}(2\epsilon)^{\12}$ for 
\begin{equation}
\label{e1.99}
S_{R}= \one+ d_{R}, \ d_{R}\in \Psi^{-1}_{\rm reg}(\Sigma; W).
\end{equation}
The identity (\ref{e1.98}) is satisfied if
\begin{equation}
\label{e1.9bb}
S_{R}^{*}J S_{R}= J+ Jc_{R}+ c_{R}^{*}J.
\end{equation}

Using  $W= W_{t}\oplus W_{\Sig}$ (see (\ref{en.4c})), we can write:
 \[
S_{R}= \mat{s_{tt,R}}{s_{t\Sig,R}}{s_{\Sig t,R}}{s_{\Sig\Sig,R}}, \ c_{R}= \mat{c_{tt,R}}{c_{t\Sig,R}}{c_{\Sig t,R}}{c_{\Sig\Sig,R}}.
\]
Let us now formulate the property that $c_{R}\in \Psi^{-1}_{\rm reg}(\Sigma; W)$ in terms of the components of $c_{R}$.  

If $\alpha, \beta$ are any of the symbols $t$ or $\scriptstyle\Sigma$,  then 
since  $h= h_{t}\oplus h_{\Sig}$, we obtain that $c_{\alpha\beta, R}\in \Psi^{-1}_{\rm reg}(\Sigma; W_{\alpha}, W_{\beta})$.  We are looking  for $s_{\alpha\beta, R}$ such that 
\[
s_{\alpha\beta, R}- \delta_{\alpha\beta}\in  \Psi^{-1}_{\rm reg}(\Sigma; W_{\alpha}, W_{\beta})
\]

Let us now suppress the index $R$ to simplify notation.
The equation  (\ref{e1.9bb}) is satisfied  iff:
\beq\label{e1.10}\left\{
\begin{aligned}
-s_{tt}^{*}s_{tt}+ s_{\Sig t}^{*}s_{\Sig t}&= \one- c_{tt}^{*}- c_{tt},\\[2mm]
-s_{tt}^{*}s_{t\Sig}+ s_{\Sig t}^{*}s_{\Sig\Sig}&= - c_{t\Sig}+ c_{\Sig t}^{*},\\[2mm]
-s_{t\Sig}^{*}s_{tt}+ s_{\Sig\Sig}^{*}s_{\Sig t}&= c_{\Sig t}- c_{t\Sig}^{*},\\[2mm]
s_{\Sig\Sig}^{*}s_{\Sig\Sig}- s_{t\Sig}^{*}s_{t\Sig}&= \one+ c_{\Sig\Sig}+ c_{\Sig\Sig}^{*}.
\end{aligned}\right.
\eeq

To solve this system we first set $s_{t\Sig}=0$. The last equation of (\ref{e1.10}) can then be solved  for $R$ large enough by
\[
s_{\Sig\Sig}= s_{\Sig\Sig}^{*}= (\one +c_{\Sig\Sig}+ c_{\Sig\Sig}^{*})^{\12}\in \one + \Psi^{-1}_{\rm reg}(\Sigma; W_\Sig, W_{\Sig}),
\]
using Lemma \ref{lem2} (3).   The second  and third equations are then solved by 
\[
 s_{\Sig t}= s_{\Sig\Sig}^{-1}(c_{\Sig t}- c_{t\Sig}^{*})\in \Psi^{-1}_{\rm reg}(\Sigma; W_{\Sig},W_{t} ),
\]
again by Lemma \ref{lem2}. Finally we solve the first equation by
\[
s_{tt}= s_{tt}^{*}= (\one + c_{tt}+ c_{tt}^{*}+ s_{\Sig t}^{*}s_{\Sig t})^{\12}\in \one + \Psi^{-1}_{\rm reg}(\Sigma; W_{t}, W_{t}).
\]
 This  completes the proof of the lemma. \qeds

We now set 
\begin{equation}
\label{e1.11}
T_{R}\defeq Z_{R}(b_{R}^{+}- b_{R}^{-})^{-1}\otimes \one_{\cc^{2}}\circ \mat{-b_{R}^{-}}{ \one}{ b_{R}^{+}}{-\one}\in \pdo{\infty}{W\oplus W},
\end{equation}
so that $T_{R}f= \lin{Z_{R} r_{R}^{+}f}{Z_{R} r_{R}^{-}f}$, where $r_{R}^{\pm}$ are defined in (\ref{e1.9bc}). We have:
\beq\label{en.7b}
T_{R}^{-1}= \mat{\one}{\one}{b_{R}^{+}}{b_{R}^{-}}\circ Z_{R}^{-1}\otimes \one_{\cc^{2}}.
\eeq
\begin{proposition}\label{1.3}
 We have:
 \beq\label{en.8}
 (T_{R}^{-1})^{*}\circ q \circ T_{R}^{-1}= \mat{J}{0}{0}{-J},
\eeq
\beq\label{en.9}
T_{R}= \frac{1}{\sqrt{2}}(\one + \Psi^{-1}_{\rm reg}(\Sigma; W\oplus W))\mat{\one}{\one}{\one}{-\one}\mat{\epsilon^{\12}}{0}{0}{\epsilon^{-\12}}.
\eeq
\end{proposition}
\proof Let us suppress again the subscript $R$ and denote $b_{R}^{\pm}$ simply by $ b^{\pm}$. Set $f^{\pm}= r^{\pm}f$, so that
\[
f^{0}= f^{+}+ f^{-}, \ f^{1}= b^{+}f^{+}+ b^{-}f^{-}. 
\]
An easy computation using that $b^{+}= b$, $b^{-}= - J b^{*}J$ yields:
\[
 \overline{f}qf= (f^{+}| (Jb+ b^{*}J)f^{+})- (f^{-}| (Jb + b^{*}J)f^{-}).
\]
By Lemma \ref{1.2} we have $Jb+ b^{*}J = Z_R^{*}JZ_R$. This implies (\ref{en.8}) by the definition of $T_R$.

Let us now prove (\ref{en.9}). From Lemma \ref{1.2} and  Prop. \ref{p1.1}  we have
\[
 \begin{array}{rl}
&Z_{R}= (\one + \Psi^{-1}_{\rm reg}(\Sigma; W)) (2\epsilon)^{\12},\\[2mm]
& (b_{R}^{+}- b_{R}^{-})^{-1}= (b_{R}+ J b_{R}^{*}J)^{-1}= (2\epsilon)^{-\12}(\one + \Psi^{-1}_{\rm reg}(\Sigma; W))(2\epsilon)^{-\12}.
\end{array}
\]
Similarly we have
\[
\begin{array}{rl}
&\mat{-b_{R}^{-}}{ \one}{ b_{R}^{+}}{-\one}\mat{\epsilon^{-\12}}{0}{0}{\epsilon^{\12}}
= \mat{J b_{R}^{*}J\epsilon^{-\12}}{ \epsilon^{\12}}{ b\epsilon^{-\12}}{-\epsilon^{\12}}\\[2mm]
=&\epsilon^{\12}(\one + \Psi^{-1}_{\rm reg}(\Sigma; W\oplus W))\mat{\one}{\one}{\one}{-\one}.
\end{array} 
\]
Then (\ref{en.9}) follows by applying formula (\ref{e1.11}). \qeds

\subsection{Hadamard two-point functions}
In this subsection we construct pairs of Hadamard two-point functions. 

\begin{proposition}\label{1.4}
Let  us define $c^{\pm}: \cH(\Sigma; W\oplus W)\to \cH(\Sigma; W\oplus W)$ by:
\beq\label{eq:defcpm}
c^{+}\defeq  T^{-1}_R\circ \mat{\one}{0}{0}{0}\circ T_R, \quad c^{-}\defeq  T^{-1}_R\circ \mat{0}{0}{0}{\one}\circ T_R,
\eeq
Then the following holds:
\ben
\item One has
  \[
c^{\pm}f =  \lin{r^{\pm}f}{b^{\pm}r^{\pm}f}, \ f\in \cH(\Sigma; W\oplus W),
\]
\item 
\[
\begin{array}{rl}
i)&c^{+}+ c^{-}= \one,  \ (c^{\pm})^{2}= c^{\pm},\\[2mm]
ii)& (c^{\pm})^{\dag}= c^{\pm}, \\[2mm]
iii)&r^{\pm}\circ c^{\pm}=r^{\pm}.
\end{array}
\]
\een
\end{proposition}
\proof  (1) is a routine computation using (\ref{e1.11}), (\ref{en.7b}). (2) follows from (\ref{en.8}). \qed

\begin{theoreme}\label{thn.1}
 Let  $c^\pm$ be defined by (\ref{eq:defcpm}) and set
 \beq\label{en.10}
  \lambda_{\Sig}^{\pm}\defeq  \pm q\circ c^{\pm}\in B(\cH(\Sigma; W\oplus W), \cH'(\Sigma; W\oplus W)).
 \eeq
  Then
 \ben
 \item $\lambda_{\Sig}^{\pm}$ is a pair of Hadamard Cauchy surface two-point functions;
 \item one has:
 \beq\label{en.10b}
\lambda_{\Sig}^{+}= T^{*}_R\mat{J}{0}{0}{0}T_R, \quad \lambda_{\Sig}^{-}= T^{*}_R\mat{0}{0}{0}{J}T_R.
\eeq
\een
\end{theoreme}
\proof The proof of (1) is  identical to the proof of \cite[Thm. 7.1]{GW}. Note that only the proof of the  
implication $\Rightarrow$ in \cite[Thm. 7.1]{GW} needs to be copied.  (2)  follows from (\ref{en.8}), (\ref{eq:defcpm}). \qeds

\begin{remark}\label{rem1}
 Statement (1) of Thm. \ref{thn.1} still holds if we replace $c^{\pm}$ by $c^{\pm}\pm r_{-\infty}$, for $r_{-\infty}\in \pdo{-\infty}{W\oplus W}$. 
\end{remark}
\subsection{Non-existence of Hadamard states for vector Klein-Gordon equations}\label{ss:non-existence}
In this subsection we consider a  vector Klein-Gordon operator $D$ as above, assuming that $J\neq \one$, i.e. that the hermitian form on $W$ is not positive definite. We show that under a mild additional condition on its two-point functions, there {\em does not exist} any Hadamard state, but only Hadamard {\em pseudo-states}.
\begin{theoreme}\label{th:non-existence}
  Assume that $J\neq \one$. Then there does not exist  spacetime two-point functions $\tilde{\lambda}^{\pm}$ for $D$ satisfying $\musc$ and $\pos$ such that additionally the Cauchy surface two-point functions $\tilde{\lambda}^{\pm}_{\Sig}$ map  continuously $\cH(\Sigma; W\oplus W)$ into itself.
\end{theoreme}
\proof Let $\tilde{\lambda}_{\Sig}^{\pm}$ the Cauchy surface two-point functions of the state $\omega$. Since by assumption $\tilde{\lambda}^{\pm}_{\Sig}$ preserve $\cH(\Sigma; W\oplus W)$ we can apply \cite[Thm. 7.1]{GW}, which generalizes directly to the vector case. We obtain that if $\musc$ holds then $\tilde{\lambda}^{\pm}_{\Sig}- \lambda^{\pm}_{\Sig}$ is smoothing.  
Let us set
\[
\tilde{A}\defeq(T^{*}_R)^{-1}\left(\tilde{\lambda}^{+}_{\Sig}+ \tilde{\lambda}^{-}_{\Sig}\right)T^{-1}_R, \quad A\defeq \mat{J}{0}{0}{J}.
\]
By (\ref{en.10b}) we obtain that $\tilde{A}= A+ R_{\infty}$
where $R_{\infty}$ is smoothing.   We may  choose  a sequence $f_{n}\in L^{2}(\Sigma; W\oplus W)$ with $\| f_{n}\|=1$,  $(f_{n}| A f_{n})=-1$, $\wlim f_{n}=0$, with support in some fixed compact $K\subset \Sigma$. Let us denote $\bbbone_K$ the characteristic function of $K$, understood as a multiplication operator. Since $\bbbone_{K}R_{\infty}\bbbone_{K}$ is compact we obtain that $\lim_{n\to \infty}(f_{n}| \tilde{A}f_{n})=-1$. But this contradicts the   positivity condition $\pos$, which implies that  $\tilde{A}\geq 0$. \qed

\subsection{Positivity of Hadamard two-point functions on subspaces}
We saw in Thm. \ref{th:non-existence} that it is impossible to construct  Hadamard  two-point functions for $D_1$, since in this case $J\neq\one$. However there  exist subspaces of  $\cH(\Sigma; W\oplus W)$ on which $\lambda^{\pm}_{1\Sig}$ are positive. This will follow from the fact that $J$ is positive on  $W_{\Sig}=(\Ker(J-\one))\otimes \fg$. 

\begin{proposition}\label{pn.1}
Let $\lambda^{1\pm}$ be defined in (\ref{en.10}), for $D=D_1$. Then there exists $r_{-1, R}\in \Psi^{-1}_{\rm reg}(\Sigma; W\oplus W)$ such that:
\[
\lambda^{\pm}_1 \geq 0\hbox{ on }(\one+ r_{-1,R})\cH(\Sigma; W_{\Sig}\oplus W_{\Sig}).
\]
\end{proposition}
\proof From (\ref{en.9})  we obtain that
\beq\label{en.11}
T_{R}= \frac{1}{\sqrt{2}}\mat{\one}{\one}{\one}{-\one}\mat{\epsilon^{\12}}{0}{0}{\epsilon^{-\12}}(\one + \Psi^{-1}_{\rm reg}(\Sigma; W\oplus W)).
\eeq
This implies, using also Lemma \ref{lem2} (3) that for $R$ large enough there exists $r_{-1, R}\in  \Psi^{-1}_{\rm reg}(\Sigma; W\oplus W)$ such that
\[
T_{R}= \frac{1}{\sqrt{2}}\mat{\one}{\one}{\one}{-\one}\mat{\epsilon^{\12}}{0}{0}{\epsilon^{-\12}}(\one +r_{-1, R})^{-1}.
\]
We note next that $\mat{J}{0}{0}{0}$ and $\mat{0}{0}{0}{J}$ are positive on $\cH(\Sigma; W_{\Sig}\oplus W_{\Sig})$, since $J$ is positive on $W_{\Sig}$.  The operators $\mat{\one}{\one}{\one}{-\one}$ and $\mat{\epsilon^{\12}}{0}{0}{-\epsilon^{\12}}$ preserve the space $\cH(\Sigma; W_{\Sig}\oplus W_{\Sig})$, since $\epsilon= \epsilon_{t}\oplus \epsilon_{\Sig}$. The proposition follows then from (\ref{en.10b}) and  (\ref{en.11}). \qeds

\section{Pair of Hadamard pseudo-covariances}\init\label{sec5}\init

In this section we consider the  pair of operators $D_{0}= \bdelta \bardel $, $D_{1}=\bardel\bdelta + \bdelta \bardel + \bF\tnI\, $ as in  Subsect. \ref{lnym}.  After going to the temporal gauge,  we may  assume that both operators fit into the framework of Sect. \ref{sec2}, i.e. that:
\[
D_{i}= \p_{t}^{2}+ a_{i}(t, x, D_{x}),
\]
where $a_{i}(t)\in \cinf(\rr; \Psi^{2}(\Sigma; W_{i}))$ for  $W_{1}= V_{1}\otimes \fg$, and $W_{0}= \fg$.  The operator $K= \bardel$ becomes in this framework:
\beq\label{e1.12}
 K= K_{0}(t)\p_{t}+ K_{1}(t),
\eeq
where $K_{j}(t)\in \cinf(\rr, {\rm Diff}^{j}(\Sigma; W_{0}, W_{1}))$ is a differential operator  in  $x$,  such that
\begin{equation}
\label{e1.13}
(\p_{t}^{2}+ a_{1}(t))\circ K =  K\circ( \p_{t}^{2}+ a_{0}(t)).
\end{equation}
It is easy to check that 
\begin{equation}
\label{e1.13b}
K_{0}(t,x)\in L(W_{0}, W_{1})\neq 0, \ \forall \ (t,x)\in\rr\times \Sigma.
\end{equation}
We recall that 
\[
K_{\Sig}\defeq  \rho_{1}\circ K \circ U_{0}\in {\rm Diff}(W_{0}\oplus W_{0}, W_{1}\oplus W_{1}),
\] where $\rho_{i}$, $U_{i}$ are the trace and  Cauchy evolution operators.

\subsection{Some preparations}
Let us denote by $u_{i}^{\pm}(t)$, $i=0,1$ the operators  constructed in Thm. \ref{1.1}.
\begin{lemma}\label{1.5}
 There exist $m_{1}^{\pm}\in \Psi^{1}(\Sigma;W_{0}, W_{1})$ and $r^{\pm}_{-\infty}(t)\in \tpdos{-\infty}{W_{0}}{W_{1}}$ such that:
 \[
K\circ u_{0}^{\pm}(t)= u_{1}^{\pm}(t)m_{1}^{\pm}+ r_{-\infty}^{\pm}(t).
\]
\end{lemma}
\proof 
We consider only the $+$ case and suppress the $+$ superscripts to simplify notation.  We also denote by $r_{-\infty}(t)$ a generic operator in $\tpdos{-\infty}{V_{1}}{V_{2}}$ for appropriate $V_{1}$, $V_{2}$. We will use repeatedly the following consequence of Prop. \ref{1.2opd}:
the map
\begin{equation}
\label{toto8}
m(t)\mapsto u_{i}(t)m(t)u_{j}(t)^{-1}\hbox{ is bijective on  }\tpdos{p}{W_{j}}{W_{i}}.
\end{equation}
This follows from the fact that  $b_{i}(t)$ have a scalar principal symbol equal to $(k_{i}h^{ij}(t,x)k_{j})^{\12}$. 

We recall the following  equivalent identities from Prop. \ref{p1.1}:
\begin{equation}
\label{toto6}
\begin{array}{rl}
i)&\i\p_{t}b_{i}(t)- b_{i}^{2}(t)+ a_{i}(t)+ r_{-\infty}(t)=0, \\[2mm]
ii)&(\p_{t}+ \i b_{i}(t))\circ(\p_{t}- \i b_{i}(t))= \p_{t}^{2}+ a_{i}(t)+ r_{-\infty}(t), \ i=0,1.
\end{array}
\end{equation}
 Since $u_{0}(t)= \Texp(\i \int_{0}^{t}b_{0}(s)ds)$, we obtain from
(\ref{e1.12}) that:
\[
K \circ u_{0}(t)= (\i K_{0}b_{0}(t)+ K_{1})\circ  u_{0}(t).
\]
Composing this identity to the left with $\p_{t}- \i b_{1}$ and using \eqref{toto6} {\it i)} we obtain:
\beq\label{e1.13c}
\begin{aligned}
&(\p_{t}-\i b_{1}) \circ K \circ u_{0}(t)\\[2mm]
&= (-K_{0}(a_{0}+r_{-\infty,0})+ \p_{t}K_{1}+ \i(\p_{t}K_{0}+ K_{1})b_{0} + b_{1}( K_{0}b_{0}-\i K_{1})) \circ u_{0}(t)\\[2mm]
&=m_{2}(t) \circ u_{0}(t), \hbox{ for }m_{2}(t)\in \tpdos{2}{W_{0}}{W_{1}}.
\end{aligned}
\eeq
By \eqref{toto8} we  obtain that:
\beq\label{e1.15}
m_{2}(t) \circ u_{0}(t)= u_{1}(t)\circ \tilde{m}_{2}(t), \hbox{ where }\tilde{m}_{2}(t)\in \tpdos{2}{W_{0}} {W_{1}}.
\eeq
Combining (\ref{e1.13c}) and (\ref{e1.15}), we obtain that:
\[
(\p_{t}- \i b_{1})\circ K\circ u_{0}(t)= u_{1}(t)\circ \tilde{m}_{2}(t).
\]
We compose the above  identity with $\p_{t}+ \i b_{1}(t)$, using again \eqref{toto6} and obtain:
\[
\begin{aligned}
(\p_{t}^{2}+ a_{1})\circ K \circ u_{0}(t)= &(\p_{t}+ \i b_{1})\circ u_{1}(t) \circ \tilde{m}_{2}(t)+  r_{-\infty}(t)Ku_{0}(t)\\[2mm]
&=2\i b_{1}\circ u_{1}(t)\circ \tilde{m}_{2}(t)+ u_{1}(t)\p_{t}\tilde{m}_{2}(t)+ r_{-\infty}(t)Ku_{0}(t)\\[2mm]
&=u_{1}(t)\circ\left( \p_{t}\tilde{m}_{2}(t)-\i\tilde{b}_{1}(t)\tilde{m}_{2}(t)+ r_{-\infty}(t)\right),
\end{aligned}
\]
where in the last line we use  \eqref{toto8}, and $\tilde{b}_{1}(t)\in \tpdo{1}{W_{1}}$ is again elliptic with a real principal symbol.

On the other hand since $(\p_{t}^{2}+ a_{1})\circ K = K\circ (\p_{t}^{2}+ a_{0})$, we have by \eqref{toto6}
\[
(\p_{t}^{2}+ a_{1})\circ K \circ u_{0}(t)= K\circ  r_{-\infty}(t)u_{0}(t)= u_{1}(t)\circ r_{-\infty}(t),
\]
again by \eqref{toto8}.
Summarizing we obtain that $\tilde{m}_{2}(t)$ solves
\[
\p_{t}\tilde{m}_{2}(t)+ \i  \tilde{b}_{1}(t)\tilde{m}_{2}(t)= r_{-\infty}(t),
\]
hence
\beq\label{toto9}
\begin{aligned}
\tilde{m}_{2}(t)&= \Texp(\i \int_{0}^{t}\tilde{b}_{1}(\sigma)d\sigma)\tilde{m}_{2}(0)+\Texp(\i\int_{s}^{t}\tilde{b}_{1}(\sigma)d\sigma)r_{-\infty}(s)ds\\[2mm]
&=\Texp(\i \int_{0}^{t}\tilde{b}_{1}(\sigma)d\sigma)\tilde{m}_{2}(0)+ r_{-\infty}(t).
\end{aligned}
\eeq
 By Lemma \ref{1.6} below this implies that $\tilde{m}_{2}(t)\in \tpdos{-\infty}{W_{0}}{W_{1}}$, hence  by Lemma \ref{budaopd}   that $m_{2}(t)\in \tpdos{-\infty}{W_{0}}{W_{1}}$. The identity (\ref{e1.13c}) becomes 
\[
(\p_{t}- \i b_{1})\circ K\circ u_{0}(t)= r_{-\infty}(t).
\]
As in \eqref{toto9} this implies that
\[
K\circ u_{0}(t) =  u_{1}(t)\circ (K\circ u_{0})(0)+ r_{-\infty}(t),
\]
and  completes the proof of the lemma. \qeds

\begin{lemma}\label{1.6}
 Let $b_{1}(t)\in \tpdo{1}{W_{1}}$ satisfying the assumptions of Prop. \ref{1.2opd} and $m(t)\in \tpdos{p}{W_{0}}{W_{1}}$, $p\in \rr$ such that:
 \[
m(t)= \Texp(\i \textstyle\int_{0}^{t}b_{1}(s)ds)m(0)+ r_{-\infty}(t), \ r_{-\infty}(t)\in \tpdos{-\infty}{W_{0}}{W_{1}}.
\]
Then $m(t)\in \tpdos{-\infty}{W_{0}}{W_{1}}$.
\end{lemma}
\proof We have $\p_{t}m(t)- \i  b_{1}(t)m(t)\in \cinf(\rr, \Psi^{-\infty})$. By induction we obtain \[
\p_{t}^{k}m(t)- p_{k}(t)m(t)\in \cinf(\rr, \Psi^{-\infty}),  k\in \nn,
\]
 where $p_{k}(t)\in \tpdo{k}{W_{1}}$,  $\sigma_{\rm pr}(p_{k})= (\i\sigma_{\rm pr}(b_{1}))^{k}$.
Note that $b_{1}$ is elliptic in $ \Psi^{1}(\Sigma; W_{1})$ hence  $p_{k}$ is elliptic in $\Psi^{k}(\Sigma;W_{1})$ and since $\p^{k}_{t}m(t)$ belongs to $\tpdos{p}{W_{0}}{W_{1}}$ by assumption we obtain that $m(t)\in \tpdos{p-k}{W_{0}}{W_{1}}$. This completes the proof. \qeds

\subsection{Compatibility of Hadamard pseudo-covariances}
We prove now the main result of this section, which will be important later on.
\begin{theoreme}\label{th:diff-cov}
 Let $c_{i}^{\pm}\in B(\cH(\Sigma; W_{i}\oplus W_{i}))$, $i=0,1$  be as in Prop. \ref{1.4}. Then
 \[
 c_{1}^{\pm}K_{\Sig}- K_{\Sig}c_{0}^{\pm}\in \Psi^{-\infty}(\Sigma; W_{0}\oplus W_{0}, W_{1}\oplus W_{1}).
\]
\end{theoreme}
\proof 
Since $c_{i}^{+}+ c_{i}^{-}= \one$, it suffices to prove the $+$ case, which amounts 
 to show that
\beq\label{e2.1}
c_{1}^{-}K_{\Sig}c_{0}^{+}\in \Psi^{-\infty}(\Sigma; W_{0}\oplus W_{0}, W_{1}\oplus W_{1}).
\eeq
In the sequel we denote simply by $r_{-\infty}(t)$ an error term in $\tpdos{-\infty}{V_{1}}{V_{2}}$ for 
appropriate  $V_{1}$, $V_{2}$.
We recall  from Thm. \ref{1.1} and Prop. \ref{1.4} that:
\[
 U_{i}(t)= u_{i}^{+}(t)r_{i}^{+}+ u_{i}^{-}(t)r_{i}^{-}+ r_{-\infty}(t),\ r_{i}^{\pm}c_{i}^{\pm}= r_{i}^{\pm}.
 \]
 Using Lemma \ref{1.5} this gives first:
\[
U_{1}(t)K_{\Sig}c_{0}^{+}= KU_{0}(t)c_{0}^{+}= K u_{0}^{+}(t)r_{0}^{+}+ r_{-\infty}(t)= u_{1}^{+}(t) m_{1}^{+}r_{0}^{+}+r_{-\infty}(t)
\]
for some $m_{1}^{+}\in \Psi^{1}(\Sigma; W_{0}, W_{1})$. On the other hand:
\[
 U_{1}(t)K_{\Sig}c_{0}^{+}= u_{1}^{+}(t)r_{1}^{+}c_{1}^{+} K_{\Sig}c_{0}^{+}+ u_{1}^{-}(t)r_{1}^{-} c_{1}^{-}K_{\Sig}c_{0}^{+}+ r_{-\infty}(t).
\]
It follows that
\beq\label{e2.2}
u_{1}^{-}(t) r_{1}^{-}c_{1}^{-} K_{\Sig}c_{0}^{+}= u_{1}^{+}(t)\circ (m_{1}^{+} r_{0}^{+}- r_{1}^{+}c_{1}^{+}K_{\Sig}c_{0}^{+})+ r_{-\infty}(t).
\eeq
We claim that if $n_{1}^{\pm}\in \Psi^{p}(\Sigma; W_{0}\oplus W_{0}, W_{1})$ satisfy 
\[
 u_{1}^{+}(t)n_{1}^{+}- u_{1}^{-}(t) n_{1}^{-}\in \tpdos{-\infty}{W_{0}\oplus W_{0}}{W_{1}},
\]
then $ n_{1}^{\pm}\in \Psi^{-\infty}(\Sigma; W_{0}\oplus W_{0}, W_{1})$.  Taking first $t=0$ we obtain that $n_{1}^{+}-n_{1}^{-}\in \Psi^{-\infty}(\Sigma;W_{0}\oplus W_{0}, W_{1})$. Next taking derivatives in $t$ at $t=0$ we obtain that  $(b_{1}^{+}(0)- b_{1}^{-}(0)) n_{1}^{+}\in\Psi^{-\infty}(\Sigma;W_{0}\oplus W_{0}, W_{1})$, hence $n_{1}^{+}\in \Psi^{-\infty}(\Sigma;W_{0}\oplus W_{0}, W_{1})$ by the ellipticity of $ b_{1}^{+}(0)- b_{1}^{-}(0)$. This also implies that $n_{1}^{-}\in \Psi^{-\infty}(\Sigma;W_{0}\oplus W_{0}, W_{1})$. 

Applying this remark to (\ref{e2.2}) we obtain that $r_{1}^{-}c_{1}^{-}K_{\Sig}c_{0}^{+}\in\Psi^{-\infty}(\Sigma;W_{0}\oplus W_{0}, W_{1})$. This implies (\ref{e2.1}) since  from Prop. \ref{1.4} and $r_{1}^{+}c_{1}^{-}=0$ we obtain:
\[
 c_{1}^{-}= \lin{r_{1}^{-}}{b_{1}^{-}(0)r_{1}^{-}}\circ c_{1}^{-}. 
\]
This completes the proof of the theorem. \qeds

\section{Proof of Thm. \ref{maintheo2}}\init\label{sec4}\init

As before, $\Sigma$ is assumed to be compact or equal to $\rr^{d}$. If $\Sigma= \rr^{d}$ we assume Hypothesis \ref{as:background}.

In this case it follows from Prop. \ref{prop.hard} that $h_{t}$ satisfies a Hardy inequality:
\beq\label{hardipetit}
h_{t}= \bardels\bards \geq C \x^{-2},
\eeq
which will be very important in the sequel.

Our goal in this section is to construct a projection $\Pi$ acting on Cauchy data with the following two properties:
\[
\begin{array}{rl}
i)&\Ker \Pi= \Ran K_{\Sig}\\[2mm]
ii)& \lambda^{\pm}_{1\Sig}\hbox{ are positive on }\Ran \Pi\cap \Ker K_{\Sig}^{\dag}.
\end{array}
\]
We will ensure {\it ii)} by choosing $\Pi$ in such a way that 
\beq\label{en.12}
\Ran \Pi\cap \Ker K_{\Sig}^{\dag}\subset (\one+ r_{-1,R})\cH(\Sigma; W_{\Sig}\oplus W_{\Sig}),
\eeq
where the operator $r_{-1, R}$ appears in Prop. \ref{pn.1}.

\subsection{Notations}\label{notata} - As before, if $E, F$ are two topological vector spaces, we write $A: E\to F$ if $A$ is linear continuous from $E$ to $F$. We write $A: E  \tarrow F$ if additionally $A$ is bijective and both $A^{-1}$ is linear continuous.

- We denote $\bra x \ket H^m(\Sigma;V)$ the Sobolev space of order $m$ with weight $\bra x \ket =(1+|x|)^{\12}$ (of course this is just the same as $H^m(\Sigma;V)$ if $\Sigma$ is compact) and $\bra x \ket L^2(\Sigma;V)=\bra x \ket H^0(\Sigma;V)$ the weighted $L^2$ space.

- We will denote $B^{-\infty}(\Sigma; V_{1}, V_{2})$ the space of  operators that are bounded from $H^{-m}(\Sigma; V_{1})$ to $H^{m}(\Sigma; V_{2})$ for any $m\in \rr$.

\subsection{The reference projection for $\Sigma= \rr^{d}$}\label{ss:refproj}
In this subsection  we assume that  $\Sigma= \rr^{d}$.  We  define a reference projection $\Proj_{0}$, which will be used to construct the projection $\Proj$.
We first state an easy consequence of the Hardy inequality. 
\begin{lemma}\label{stup1}
The following operators are bounded:
 \[
\begin{array}{rl}
i)&h_{t}^{-\12}\bardels: L^{2}(\Sigma; W_{\Sig})\to  L^{2}(\Sigma; W_{t}), \\[2mm]
 ii)&\bards h_{t}^{-\12}:  L^{2}(\Sigma; W_{t})\to  L^{2}(\Sigma; W_{\Sig}),\\[2mm]
 iii)&h_{t}^{-\12}\x^{-1}:  L^{2}(\Sigma; W_{t})\to  L^{2}(\Sigma; W_{t})
\end{array}
\]
\end{lemma}
\proof  {\it i)} and {\it ii)} follow from the definition of $h_{t}$. To prove {\it iii)} we use the Hardy inequality (\ref{hardipetit}) and the Kato-Heinz theorem which yield $h_{t}^{-1}\leq C \x^{-2}$. \qeds

\begin{definition}
We set:
\[
\begin{array}{rl}
&\pi\defeq  \bards h_{t}^{-1}\bardels: L^{2}(\Sigma; W_{\Sig})\to L^{2}(\Sigma; W_{\Sig}),\\[2mm]
&\varb\defeq h_{t}^{-1}\bardels: \  L^{2}(\Sigma; W_{\Sig})\to\x L^{2}(\Sigma; W_{t}),\\[2mm]
&\vara\defeq \wbar{F}_{t}\wedge\,\cdot\,: \ \x L^{2}(\Sigma; W_{t})\to L^{2}(\Sigma; W_{\Sig}).
\end{array}
\]\end{definition}
The above operators are well defined by Lemma \ref{stup1} and Hypothesis  \ref{as:background}.

Clearly $\pi$ is the orthogonal projection on $\Ran\bards$, where $\bards$ is considered as a closed operator on $L^{2}(\Sigma; W_{t})$ with domain  $H^{1}(\Sigma; W_{t})$.   Moreover one has:
\begin{equation}
\label{en.14}
\bards\circ \varb= \pi, \ \varb\circ \bards= \one.
\end{equation}

We will construct $\Proj$ by modifying a reference projection $\Proj_{0}$.  We denote by $\Proj_{0}$ the operator defined in the adapted Cauchy data by the matrix:
\begin{equation}
\label{en.16}
\Proj_{0}\defeq   \left(\begin{array}{cccc}
0&0&0&0\\
0&\one -\pi&0&0\\
0&0&\one&0\\
0&\i \vara\circ \varb&0&\one
\end{array}\right).
\end{equation}
Since $\vara\x: L^{2}(\Sigma; W_{t})\to L^{2}(\Sigma; W_{\Sig})$ by Hypothesis \ref{as:background} we see that  
\[
\Proj_{0}: L^{2}(\Sigma; W\oplus W)\to L^{2}(\Sigma; W\oplus W).
\]
Let us consider the operator $K_{\Sig}$  given in Lemma \ref{ln.1} as an unbounded operator
\[
\begin{array}{rl}
&K_{\Sig}: L^{2}(\Sigma; W_{t}\oplus W_{t})\to L^{2}(\Sigma; W_{\Sig}\oplus W_{\Sig}), \\[2mm]
& \Dom K_{\Sig}=H^{1}(\Sigma; W_{t})\oplus L^{2}(\Sigma; W_{t}).
\end{array}
\]
\begin{lemma}\label{ln.6}
  $\Proj_{0}$ is a  bounded projection on $L^{2}(\Sigma; W\oplus W)$ with  $\Ker \Proj_{0}= \Ran K_{\Sig}$.
\end{lemma}
\proof 
The fact that $\Proj_{0}$ is a projection is a routine computation, using that $\varb(\one - \pi)=0$. 
 Since $\vara\varb$ is bounded by Lemma \ref{stup1} and Hypothesis \ref{as:background}\ we see that $\Proj_{0}$ is bounded.
To prove the second statement we 
note first that  $\Proj_{0}K_{\Sig}=0$, using (\ref{en.14}).
This implies that $\Ran K_{\Sig}\subset \Ker \Proj_{0}$. Conversely let $g\in \Ker \Proj_{0}$, i.e.
\[
g^{0}_{\Sig}= \pi g^{0}_{\Sig}, \ g^{1}_{t}=0, \ g^{1}_{\Sig}= -\i \vara\varb g^{0}_{\Sig}.
\]
From the first equation we get $g^{0}_{\Sig}= \bards u^{0}$ for $u^{0}= b g^{0}_{\Sig}\in H^{1}(\Sigma; \fg)$, and hence $g^{1}_{\Sig}= -\i \vara u^{0}$, i.e. $g= K_{\Sig}u$, for $u= (u^{0}, \i ^{-1}g^{0}_{t})$.  \qeds

We end this subsection by constructing an operator $B_{0}$ such that $(\one - \Proj_{0})= K_{\Sig}B_{0}$ (see the discussion at the end of Subsect. \ref{ss:psecauchy}).
\begin{lemma}\label{biziro}
 Let $B_{0}: L^{2}(\Sigma; W\oplus W)\to \x L^{2}(\Sigma; W_{t})\oplus L^{2}(\Sigma; W_{t})$ be given by:
\begin{equation}
\label{en.100}
B_{0}\defeq  \left(\begin{array}{cccc}
0&\varb&0&0\\
-\i&0&0&0
\end{array}\right).
\end{equation}
Then one has 
\[
 (\one - \Proj_{0})= K_{\Sig}B_{0}, \quad{  B_{0}K_{\Sig}=\one }.
\]
\end{lemma}
\proof The proof is a direct computation that uses $\bards \varb= \pi$. \qed

\subsection{The reference projection for $\Sigma$ compact} In this subsection, we assume that 
 $\Sigma$ is compact.  This implies that $\Ker h_{t}= \Ker \bards$ is not necessarily trivial. Therefore we need to change the definition of $\pi$, $\varb$ and $\Proj_{0}$. We set now:
\begin{definition}\label{defocompact}
 \[
\begin{array}{rl}
&\pi\defeq  \bards h_{t}^{-1}\bbbone_{\rr\backslash\{0\}}(h_{t})\bardels: L^{2}(\Sigma; W_{\Sig})\to L^{2}(\Sigma; W_{\Sig}),\\[2mm]
&\varb\defeq h_{t}^{-1}\bbbone_{\rr\backslash\{0\}}(h_{t})\bardels: \  L^{2}(\Sigma; W_{\Sig})\to L^{2}(\Sigma; W_{t}),\\[2mm]
&\vara\defeq \wbar{F}_{t}\wedge\,\cdot\,: \ L^{2}(\Sigma; W_{t})\to L^{2}(\Sigma; W_{\Sig}),
\end{array}
\]
where $\bbbone_{\rr\backslash\{0\}}$ stands for the characteristic function of $\rr\backslash\{0\}$. 
\end{definition}
Note that since $h_{t}$ has  compact resolvent, we know that 
\begin{equation}
\label{pade}
\pi\in \Psi^{0}(\Sigma; W_{\Sig}), \ \varb\in \Psi^{-1}(\Sigma; W_{\Sig}, W_{t}), \ \vara\in \Psi^{0}(\Sigma; W_{t}, W_{\Sig}). 
\end{equation}

We also denote by $\pi_{1}: L^{2}(\Sigma; W_{\Sig})\to L^{2}(\Sigma; W_{\Sig})$ a bounded projection with 
\begin{equation}
\label{kerpi}
\Ker\,\pi_{1}= \vara(\Ker h_{t}),
\end{equation}
like for example the orthogonal projection for the natural Hilbertian scalar product on $L^{2}(\Sigma; W_{\Sig})$ along $\vara\Ker h_{t}$. By the ellipticity of $h_{t}$, we know that $\Ker h_{t}\subset \cinf(\Sigma; W_{t})$, hence $\vara \Ker h_{t}\subset \cinf(\Sigma; W_{\Sig})$ and these two spaces are  finite dimensional. 

This implies first that there exists  a right inverse $\vara^{-1}\in L(\Ker\,\pi_{1}, \Ker h_{t})$ such that
\begin{equation}
\label{rightinv}
 \vara \circ \vara^{-1}=\one  \hbox{ on }\Ker\,\pi_{1}.
\end{equation}
Moreover since $\Ker\,\pi_{1}$ is a finite dimensional subspace of $\cinf(\Sigma; W_{\Sig})$ we have:
\begin{equation}
\label{pide}
\pi_{1}\in \one + \Psi^{-\infty}(\Sigma; W_{\Sig}), \ \vara^{-1}(\one - \pi_{1})\in \Psi^{-\infty}(\Sigma; W_{\Sig}, W_{t}). 
\end{equation}
We set now:
\begin{equation}
\label{en.16comp}
\Proj_{0}\defeq   \left(\begin{array}{cccc}
0&0&0&0\\
0&\one -\pi&0&0\\
0&0&\one&0\\
0&\i\pi_{1} \vara\circ \varb&0&\pi_{1}
\end{array}\right).
\end{equation}
\begin{lemma}\label{ln.66}
  $\Proj_{0}$ is a  bounded projection on $L^{2}(\Sigma; W\oplus W)$ with  $\Ker\,\Proj_{0}= \Ran K_{\Sig}$. Moreover $\Proj_{0}\in \Psi^{0}(\Sigma; W\oplus W)$. 
\end{lemma}

\proof The fact that $\Proj_{0}$ is bounded follows from the properties of $\pi$, $\vara$, $\varb$ stated in Def. \ref{defocompact} and from (\ref{pide}).  Again the fact that $\Proj_{0}$ is a projection follows from $\varb(\one - \pi)=0$.   Let us now prove that $\Proj_{0}K_{\Sig}=0$ hence $\Ran K_{\Sig}\subset \Ker\,\Proj_{0}$. By a  routine computation  this amounts to show that $(\one - \pi)d_{\Sig}=0$ and that $\pi_{1}\vara(\varb\bards-\one)=0$. The first identity is immediate. To prove the second, we use that 
$\varb\bards-\one=\bbbone_{\{0\}}(h_{t})$. Then $\pi_{1}\vara\bbbone_{\{0\}}(h_{t})=0$ since $\Ker\,\pi_{1}= \vara(\Ker h_{t})$.

Let us now prove that $\Ker\,\Proj_{0}\subset \Ran K_{\Sig}$.  Let $g\in \Ker\,\Proj_{0}$ i.e. 
\[
g^{0}_{\Sig}= \pi g^{0}_{\Sig}, \ g^{1}_{t}=0, \ \pi_{1}( g^{1}_{\Sig}+ \i \vara\varb g^{0}_{\Sig})=0.
\]
Then $g= K_{\Sig}u$ for $u= (u^{0}, u^{1})$ if
\beq\label{condit}
\i u^{1}= g^{0}_{t}, \ \bards u^{0}= g^{0}_{\Sig}, \ -\i \vara u^{0}= g^{1}_{\Sig}.
\eeq
We take $u^{1}= \i^{-1}g^{0}_{t}$ and $u^{0}= \varb g^{0}_{\Sig}+ v^{0}$ for $v^{0}\in \Ker h_{t}$, so that $\bards u^{0}=  \bards \varb g^{0}_{\Sig}= \pi g^{0}_{\Sig}= g^{0}_{\Sig}$.  It remains to satisfy the third identity in (\ref{condit}), which yields $-\i \vara v^{0}= g^{1}_{\Sig}+ \i \vara\varb g^{0}_{\Sig}$. Since $\pi_{1}( g^{1}_{\Sig}+ \i \vara\varb g^{0}_{\Sig})=0$, we can find $v^{0}\in \Ker h_{t}$ satisfying the above condition, using that $\Ker\,\pi_{1}= \vara \Ker h_{t}$.  
The fact that   $\Proj_{0}\in \Psi^{0}$ follows from  (\ref{pade}) and  (\ref{pide}). \qeds

We need the analog of Lemma \ref{biziro} in the compact case.
\begin{lemma}\label{bizirobis}
  Let $B_{0}: L^{2}(\Sigma; W\oplus W)\to  L^{2}(\Sigma; W_{t})\oplus L^{2}(\Sigma; W_{t})$ be given by:
\begin{equation}
\label{en.1001}
B_{0}\defeq  \left(\begin{array}{cccc}
0&\varb- \vara^{-1}(\one - \pi_{1})\vara\varb&0&\i \vara^{-1}(\one - \pi_{1})\\
-\i&0&0&0
\end{array}\right),
\end{equation}
where  $\vara^{-1}: \Ker\,\pi_{1}\to \Ker h_{t}$ is defined in (\ref{rightinv}). 
Then one has 
\beq\label{taratata}
 (\one - \Proj_{0})= K_{\Sig}B_{0}, \quad {  B_{0}K_{\Sig}=\one }.
\eeq
Moreover $B_{0}\in \Psi^{\infty}(\Sigma; W\oplus W, W_{t}\oplus W_{t})$.
 \end{lemma}
\proof Again the first property of $B_{0}$ is a direct computation, the fact that $B_{0}\in \Psi^{\infty}$ follows from (\ref{pade}), (\ref{pide}). \qeds

\subsection{Change of Cauchy data}
In this section we systematically work with the adapted Cauchy data, in which the operators $K_{\Sig}$ and $K_{\Sig}^{\dag}$ take simple forms.  Therefore the operator $r_{-1, R}\in \Psi^{-1}_{\rm reg}(\Sigma; W\oplus W)$ appearing in Prop. \ref{pn.1} is  replaced by  $R_{\rm F}\circ r_{-1, R}\circ R_{\rm F}^{-1}$.

Moreover it is convenient to perform another change of Cauchy data, corresponding to putting different weights on the two components $f^{0}, f^{1}$ or $g^{0}, g^{1}$ of a set of Cauchy data.  The need for these weights is already apparent  from the presence of the matrix
\beq\label{defdeS}
S\defeq  \mat{\epsilon^{\12}}{0}{0}{\epsilon^{-\12}},
\eeq
in the expression of the operator $T_{R}$  in Prop. \ref{1.3}.  It can also be seen from the fact that the natural space of Cauchy data appearing for example in the quantization of the scalar Klein-Gordon equation  is $H^{\12}(\Sigma)\oplus H^{-\12}(\Sigma)$. It is convenient to treat the two components of the Cauchy data as follows: If $f\in \cH(\Sigma; W\oplus W)$  and $g= R_{\rm F}f$ we will set
\beq\label{turut}
\tf\defeq  Sf, \quad \tg\defeq  Sg. 
\eeq
 Note that $S$ maps $H^{\12}(\Sigma; W)\oplus H^{-\12}(\Sigma; W)$ into $L^{2}(\Sigma; W\oplus W)$.
Let us now collect a few properties of $S$. Clearly
\[
 S^{*}q_{1}S= q_{1},
\] i.e. $S$ is symplectic. Moreover:
\beq\label{en.13b}
\begin{array}{rl}
&S \Psi^{p}_{\rm as}(\Sigma; W\oplus W)S^{-1}= \Psi^{p}_{\rm as}(\Sigma; W\oplus W), \\[2mm]
&S \Psi^{p}_{\rm reg}(\Sigma; W\oplus W)S^{-1}= \Psi^{p}_{\rm reg}(\Sigma; W\oplus W).
\end{array}
\eeq
If $\tilde{f}, \tilde{g}$ are as in (\ref{turut}), then  $\tilde{g}= \tilde{R}_{\rm F}\tilde{f}$ for 
\beq\label{defrtil}
\tilde{R}_{\rm F}\defeq  S R_{\rm F}S^{-1}= \left(\begin{array}{cccc}
\one&0&0&0\\
0&\one&0&0\\
0&-\i\tilde{\delta}_{\Sig}&\one&0\\
\i\tilde{d}_{\Sig}&0&0&\one
\end{array}\right)\in \Psi^{0}(\Sigma; W\oplus W),
\eeq
and
\begin{equation}
\label{defdetilde}
\tilde{\delta}_{\Sig}\defeq  \epsilon_{t}^{-\12}{\bdeltas}\epsilon_{\Sig}^{-\12}, \ \tilde{d}_{\Sig}\defeq   \epsilon_{\Sig}^{-\12}{\bds}\epsilon_{t}^{-\12}.
\end{equation}
Finally let us express the transformed reference projection. If $\Sigma= \rr^{d}$ then:
\beq\label{transfo}
\tilde{\Proj}_{0}\defeq  S\Proj_{0}S^{-1}= \left(
\begin{array}{cccc}
0&0&0&0\\
0&\one - \epsilon_{\Sig}^{\12}\pi\epsilon_{\Sig}^{-\12}&0&0\\
0&0&\one&0\\
0&\i \epsilon_{\Sig}^{-\12} \vara\circ \varb \epsilon_{\Sig}^{-\12}&0&\one
\end{array}\right),
\eeq
and if $\Sigma$ is compact:
\beq\label{transfol}
\tilde{\Proj}_{0}\defeq  S\Proj_{0}S^{-1}= \left(
\begin{array}{cccc}
0&0&0&0\\
0&\one - \epsilon_{\Sig}^{\12}\pi\epsilon_{\Sig}^{-\12}&0&0\\
0&0&\one&0\\
0&\i \epsilon_{\Sig}^{-\12} \pi_{1}\vara\circ \varb \epsilon_{\Sig}^{-\12}&0&\epsilon_{\Sig}^{-\12}\pi_{1}\epsilon_{\Sig}^{\12}
\end{array}\right).
\eeq
%ici
\subsection{Operator classes for adapted Cauchy data}\label{foutoir}
It follows from the above discussion that after going to the adapted Cauchy data and conjugating by $S$,  the  class 
$\Psi^{-1}_{\rm reg}(\Sigma; W\oplus W)$ appearing in Sect. \ref{sec2} should be  replaced by $
\tilde{R}_{\rm F}\Psi^{-1}_{\rm reg}(\Sigma; W\oplus W) \tilde{R}_{\rm F}^{-1}$, 
which is different from $\Psi^{-1}_{\rm reg}(\Sigma; W\oplus W)$. 
In this subsection we introduce  classes of pseudodifferential operators in which the operator equation
$\tilde{\delta}_{\Sig}\circ v=r$ can be solved in $v$ (see Lemma \ref{tecnico}) and which contain the class  $
\tilde{R}_{\rm F}\Psi^{-1}_{\rm reg}(\Sigma; W\oplus W) \tilde{R}_{\rm F}^{-1}$.
We first introduce some  notation.

In the sequel $i, j$ are indices equal to either $0$ or $1$, and  $\alpha, \beta$ are indices equal to either $t$ or $\scriptstyle\Sigma$.  
 If $\alpha=t$, resp. $\scriptstyle\Sigma$, we set $\bar{\alpha}=\, \scriptstyle\Sigma$, resp. $t$ and:
 \[
s_{\alpha}=\begin{cases}
\tilde{d}_{\Sig},\hbox{ if }\alpha=t,\\
\tilde{\delta}_{\Sig},\hbox{ if }\alpha={\scriptstyle\Sigma},
\end{cases}
\]
so that $s_{\alpha}\in \Psi^{0}(\Sigma; W_{\alpha}, W_{\bar\alpha})$.

If $c\in \Psi^{p}(\Sigma; W\oplus W)$ we denote by  $c_{i\alpha, j \beta}$ its matrix entries according to the decomposition
\[
W\oplus W= (W_{t}\oplus W)\oplus (W_{t}\oplus W_{\Sig})= W_{0t}\oplus W_{0\Sig}\oplus W_{1t}\oplus W_{1\Sig}.
\]
Recall also that $\chi_{\gs}$ denotes a cutoff function as in (\ref{cutoffs}).
\begin{definition}
Let $p\in \rr$. \ben 
\item We set
\[
\begin{aligned}
\widetilde\Psi^{p}_{\rm reg, r}(\Sigma; W_{\beta}, W_{\alpha})\defeq & \ \Psi^{p}_{\rm as}(\Sigma; W_{\beta},W_{\alpha}) \chi_{\gs}(h_{\beta})+ 
\Psi^{p}_{\rm as}(\Sigma; W_{\bar \beta},W_{\alpha}) s_{\beta},\\[2mm]
\widetilde\Psi^{p}_{\rm reg, l}(\Sigma; W_{\beta}, W_{\alpha})\defeq & \ \chi_{\gs}(h_{\alpha})\Psi^{p}_{\rm as}(\Sigma; W_{\beta},W_{\alpha})+ s_{\bar \alpha}\Psi^{p}_{\rm as}(\Sigma; W_{\beta}, W_{\bar \alpha}),\\[2mm]
\widetilde{\Psi}^{p}_{\rm reg}(\Sigma; W_{\beta}, W_{\alpha})\defeq & \ \chi_{\gs}(h_{\alpha})  \Psi^{p}_{\rm as}(\Sigma; W_{\beta},W_{\alpha})\chi_{\gs}(h_{\beta})+ s_{\bar \alpha} \Psi^{p}_{\rm as}(\Sigma; W_{\beta},W_{\bar\alpha})\chi_{\gs}(h_{\beta})\\[2mm]
&+s_{\bar \alpha} \Psi^{p}_{\rm as}(\Sigma; W_{\beta},W_{\bar\alpha})\chi_{\gs}(h_{\beta})+ s_{\bar \alpha} \Psi^{p}_{\rm as}(\Sigma; W_{\bar \beta},W_{\bar \alpha})s_{\alpha}.
\end{aligned}
\]
\item
We say that $c\in \widetilde{\Psi}_{{\rm reg}, \sharp}^{p}(\Sigma; W\oplus W)$ for $\sharp= {\rm l, r, }$ if $c_{i\alpha, j\beta}\in\widetilde{\Psi}^{p}_{{\rm reg}, \sharp}(\Sigma; W_{\alpha}, W_{\beta})$ for all $i, \alpha$, $j, \beta$.
\een
\end{definition}
The next lemma shows that the above classes have similar properties to $\Psi^{p}_{\rm reg}(\Sigma; W\oplus W)$.
\begin{lemma}\label{lem4} The following properties hold:
\ben
\item  $\tilde{R}_{\rm F} \Psi^{p}_{\rm as}(\Sigma; W\oplus W)\tilde{R}_{\rm F}^{-1}= \Psi^{p}_{\rm as}(\Sigma; W\oplus W)$,

\item $\tilde{R}_{\rm F} \Psi^{p}_{\rm reg}(\Sigma; W\oplus W)\tilde{R}_{\rm F}^{-1}\subset\widetilde{\Psi}^{p}_{{\rm reg}}(\Sigma; W\oplus W)\subset \Psi^{p}_{\rm as}(\Sigma; W\oplus W)$,

\item Let $c_{R}\in \Psi^{-\varepsilon}_{{\rm reg}, \sharp}(\Sigma; W\oplus W)$ for $\varepsilon>0$ and let $\alpha\in \rr$. Then for $R\geq R_{0}$ we have
\[
(\one + c_{R})^{\alpha}\in \one + \Psi^{-\varepsilon}_{{\rm reg}, \sharp}(\Sigma; W\oplus W).
\]
\een 
\end{lemma}
\proof (1) follows from the fact that  the class $ \Psi^{p}_{\rm as}$ is invariant under left or right composition with elements of $\Psi^{0}$. (2) is  a routine computation, introducing the matrix entries of some $c\in \Psi^{p}_{\rm reg}(\Sigma; W\oplus W)$ and using  (\ref{defrtil}). To prove (3) we use the identity $(\one -a)^{-1}= \one + a + a(\one -a)^{-1}a$ and the following easy observations:
\[
\Psi^{0}_{\rm as}\widetilde{\Psi}^{-\varepsilon}_{\rm reg, r}\subset \widetilde{\Psi}^{-\varepsilon}_{\rm reg, r},\ \widetilde{\Psi}^{-\varepsilon}_{\rm reg, l}\Psi^{0}_{\rm as}\subset \widetilde{\Psi}^{-\varepsilon}_{\rm reg, l},\ \widetilde{\Psi}^{-\varepsilon}_{\rm reg, l}\widetilde{\Psi}^{-\varepsilon}_{\rm reg, r}\subset \widetilde{\Psi}_{\rm reg}^{-2\varepsilon}. \ \Box
\]
We end this subsection with another technical lemma, which will  motivate the introduction of the above operator classes.
\begin{lemma}\label{tecnico}
 Let $r\in \widetilde{\Psi}^{p}_{\rm reg}(\Sigma; W_{\alpha}, W_{t})$ for $\alpha= t, {\scriptstyle\Sigma}$. Then there exists $v\in \widetilde{\Psi}^{p}_{\rm reg, r}(\Sigma; W_{\alpha}, W_{\Sig})$ such that
 \[
\tilde{\delta}_{\Sig}\circ v=r.
\]
\end{lemma} 
\proof Since $r\in \widetilde{\Psi}^{p}_{\rm reg}(\Sigma; W_{\alpha}, W_{t})$ we can write
\[
r= \chi_{\gs}(h_{t})m_{1}+\tilde{\delta}_{\Sig}m_{2},\ m_{1}\in \widetilde{\Psi}^{p}_{\rm reg, r}(\Sigma; W_{\alpha}, W_{t}), \ m_{2}\in\widetilde{\Psi}^{p}_{\rm reg, r}(\Sigma; W_{\alpha}, W_{\Sig}).
\]
If follows that
\[
v= \epsilon_{\Sig}^{\12}\bards h_{t}^{-\12}\epsilon_{t}^{\12}\chi_{\gs}(h_{t})m_{1}+ m_{2}\in \widetilde{\Psi}^{p}_{\rm reg, r}(\Sigma; W_{\alpha}, W_{\Sig})
\]
solves $\tilde{\delta}_{\Sig}\circ v=r$. \qeds
\subsection{Technical estimates for $\Sigma= \rr^{d}$}
In this subsection we collect some delicate technical estimates on the operators $\pi,\varb$ in the case $\Sigma= \rr^{d}$. 
It is convenient to introduce some notation related to Hypothesis \ref{as:background}: if $V$ is a finite dimensional vector space we set:
 \[
S^{m}_0(\Sigma; V)\defeq\{f\in \cinf(\Sigma; V) : \ 
\p_{x}^{\alpha}f(x)\in O(\langle x\rangle^{m}), \ \alpha\in \nn^{d}\}.
\]
Abusing notation we see that Hypothesis \ref{as:background} implies  that 
\[
\wbar{A}_{\Sig}\in S_0^{0}, \ \wbar{\delta}_{\Sig}\wbar{F}_{\Sig}\in S_0^{-1}, \ \wbar{F}_{t}\in S_0^{-2}.
\]
Recall that $B^{-\infty}(\Sigma;V_1,V_2)$ denotes the space of operators that map $H^{-m}(\Sigma;V_1)\to H^m(\Sigma;V_2)$ for all $m$.
\begin{lemma}\label{keylemma}
 Assume  that  $\Sigma= \rr^{d}$. Then:
 \ben
 \item $\bards \chi_{\ls}(h_{t})h_{t}^{-1}\bardels\in B^{-\infty}(\Sigma;W_{\Sig})$, 
 \item $\x^{-1} \chi_{\ls}(h_{t})h_{t}^{-1}\bardels\in B^{-\infty}(\Sigma;W_{\Sig})$
 \item $\pi \in \Psi^{0}(\Sigma; W_{\Sig})+ B^{-\infty}(\Sigma; W_{\Sig})$,
 \item $\varb\in \Psi^{-1}(\Sigma; W_{\Sig}, W_{t})+ \x B^{-\infty}(\Sigma; W_{\Sig}, W_{t})$,
 \item $\chi_{\gs}(h_{\Sig})\pi\in \Psi^{0}(\Sigma; W_{\Sig})+
  \x^{-1}B^{-\infty}(\Sigma; W_{\Sig})$,
 \item $\vara\circ \varb\in \x^{-1}\Psi^{-1}(\Sigma; W_{\Sig})+ \x^{-1}B^{-\infty}(\Sigma; W_{\Sig})$.
 \een
\end{lemma}
\proof 
(1): let $A=\bards \chi_{\ls}(h_{t})h_{t}^{-1}\bardels$. We need to prove that 
\[
(h_{\Sig}^{n}+ \i )A(h_{\Sig}^{n}+ \i): L^{2}\to L^{2}, \ \forall n\in \nn,
\]  which will follow from
\[
\begin{array}{rll}
&i):\ A: L^{2}\to L^{2},  & ii): \ Ah_{\Sig}: H^{-n}\to L^{2}, \\[2mm]
&iii):\  h_{\Sig}A: L^{2}\to H^{n}, & iv):\ h_{\Sig}Ah_{\Sig} :H^{-n}\to H^{n}.
\end{array}
\]
{\it i)} is straightforward by Lemma \ref{stup1}. Let us now prove {\it ii)}. By Lemma \ref{ln.2} (3), we have:
\[
Ah_{\Sig}= \bards \chi_{\ls}(h_{t})h_{t}^{-1}\bardels h_{\Sig}= \bards \chi_{\ls}(h_{t})\bardels+ \bards \chi_{\ls}(h_{t})h_{t}^{-1}R,
\]
for $R= \bardels \wbar{F}_{\Sig}\lrcorner\cdot$. The first term on the right belongs to $\Psi^{-\infty}$. We write the second term as $\bards h_{t}^{-1}\x^{-1}\circ \x \chi_{\ls}(h_{t})R$. The first factor is bounded on $L^{2}$ by Lemma \ref{stup1}, the second belongs to $\Psi^{-\infty}$, since $\bardels \wbar{F}_{\Sig}\in S_0^{-1}$. This implies {\it ii)} and hence {\it iii)} by duality. To prove {\it iv)} we write
\[
\begin{array}{rl}
h_{\Sig}Ah_{\Sig}=&h_{\Sig}\bards \chi_{\ls}(h_{t})\bardels + h_{\Sig} \bards \chi_{\ls}(h_{t}) h_{t}^{-1}R\\[2mm]
=&h_{\Sig}\bards \chi_{\ls}(h_{t})\bardels+  \bards \chi_{\ls}(h_{t})R+ R^{*} \chi_{\ls}(h_{t}) h_{t}^{-1}R.
\end{array}
\]
The first two terms belong to $\Psi^{-\infty}$. We factor the third term as:
\[
R^{*}\chi_{\ls}(h_{t})\x\circ \x^{-1}h_{t}^{-1}\x^{-1}\circ \x \tilde{\chi}_{\ls}(h_{t})R,
\]
for some cutoff function $\tilde{\chi}_{\ls}$ with the same properties as $\chi_{\ls}$ and $\tilde{\chi}_{\ls}\chi_{\ls}= \chi_{\ls}$.
The first and last factor belong to $\Psi^{-\infty}$, the middle one is bounded on $L^{2}$ by Lemma \ref{stup1}. This proves 
{\it iv)} and completes the proof of (1).

(2): the proof of (2) is completely analogous to the proof of (1) and left to the reader.

(3): we write 
\[
\pi= \bards \chi_{\gs}(h_{t}) h_{t}^{-1}\bardels + \bards \chi_{\ls}(h_{t}) h_{t}^{-1}\bardels.
\]
The first term belongs to $\Psi^{0}$, the second to $B^{-\infty}$ by (1). This proves (3).

(4):  we write 
\[
\varb= \chi_{\gs}(h_{t})h_{t}^{-1} \bardels+ \chi_{\ls}(h_{t})h_{t}^{-1}\bardels,
\]
the first term belongs to $\Psi^{-1}$, the second to $\x B^{-\infty}$, by (2).

(5): We write as before:
\[
\chi_{\gs}(h_{\Sig})\pi= \chi_{\gs}(h_{\Sig})\bards \chi_{\gs}(h_{t}) h_{t}^{-1}\bardels + \chi_{\gs}(h_{\Sig})\bards \chi_{\ls}(h_{t}) h_{t}^{-1}\bardels.
\]
The first term belongs to $\Psi^{0}$.  We write the second term as
\[
\chi_{\gs}(h_{\Sig})h_{\Sig}^{-1}h_{\Sig}\bards \chi_{\ls}(h_{t}) h_{t}^{-1}\bardels= \chi_{\gs}(h_{\Sig})h_{\Sig}^{-1}\bards \chi_{\ls}(h_{t})\bardels+ \chi_{\gs}(h_{\Sig})h_{\Sig}^{-1}R^{*} \chi_{\ls}(h_{t}) h_{t}^{-1}\bardels.
\]
The first term belongs to $\Psi^{-\infty}$.  We factor the second term as:
\[
\x^{-1}\circ \x\chi_{\gs}(h_{\Sig})h_{\Sig}^{-1}R^{*}\x\circ \x^{-1}\chi_{\ls}(h_{t}) h_{t}^{-1}\bardels.
\]
Now $ \x\chi_{\gs}(h_{\Sig})h_{\Sig}^{-1}R^{*}\x\in \Psi^{0}$ since $\bardels \wbar{F}_{\Sig}\in S_0^{-2}$ and $\x^{-1}\chi_{\ls}(h_{t}) h_{t}^{-1}\bardels\in B^{-\infty}$ by (2). This proves that the second term belongs to $\x^{-1}B^{-\infty}$ and completes the proof of (5).

(6):  we write once again:
\[
\vara\circ \varb= \vara\chi_{\gs}(h_{t})h_{t}^{-1}\bardels + \vara\circ \chi_{\ls}(h_{t})h_{t}^{-1}\bardels.
\]
The first term belongs to $\x^{-1}\Psi^{-1}$, since $\wbar{F}_{t}\in S_0^{-1}$. The second term belongs to $\x^{-1}B^{-\infty}$, using (2) and the fact that $\wbar{F}_{t}\in S_0^{-2}$. \qeds

\subsection{Construction of the projection $\Proj$}\label{copilo}
In this subsection we construct the projection $\Proj$.   The first step consists in determining its range.

\begin{proposition}\label{pn.2}
 There exists $s_{-1, R}\in \Psi^{-1}_{\rm as}(\Sigma; W\oplus W)$ such that:
 \[
(\one + s_{-1, R}) \Ran \Proj_{0}\cap \Ker K_{\Sig}^{\dag}\subset (\one + r_{-1, R})L^{2}(\Sigma; W_{\Sig}\oplus W_{\Sig}),
\]
where $r_{-1, R}\in \Psi^{-1}_{\rm reg}(\Sigma; W\oplus W)$ is the operator in Prop. \ref{pn.1}.
\end{proposition}
\proof 
We set $g= R_{\rm F}f$. It is easy to check that for $\Proj_{0}$ given either by (\ref{en.16}) or (\ref{en.16comp}):
\begin{equation}
\label{en.17}
\begin{array}{rl}
f\in \Ker \kasd&\Rightarrow \ g^{1}_{t}=0,\\[2mm]
f\in \cH'(\Sigma; W_{\Sig}\oplus W_{\Sig})& \Leftrightarrow \ g^{0}_{t}=0, \ g^{1}_{t}+ \i \bardels g^{0}_{\Sig}=0,\\[2mm]
f\in \Ran \Proj_{0} &\Rightarrow \ g^{0}_{t}=0, \ \bardels g^{0}_{\Sig}=0.
\end{array}
\end{equation}
As explained in Subsect. \ref{foutoir} it is convenient to work with  $\tg= Sg$, which  amounts to replace 
 $r_{-1, R}$ by  $\tilde{R}_{\rm F}r_{-1, R}\tilde{R}_{\rm F}^{-1}\eqdef  \tilde{r}$,   and $s_{-1, R}$ by $\tilde{R}_{\rm F}r_{-1, R}\tilde{R}_{\rm F}^{-1}\eqdef  \tilde{s}$.

 By Lemma \ref{lem4} we know that $\tilde{r}\in \widetilde{\Psi}^{-1}_{\rm reg}(\Sigma; W\oplus W)$, and we will look for $\tilde{s}\in \widetilde{\Psi}^{-1}_{\rm reg, r}(\Sigma; W\oplus W)$. Again by Lemma \ref{lem4} it will follow that $s\in \Psi^{-1}_{\rm as}(\Sigma; W\oplus W)$.

  Expressed in terms of $\tilde{g}$, the statements in  (\ref{en.17}) become:
 \begin{equation}
\label{en.17bb}
\begin{array}{rl}
f\in \Ker \kasd&\Rightarrow \ \tg^{1}_{t}=0,\\[2mm]
f\in \cH'(\Sigma; W_{\Sig}\oplus W_{\Sig})& \Leftrightarrow \  \tg^{0}_{t}=0, \ \tg^{1}_{t}+ \i \tilde{\delta}_{\Sig} \tg^{0}_{\Sig}=0,\\[2mm]
f\in \Ran \Proj_{0} &\Rightarrow \ \tg^{0}_{t}=0, \ \tilde{\delta}_{\Sig} \tg^{0}_{\Sig}=0,
\end{array}
\end{equation}
where $\tilde{\delta}_{\Sig}=  \epsilon_{t}^{-\12}\bardels\epsilon_{\Sig}^{-\12}$ was defined in (\ref{defdetilde}).
We set: 
\[
A_{1}= \left(\begin{array}{cccc}
\one&0&0&0\\
0&\tilde{\delta}_{\Sig}&\i^{-1}&0
\end{array}\right), \ A_{2}= \left(\begin{array}{cccc}
\one&0&0&0\\
0&\tilde{\delta}_{\Sig}&0&0
\end{array}\right),
\]
so that
\beq\label{en.17b}
\begin{array}{rl}
f\in (\one + r)\cH'(\Sigma; W_{\Sig}\oplus W_{\Sig})& \Leftrightarrow \ \tg\in \Ker\big( A_{1}\circ (\one + \tilde{r})^{-1}\big),\\[2mm]
f\in (\one + s)\Ran \Proj_{0} &\Rightarrow \ \tg\in \Ker \big(A_{2}\circ(\one + \tilde{s})^{-1}\big).
\end{array}
\eeq
To prove the proposition it suffices  to  find $\tilde{s}\in \Psi^{-1}_{\rm reg, r}(\Sigma; W\oplus W)$ such that 
\begin{equation}
\label{en.17c}
 \tg\in \Ker\big( A_{2}\circ(\one + \tilde{s})^{-1}\big), \  \tg^{1}_{t}=0 \Rightarrow \ \tg\in \Ker \big(A_{1}\circ (\one + \tilde{r})^{-1}\big).
\end{equation}
Again by Lemma \ref{lem4} (3), we know that  for $R$ large enough $(\one + \tilde{r})^{-1}= \one + \hat{r}$ for $\hat{r}\in \widetilde{\Psi}^{-1}_{\rm reg}$. Let assume that we have found $\hat{s}\in \widetilde{\Psi}^{-1}_{\rm reg, r}$ such that
\begin{equation}
\label{en.17d}
 \tg\in \Ker \big(A_{2}\circ(\one + \hat{s})\big), \  \tg^{1}_{t}=0 \Rightarrow \ \tg\in \Ker \big(A_{1}\circ (\one + \hat{r})\big).
\end{equation}
Then setting $\one + \tilde{s}\defeq (\one + \hat{s})^{-1}$, we know that $\tilde{s}\in \widetilde{\Psi}^{-1}_{\rm reg, r}$ by Lemma \ref{lem4} and that $\tilde{s}$ solves (\ref{en.17c}). Hence to complete the proof of the proposition, it remains to solve (\ref{en.17d}).

We have
\[
A_{1}= A_{2}+ A_{3} \ \hbox{ for } A_{3}= \left(\begin{array}{cccc}
0&0&0&0\\
0&0&\i^{-1}&0
\end{array}\right).
\]
Therefore we look for $\hat{s}= \hat{r}+ \hat{v}$ and need to find $\hat{v}\in \widetilde{\Psi}^{-1}_{\rm reg, r}$ such that:
\[
A_{2}\hat{v}= A_{3}(\one + \hat{r})\hbox{ on }\{\tilde{g}^{1}_{t}=0\}.
\]
Since $A_{3}=0$ on $\{\tilde{g}^{1}_{t}=0\}$, we finally need to find $\hat{v}$ such that 
\[
A_{2}\hat{v}= A_{3}\hat{r}\hbox{ on }\{\tilde{g}^{1}_{t}=0\}.
\]
A routine computation yields the following equations for the entries of $\hat{v}$:
\beq\label{en.i1}
\begin{array}{rl}
&\hat{v}_{0t, j \beta}=0, \ \forall \ \scriptstyle{j\beta},\\[2mm]
 & \tilde{\delta}_{\Sig} \hat{v}_{0\Sig, j \beta}= \i^{-1}\hat{r}_{1t, j\beta}\hbox{ for }\scriptstyle{j\beta= 0t, \ 0\Sig, \ 1\Sig}.
\end{array}
\eeq
We can set all the other entries of $\hat{v}$ to $0$. It remains to solve the  equations in the second line of (\ref{en.i1}).
This can be done by applying Lemma \ref{tecnico}.
  This completes the proof of the proposition.  \qeds

In the proof of Prop. \ref{pn.2}, we use the assumption that $(M,g)$ is ultra-static: otherwise the expression in the second line of (\ref{en.17bb}) becomes more complicated and it is not clear how to choose the reference projection $\Proj_0$.

If $\Sigma= \rr^{d}$ we will need  some further properties of the operator $s_{-1,R}$ constructed in Prop. \ref{pn.2}.
\begin{proposition}\label{pn.3}
Assume that $\Sigma= \rr^{d}$. 
 Then  there exists $R_{0}$ such that for $R\geq R_{0}$ and for any $m\in \rr$:
\[
\begin{array}{rl}
i) & \one + s_{-1,R}\Proj_{0}:\  H^{m+\12}(\Sigma; W)\oplus H^{m-\12}(\Sigma;W)\tarrow H^{m+\12}(\Sigma; W)\oplus H^{m-\12}(\Sigma;W),\\[2mm]
ii)&\x(\one + s_{-1,R}\Proj_{0})\x^{-1} : \ H^{m+\12}(\Sigma; W)\oplus H^{m-\12}(\Sigma;W)\tarrow H^{m+\12}(\Sigma; W)\oplus H^{m-\12}(\Sigma;W).%\\[2mm]
%iii) & \one + Ps_{-1,R}P : \ H^{m}(\Sigma; W\oplus W)\tarrow H^{m}(\Sigma; W\oplus W),\\[2mm]
%iv)&\one +  Ps_{-1,R}P: \ \x^{-1} L^{2}(\Sigma; W\oplus W)\tarrow \x^{-1}L^{2}(\Sigma; W\oplus W).
\end{array}
\]
\end{proposition}
\proof 
As before we conjugate all operators by $\tilde{R}_{\rm F}$, which amounts to replace  $s_{-1,R}$ by $\tilde{s}_{-1,R}= \tilde{R}_{\rm F}s_{-1, R}\tilde{R}_{\rm F}^{-1}$, $\Proj_{0}$ by $\tilde{\Proj}_{0}= \tilde{R}_{\rm F}\Proj_{0}\tilde{R}_{\rm F}^{-1}$ and $H^{m+\12}\oplus H^{m-\12}$ by $H^{m}\oplus H^{m}$. From the expression (\ref{transfo}) of $\tilde{\Proj}_{0}$ we see that  the entries of $\tilde{s}_{-1,R}\tilde{\Proj}_{0}$ are of one of these three types:
\[
1)\ \Psi^{-1}_{\rm reg, r}, \quad 2)\ \Psi^{-1}_{\rm reg, r}(\one -\pi), \quad 3)\ \Psi^{-1}_{\rm reg, r}\vara\circ  \varb.
\]
Terms of type 1) are simply considered as belonging to $\Psi^{-1}_{\rm as}$. To control terms of type 2) we recall that $\Psi^{-1}_{\rm reg, r}= \Psi^{-1}_{\rm as} \chi_{\gs}(h_{\Sig})+ \Psi^{-1}_{\rm as} \tilde{\delta}_{\Sig}$.  By Lemma \ref{keylemma} (5) we know that 
$\Psi^{-1}_{\rm as} \chi_{\gs}(h_{\Sig})\pi\in \Psi^{-1}_{\rm as} + \x^{-1}\Psi^{-1}_{\rm as} B^{-\infty}$.
The terms of type 3) belong to $\Psi^{-1}_{\rm as} + \x^{-1}\Psi^{-1}_{\rm as} B^{-\infty}$, by Lemma \ref{keylemma} (6).
It follows that 
\begin{equation}
\label{toto.ert}
\tilde{s}_{-1,R}\tilde{\Proj}_{0}\in \Psi^{-1}_{\rm as}+ \x^{-1}\Psi^{-1}_{\rm as} B^{-\infty}.
\end{equation}
Let us now  prove {\it i)}. From (\ref{toto.ert}) we first deduce that $\|\tilde{s}_{-1,R}\tilde{\Proj}_{0}\|_{B(L^{2})}\in o(R^{0})$, hence we can find $R_{0}$ such that
\[
\one+\tilde{s}_{-1,R}\tilde{\Proj}_{0}:\ L^{2}(\Sigma; W\oplus W)\tarrow  L^{2}(\Sigma;W\oplus W).
\]
Let us first assume that $m>0$.  We apply the identity  
\[
(\one -A)^{-1}=\sum_{j=0}^{n-1} A^j + A^{n}(\one -A)^{-1}
\]
 to $A= -\tilde{s}_{-1,R}\tilde{\Proj}_{0}$. By (\ref{toto.ert}) we know that  $\tilde{s}_{-1,R}\tilde{\Proj}_{0}: H^{m}(\Sigma; W\oplus W)\to H^{m+1}(\Sigma; W\oplus W)$. We obtain taking $n$ large enough that 
\[
(\one + \tilde{s}_{-1,R}\tilde{\Proj}_{0})^{-1}: H^{m}(\Sigma; W\oplus W)\to H^{m}(\Sigma; W\oplus W), 
\]
 which proves {\it i)} for $m>0$. The same argument shows that for $m>0$
\[
\one + (\tilde{s}_{-1,R}\tilde{\Proj}_{0})^{*}: \ H^{m}(\Sigma; W\oplus W)\tarrow H^{m}(\Sigma; W\oplus W),
\]
which by duality proves {\it i)} for $m<0$.

To prove {\it ii)} we split $\tilde{s}_{-1,R}\tilde{\Proj}_{0}$ as $m_{1,R}+m_{2,R}$, where $m_{1,R}\in \Psi^{-1}_{\rm as}$ and $m_{2,R}\in \x^{-1}\Psi^{-1}_{\rm as}B^{-\infty}$. We can choose $R_{0}$ above large enough such that $(\one + m_{1, R})^{-1}\in \Psi^{0}$ for $R\geq R_{0}$.  We have
\[
(\one + \tilde{s}_{-1,R}\tilde{\Proj}_{0})^{-1}= (\one + m_{1, R})^{-1}(\one - m_{2,R}(\one + \tilde{s}_{-1,R}\tilde{\Proj}_{0})^{-1}).
\]
Now  $m_{2, R}:\  H^{m}\to \x^{-1}H^{m}$ and $(\one + m_{1, R})^{-1}: \x^{-1}H^{m}\to \x^{-1}H^{m}$ by pdo calculus, which implies that
$(\one + \tilde{s}_{-1,R}\tilde{\Proj}_{0})^{-1}: \x^{-1}H^{m}\to \x^{-1}H^{m}$.  This completes the proof of the proposition.
\qeds
\subsection{The projection $\Proj$ and the right inverse $B$}
We now define a projection $\Proj$ and a right inverse $B$ to $K_{\Sig}$ as in \ref{sss:proj}, \ref{sss:rightinverse}.

\bet\label{th.n2}
 Let $\Proj_{0}$ be given by (\ref{en.16}) if $\Sigma= \rr^{d}$ and (\ref{en.16comp}) if $\Sigma$ is compact. Let also $s_{-1, R}$ be the operator constructed in Prop. \ref{pn.2}. Then  there exists $R_{0}$ such that for all $R\geq R_{0}$:
\ben
\item the operator
 \[
\Proj\defeq  (\one + s_{-1, R})\Proj_{0}(\one + \Proj_{0} s_{-1, R}\Proj_{0})^{-1}
\]
is  a bounded projection on  $L^{2}(\Sigma; W\oplus W)$. 
\item  moreover
\[
\one - \Proj= (\one -\Proj_{0})(\one + s_{-1, R}\Proj_{0})^{-1}.
\]
\item one has
\[
\begin{array}{rl}
a)&\Ker\,\Proj= {\rm Ran}K_{\Sig}, \\[2mm]
b)&\lambda_{1\Sig}^{\pm}\hbox{ are positive on }\Ran \Proj\cap \Ker K_{\Sig}^{\dag}.
\end{array}
 \]
 \item $\Proj: \cH(\Sigma; W)\to \cH(\Sigma; W)$, $\Proj: \cH'(\Sigma; W)\to \cH'(\Sigma; W)$.
 \item if $\Sigma$ is compact then $\Proj\in \Psi^{\infty}(\Sigma; W\oplus W)$.
 \een
\eet
\proof If $\Proj_{0}$ is a bounded projection on a Hilbert space $\cH$ and $\|r\|\ll 1$, then $\Ker\,\Proj_{0}$ and $(\one + r)\Ran \Proj_{0}$ are supplementary subspaces and it is easy to show that the projection $\Proj$ with $\Ker\,\Proj=\Ker\,\Proj_{0}$ and $\Ran \Proj= (\one +r)\Ran \Proj_{0}$ is given by the formulas in (1) and (2).  Statement (3a) follows  from $\Ker\,\Proj= \Ker\,\Proj_{0}= \Ran K_{\Sig}$. Statement (3b)  follows from  $\Ran \Proj= (\one + s_{-1, R})\Ran \Proj_{0}\subset (\one + r_{-1, R})\cH(\Sigma; W_{\Sig}\oplus W_{\Sig})$ by Prop. \ref{pn.2}, and from Prop. \ref{pn.1}. 

Let us now prove (4). It suffices to prove the corresponding statements for $\one - \Proj$. Using that by Prop. \ref{pn.3} $(\one + s_{-1, R}\Proj_{0})^{-1}$ maps  $\cH(\Sigma; W)$ and $\cH'(\Sigma; W)$ into themselves, we can replace $\one - \Proj$ by $\one -\Proj_{0}$. The result follows then from the expression of $\Proj_{0}$ in (\ref{en.16}) and statements (3), (6) of Lemma \ref{keylemma}. 
Finally the fact that  $\Proj\in \Psi^{\infty}$ if $\Sigma$ is compact, follows from the same property of $\Proj_{0}$, see Lemma \ref{ln.66}.   This proves (5). \qeds

Let us now define the right inverse $B$ to $K_{\Sig}$. 
\begin{proposition}\label{bizirofinal}
 Let $B_{0}$ be given by (\ref{en.100}) if $\Sigma= \rr^{d}$ or by (\ref{en.1001}) if $\Sigma$ is compact.  Let
\begin{equation}
\label{en.101}
B\defeq  B_{0}(\one+ s_{-1, R}\Proj_{0})^{-1}.
\end{equation}
Then 
\begin{equation}
\label{en.102}
K_{\Sig}B= \one- \Proj, \quad {  B K_\Sig =\one}.
\end{equation}
Moreover 
\ben
\item if $\Sigma= \rr^{d}$ then $B: \cH(\Sigma; W)\to \x\cH(\Sigma; W)$, $B :\cH'(\Sigma; W)\to \x\cH'(\Sigma; W)$.
\item if $\Sigma$ is compact then $B\in \Psi^{\infty}(\Sigma; W\oplus W, W_{t}\oplus W_{t})$.
\een
\end{proposition}
\proof The fact that $K_{\Sig}B= \one - \Proj$ follows from the definitions of $B$, $\Proj$ and the  fact that $K_{\Sig}B_{0}= \one - \Proj_{0}$.  {  The identity $B K_\Sig =\one$ follows from $B_0 K_\Sig =\one$ and $\Proj K_\Sig=0$.} To prove (2) we can as in the proof of Thm. \ref{th.n2} replace $B$ by $B_{0}$. The statement follows then for the expression (\ref{en.100}) of $B_{0}$ and from (4) of Lemma \ref{keylemma}. Finally, (2) follows from the fact that $B_{0}$, $\Proj_{0}$ belong to $\Psi^{\infty}$, see Lemmas  \ref{ln.66} and \ref{bizirobis}. \qeds

\subsection{Proof of Thm. \ref{maintheo2}}
 We now complete the proof of Thm. \ref{maintheo2}, by checking the assumptions of  Thm. \ref{thm:fincov}. 
 We take for  $c_{i}^{\pm}$ for $i=0,1$ the operators  constructed in Prop. \ref{1.4} for the operators $\p_{t}^{2}+ a_{i}(t)= D_{i}$.

 - $c_{i}^{\pm}$ are pseudodifferential operators,  hence $c_{i}^{\pm}$ satisfy  (\ref{eq:azert} i),  ii) and $c_{0}^{\pm}$ satisfy  (\ref{eq:fouine}) iii).
 
 -  $G_{i\Sig}$ are equal to $\i\mat{J_{i}}{0}{0}{-J_{i}}$, for $J_{i}$ given in (\ref{e1.1.00}),  hence conditions (\ref{eq:gsig1}) and (\ref{eq:fouine}) i) are satisfied.
 
-  $K_{\Sig}$ is a matrix of differential operators with coefficients bounded with all derivatives, by Hypothesis \ref{as:background}, hence  conditions (\ref{eq:terrier}) and (\ref{eq:fouine}) ii) are satisfied.

- $\Proj$ and $B$ satisfy conditions (\ref{nmis.0}) and (\ref{eq:moufette}), by Thm. \ref{th.n2}  and Prop. \ref{bizirofinal}.

- the positivity condition (\ref{D520}) is satisfied by $\Proj$, using Thm. \ref{th.n2} and the fact that $\Ran \Proj\cap \Ker K_{\Sig}^{\dag}= \Proj\Ker K_{\Sig}^{\dag}$ since $\Ker \Proj= \Ran K_{\Sig}\subset \Ker K_{\Sig}^{\dag}$.

- the two-point functions $\lambda_{1\Sig}^{\pm}$ are Hadamard, by Prop. \ref{1.4}. To prove that $\tilde{\lambda}^{\pm}_{1\Sig}$ are also Hadamard, we  need to check that $c_{1{\rm reg}}^{\pm}$ are regularizing. This delicate point is shown in Prop. \ref{keyprop} below. The proof of Thm. \ref{maintheo2} is complete.\qed

\begin{remark}
It is easy to deduce from (\ref{en.10b}) and the property $\Ker\,\Pi=\Ran\,K_\Sig$ that the two-point functions $\tilde{\lambda}^{\pm}_{1\Sig}$ we construct have the property that $\tilde{\lambda}^{+}_{1\Sig}+\tilde{\lambda}^{-}_{1\Sig}$ is injective on $\Ker K_\Sig^\dag$. This issue is related to faithfulness of the state $\omega$. 
\end{remark}

\begin{proposition}\label{keyprop}
\ben
\item  assume that $\Sigma= \rr^{d}$. Then for any $n\in \nn$ one has:
 \[
\begin{array}{rl}
i)& R_{-\infty}B: H^{-n}(\Sigma; W\oplus W)\to \x H^{n}(\Sigma; W\oplus W),\\[2mm]
ii)&(\one - \Proj^{\dag})R_{-\infty}B:   H^{-n}(\Sigma; W\oplus W)\to \x H^{n}(\Sigma; W\oplus W).
\end{array}
\]
\item assume that $\Sigma$ is compact.  Then $R_{-\infty}B$ and $(\one - \Proj^{\dag})R_{-\infty}B$ belong to $\Psi^{-\infty}(\Sigma; W\oplus W)$.
\een
\end{proposition}
\proof 
The proof of (2) is straightforward, since  if $\Sigma$ is compact we know that $B$, $(\one - \Proj^{\dag})\in \Psi^{\infty}$ and $R_{-\infty}\in \Psi^{-\infty}$. 

We now turn to the proof of (1) which is much more delicate. 
 The Sobolev spaces or pseudodifferential classes between the various vector bundles over $\Sigma$ will be abbreviated $H^{m}$, $\Psi^{p}$, $m, p\in \rr$.
 
 We will work with the adapted Cauchy data. Note that because the operators $R_{\rm F}$ and $R_{\rm F}^{-1}$ are differential operators (see Lemma \ref{n1.1}), the operator $R_{-\infty}$, expressed in term of adapted Cauchy data, i.e. $R_{\rm F} R_{-\infty}R_{\rm F}^{-1}$ belongs also to $\Psi^{-\infty}$, and will still be denoted by $R_{-\infty}$.

Let us first consider the operator $R_{-\infty}B_{0}$, which we write as a $4\times 4$ matrix.  
A routine computation shows that the entries of $R_{-\infty}B_{0}$  are of  one of the two forms
\beq\label{en.104}
r_{-\infty}, \ r_{-\infty}\varb,
\eeq
for $r_{-\infty}\in \Psi^{-\infty}$.  From Lemma \ref{keylemma} (4) we obtain that $\varb: H^{-m}\to \x H^{-m}$ for all $m\in \nn$. Since  $r_{-\infty}: \x H^{-n}\to \x H^{n}$ by pdo calculus, we obtain that  $R_{-\infty}B_{0}: H^{-n}\to \x H^{n}$. By Prop. \ref{pn.3}  {\it i)} we know that $1+ s_{-1}\Proj_{0}: H^{-n}\to H^{-n}$.  This completes the proof of {\it i)}.

The proof of  {\it ii)} is more delicate.   We  claim that it suffices to prove that:
\begin{equation}
\label{en.107}
(\one - \Proj_{0}^{\dag})R_{-\infty}B_{0}: H^{-n}\to \x H^{n}, \  \forall n\in \nn.
\end{equation}
In fact   by Thm. \ref{th.n2} we have:
\[
(\one - \Proj^{\dag})= (\one + (s_{-1}\Proj_{0})^{\dag})^{-1}(\one - \Proj_{0}^{\dag}).
\]
By Prop. \ref{pn.3} {\it i)}  $(\one + s_{-1}\Proj_{0})^{-1}: H^{-n}\to H^{-n}$, and by Prop. \ref{pn.3} {\it ii)} and duality $ (\one + (s_{-1}\Proj_{0})^{\dag})^{-1}: \x H^{n}\to \x H^{n}$. Hence {\it ii)} will follow from (\ref{en.107}). 

Let us now prove (\ref{en.107}). 
We write $R_{-\infty}$ as a $4\times 2$ matrix:
\[
R_{-\infty}= \left(\begin{array}{cc}
r_{0t, 0}&r_{0t,1}\\
r_{0\Sig, 0}&r_{0\Sig,1}\\
r_{1t, 0}&r_{1t,1}\\
r_{1\Sig, 0}&r_{1\Sig,1}\\
\end{array}\right).
\]
Using that 
\[
\one - \Proj_{0}^{\dag}= \left(\begin{array}{cccc}
0&0&0&0\\
0&0&0&0\\
0&0&\one&0\\
0&\i \varb^{*}\vara^{*}&0&\pi
\end{array}\right),
\]
we obtain that the entries of $(\one-\Proj_{0}^{\dag})R_{-\infty}B_{0}$ are  of the form (\ref{en.104}), except for (sums) of  the more singular terms
\[
\begin{aligned}
&(1) \  \pi r_{1\Sig, 1},  & (2) \  \varb^{*}\vara^{*}r_{0\Sig, 1}, \\
&(3)\  \varb^{*}\vara^{*}r_{0\Sig,0}\varb,  & (4)\ \pi r_{1\Sig, 0}\varb,
\end{aligned}
\]
where as before all the $r_{i,j}$ terms belong to $\Psi^{-\infty}$.
We will examine successively these $4$ terms.

{\it Term 1:} by Lemma \ref{keylemma} (3) we know that $\pi: H^{n}\to H^{n}$ for all $n\in \nn$, hence $\pi r_{1\Sig, 1}: H^{-n}\to H^{n}$.

{\it Term 2:}  by Lemma \ref{keylemma} (6)  and duality, we know that $\varb^{*} \vara^{*}: H^{n}\to H^{n}$, the same argument as before shows that $ \varb^{*}\vara^{*}r_{0\Sig, 1}: H^{-n}\to H^{n}$.

The terms 3 and 4 will be more delicate to estimate. We will cut them into a high and low energy part. The high energy part is not affected by the infrared problem and is easy to estimate. The low energy part will be estimated by `undoing the commutator', i.e. rewriting $R_{-\infty}$ as $c_{1}^{+}K_{\Sig}- K_{\Sig}c_{0}^{+}$.

{\it Term 3:} we write $r_{0\Sig, 0}= r_{0\Sig, 0}\chi_{\gs}(h_{t})+ r_{0\Sig, 0}\chi_{\ls}(h_{t})$. 
We know that $\chi_{\gs}(h_{t})\varb= \chi_{\gs}(h_{t})h_{t}^{-1}\bardels\in \Psi^{-1}$, hence $r_{0\Sig, 0}\chi_{\gs}(h_{t})\varb\in \Psi^{-\infty}$. This implies that $r_{0\Sig, 0}\chi_{\gs}(h_{t})\varb: H^{-n}\to H^{n}$. Since by Lemma \ref{keylemma} (6)  $\varb^{*}\vara^{*}: \x H^{n}\to \x H^{n}$ it follows that $ \varb^{*}\vara^{*}r_{0\Sig,0}\chi_{\gs}(h_{t})\varb: H^{-n}\to\x H^{n}$.  

It remains to control the term  $\varb^{*}\vara^{*}r_{0\Sig,0}\chi_{\ls}(h_{t})\varb$.  We claim that
\begin{equation}
\label{en.105}
\varb^{*}\vara^{*}r_{0\Sig,0}\chi_{\ls}(h_{t})\varb: H^{-n}\to \x H^{n}, \ \forall n\in \nn.
\end{equation} 
To prove (\ref{en.105}) we write  $R_{-\infty}$ as $c_{1}^{+}K_{\Sig}- K_{\Sig}c_{0}^{+}$.  Writing  $c_{1}^{+}$  and $c_{0}^{+}$ in matrix form, we obtain after a routine computation that:
\[
r_{0\Sig, 0}=  m_{1}\bards + m_{2}\vara+ \bards m_{3}, \ m_{i}\in\Psi^{\infty}.
\]
 We have hence to consider the three terms:
 \[
(3a) \ \varb^{*}\vara^{*}m_{1}\bards\chi_{\ls}(h_{t}) \varb,\ (3b)\ \varb^{*}\vara^{*}m_{2}\vara\chi_{\ls}(h_{t})\varb, \ (3c)\  \varb^{*}\vara^{*} \bards  m_{3}\chi_{\ls}(h_{t})\varb,
\]
and to show that each of them maps $H^{-n}$ into $\x H^{n}$.

{\it Term 3a: }  we have 
 \[
\ \varb^{*}\vara^{*}m_{1}\bards\chi_{\ls}(h_{t}) \varb= \varb^{*}\vara^{*}m_{1}\bards \chi_{\ls}(h_{t})h_{t}^{-1}\bardels.
\]
Using  Lemma \ref{keylemma} (1)  and the fact that $m_{1}\in \Psi^{\infty}$, we know that $m_{1}\bards \chi_{\ls}(h_{t})h_{t}^{-1}\bardels: H^{-n}\to H^{n}$. Next we use that   by Lemma \ref{keylemma} (6) $\varb^{*}\vara^{*}: \x H^{n}\to \x H^{n}$.

{\it Term 3b:}  by Lemma \ref{keylemma} (2) and the fact that $\wbar{F}_{t}\in S_0^{-1}$ we know that  $m_{2}\vara\chi_{\ls}(h_{t})\varb: H^{-n}\to H^{n}$ and we can conclude the proof as for term 3a).

{\it Term 3c:} we use identity (\ref{en.4}) to obtain that $ \varb^{*}\vara^{*} \bards= \varb^{*}\bardels \vara= \pi \vara$. Therefore:
\[
 \varb^{*}\vara^{*} \bards  m_{3}\chi_{\ls}(h_{t})\varb=  \pi \vara m_{3}\chi_{\ls}(h_{t})\varb.
\]
Since $\wbar{F}_{t}\in S_0^{-1}$ we deduce from Lemma \ref{keylemma} (2) that $\vara m_{3}\chi_{\ls}(h_{t})\varb: H^{-n}\to H^{n}$. Next by Lemma \ref{keylemma} (3) we know that $\pi: H^{n}\to H^{n}$.
This completes the proof of (\ref{en.105}). 

{\it Term 4:} we split  $r_{1\Sig, 0}$ as $\chi_{\gs}(h_{\Sig})r_{1\Sig, 0}+ \chi_{\ls}(h_{\Sig})r_{1\Sig, 0}$.  
By Lemma \ref{keylemma} (4) we know that $\varb: H^{-n}\to \x H^{-n}$. Since $r_{1\Sig, 0}\in \Psi^{-\infty}$  we know that $r_{1\Sig, 0}: \x H^{-n}\to \x H^{n}$. Finally by Lemma \ref{keylemma} (5) and duality $\pi \chi_{\gs}(h_{\Sig})\x H^{n}\to \x H^{n}$.

We now claim that:
\begin{equation}
\label{en.106}
\pi \chi_{\ls}(h_{\Sig})r_{1\Sig, 0}\varb : H^{-n}\to \x H^{n}.
\end{equation}
Again we write $R_{-\infty}$ as $c_{1}^{+}K_{\Sig}- K_{\Sig}c_{0}^{+}$, obtain that
\[
r_{1\Sig, 0}= m_{1}\bards  + m_{2}\vara + \vara m_{3}, \ m_{i}\in\Psi^{\infty},
\]
and have to consider the three terms:
\[
(4a)\ \pi \chi_{\ls}(h_{\Sig})m_{1}\bards \varb, \ (4b)\ \pi \chi_{\ls}(h_{\Sig})m_{2}\vara \varb, \ (4c)\ \pi \chi_{\ls}(h_{\Sig})\vara m_{3}\varb.
\]
{\it Term 4a:}  using that $\bards \varb= \pi$,  this term equals $\pi \chi_{\ls}(h_{\Sig})m_{1}\pi$, which maps $H^{-n}$ into $H^{n}$ by now standard arguments.

{\it Term 4b, 4c:}  these two terms can be treated as term 3b), using that $\wbar{F}_{t}\in S_0^{-1}$. \qeds

\subsection*{Acknowledgments} The authors are grateful to Jochen Zahn for suggesting the problem and for useful discussions.  The work of M.\,W. was partially supported by the FMJH (Governement Program:  ANR-10-CAMP-0151-02).

\appendix
 \section{Background on pseudo-differential  calculus}\label{sec1opd}\init
 In this section we recall some facts about pseudo-differential calculus. We refer to \cite[Sect. 4]{GW} for more details.  
 We need to extend slightly the situation in \cite{GW} to include matrix-valued symbols.
\subsection{Notation}\label{sec1opd.0}
 - We  denote by $\Sigma$  either $\rr^{d}$ or a smooth compact manifold.  If $\Sigma$ is compact we choose  a smooth, non-vanishing  density $\mu$ which allows to  equip $\cinf(\Sigma)$ with an Hilbertian scalar product. Typically $\mu$ will be the canonical density associated to some Riemannian metric on $\Sigma$. If $\Sigma= \rr^{d}$ we use of course the Lebesgue density $dx$. 
  
 %The only important property of $\Sigma$ is that there exists a sufficiently developed  pseudo-differential calculus on $\Sigma$.
 
 - We denote by $V$ a finite dimensional complex vector space. 
 For simplicity we assume that $V$ is equipped with a  Hilbertian scalar product,  which allows to identify $V$ and $V^{*}$.
 
 %- We extend standard notation for $\Sigma=\rr^{d}$ to the case when $\Sigma$ is compact, for example $\coinf(\Sigma; V)= \cinf(\Sigma; V)$, $\cS(\Sigma)\defeq \cinf(\Sigma)$, $\cS'(\Sigma; V)\defeq \cD'(\Sigma; V)$ etc   if $\Sigma$ is compact, 

%- If $f:\rr_{t}\times \Sigma_{x}\to \cc$ is a function, and $t\in \rr$ we denote by $f(t)$ the function:
%\[
%f(t): \ \Sigma\ni x\mapsto f(t, x)\in \cc.
%\]

- We denote by $C^{\infty}_{\rm bd}(\Sigma; V)$ the space of smooth functions $\Sigma \to V$ uniformly bounded with all derivatives. We equip $C^{\infty}_{\rm bd}(\Sigma;V)$ with its canonical Fr\'echet space structure. 

- The Sobolev space of order $m$ is denoted $H^{m}(\Sigma; V)$. Furthermore, we define the spaces
\[\textstyle
 \cH(\Sigma;V)\defeq\bigcap_{m\in \rr}H^{m}(\Sigma; V), \quad \cH'(\Sigma; V)\defeq\bigcup_{m\in \rr}H^{m}(\Sigma; V),
\]
equipped with their canonical topologies.

%- We denote by $\cH(\Sigma; V)$, resp. $\cH'(\Sigma; V)$ the space of Schwartz functions, resp. distributions on $\Sigma$.

%- If $E, F$ are two topological vector spaces and $A\in L(E, F)$ we write $A: E\to F$ if $A$ is continuous  and $A:E \tarrow F$ is $A$ is  bijective with $A^{-1}$ continuous.  

%- We set $D_{x}= \i^{-1}\p_{x}$, $\langle x\rangle= (1+x^{2})^{\12}$, $x\in \Sigma$.
\subsection{Symbol classes}\label{sec1opd.1}
We denote by $S^{m}(\co{\Sigma})$, $m\in \rr$ the usual class of poly-homogeneous symbols of order $m$ such that additionally
\beq\label{e1.1}
 \p_{x}^{\alpha}\p_{k}^{\beta}a(x, k)\in O(\langle k\rangle^{m-|\beta|}), \ \alpha, \beta\in \nn^{d}.
\eeq
Similarly we will denote by $S^{m}(\rr)$ the class of poly-homogeneous functions $f: \co{\Sigma}\to \cc$.

We denote by  $S^{m}_{\rm h}(\co{\Sigma})\subset S^{m}(\co{\Sigma})$ the subspace  of symbols homogeneous of degree $m$ in $k$ away from $0$.

 These spaces are equipped with the 
 Fr\'echet space topology given by the semi-norms:
\[
\| a\|_{m, N}\defeq \sup_{|\alpha|+ |\beta|\leq N}| \langle k\rangle^{-m+ |\beta|}\p_{x}^{\alpha}
\p_{k}^{\beta}a|.
\]
We set
\[\textstyle
S^{-\infty}(\co{\Sigma})\defeq\bigcap_{m\in \rr}S^{m}(\co{\Sigma}),  \quad  S^{\infty}(\co{\Sigma})\defeq\bigcup_{m\in \rr}S^{m}(\co{\Sigma}).
\]

Let now  $V_{1}, V_{2}$ be  finite dimensional complex vector spaces  equipped with non-degenerate hermitian sesquilinear forms. The spaces $S^{m}_{({\rm h})}(\co{\Sigma})\otimes L(V_{1}, V_{2}) $ will be denoted by $S^{m}_{({\rm h})}(\co{\Sigma}; V_{1}, V_{2})$ and by $S^{m}_{({\rm h})}(\co{\Sigma}; V)$ if $V_{1}= V_{2}= V$.

The subspace of {\em scalar} symbols $S^{m}(\co{\Sigma})\otimes \one_V$ will be denoted by $S^{m}_{\rm scal}(\co{\Sigma}; V)$. 
\subsection{Principal symbol and characteristic set}\label{sec1opd.2}

For $a\in S^{m}(\co{\Sigma}; V_{1}, V_{2})$ we denote  by $a_{\rm pr}\in S^{m}_{\rm h}(\co{\Sigma}; V_{1}, V_{2})$  the {\em principal part} of $a$, which is homogeneous of degree $m$. 

The {\em characteristic set} of $a\in S^{m}(\co{\Sigma}; V)$ is defined as
\begin{equation}
\label{e1.4opd}
{\rm Char}(a)\defeq \{(x, k)\in\cooo{\Sigma} : \  \det a_{\rm pr}(x, k)= 0\},
\end{equation}
which is  conic  in the $k$ variable.

A symbol $a\in S^{m}(\co{\Sigma}; V)$ is {\em elliptic} if  ${\rm Char}(a)= \emptyset$.

\subsection{Pseudo-differential operators}\label{sec1opd.3}

In this subsection we collect some well-known results about pseudo-differential  calculus. 

We denote by $\Op: a\mapsto \Op (a)$ a quantization procedure  assigning to a symbol in $S^{\infty}(\co{\Sigma}; V_{1}, V_{2})$ a pseudo-differential operator on $\Sigma$. If $\Sigma$ is compact, this quantization depends on the choice of a partition of unity on $\Sigma$ and of associated coordinate mappings, the difference between two choices being a smoothing operator. If $\Sigma= \rr^{d}$ w choose the Weyl quantization. One has 
\[
\Op(a): \cH(\Sigma; V_{1})\to \cH(\Sigma; V_{2}), \quad \Op (a):  \cH'(\Sigma; V_{1})\to \cH'(\Sigma; V_{2}).
\] 

We denote by  $\Psi^{m}_{({\rm scal})}(\Sigma; V_{1}, V_{2})$ the space $\Op (S^{m}_{({\rm scal})}(\Sigma; V_{1}, V_{2}))$ and set
\[\textstyle
 \Psi^{-\infty}(\Sigma; V_{1}, V_{2})= \bigcap_{m\in \rr}\Psi^{m}(\Sigma; V_{1}, V_{2}), \quad 
\Psi^{\infty}(\Sigma; V_{1}, V_{2})= \bigcup_{m\in \rr}\Psi^{m}(\Sigma; V_{1}, V_{2}). 
\]
We  equip $\Psi^{m}(\Sigma; V_{1}, V_{2})$ with the  Fr\'{e}chet space topology induced from the one  of $S^{m}(\co{\Sigma}; V_{1}, V_{2})$.

Let $s,m\in \rr$. Then   the map
\beq\label{e0.0}
S^{m}(\co{\Sigma}; V_{1}, V_{2})\ni a\mapsto \Op(a)\in B(H^{s}(\Sigma; V_{1}), H^{s-m}(\Sigma; V_{2}))
\eeq
is continuous.

We denote by $\sigma: \Psi^{\infty}(\Sigma; V_{1}, V_{2})\to S^{\infty}(\co{\Sigma}; V_{1}, V_{2})$ the inverse of $\Op$, $\sigma(a)$ being called the (full) {\em symbol} of $a$.   

 If $\Sigma$ is a compact manifold, different choices of $\Op$ lead of course to different maps $\sigma$, differing by a map from $\Psi^{m}$ to $S^{m-1}$. On the other hand, the principal symbol map:
 \[
\sigma_{\rm pr}: \Psi^{m}(\Sigma; V_{1}, V_{2})\to S^{m}_{{\rm h}}(\co{\Sigma}; V_{1}, V_{2})
\]
 is  independent on the choice of the quantization.

 An operator $\Op(a)\in \Psi^{m}(\Sigma; V)$ is {\em elliptic} if its  principal symbol $\sigma_{\rm pr}(a)(x, k)$ is elliptic in $S^{m}(\Sigma; V)$.  If $a\in \Psi^{m}$ is elliptic 
 then there exists $b\in \Psi^{-m}$, unique modulo $\Psi^{-\infty}$ such that $ab=ba=\one$ modulo $\Psi^{-\infty}$. Such an operator $b$ is called a {\em pseudo-inverse} or a {\em parametrix} of $a$. As a typical example  $\one + b$ for $b\in \Psi^{-m}$, $m>0$ is elliptic in $\Psi^{0}$.

\subsection{Functional calculus for pseudo-differential  operators}\label{sec1opd.4}

We recall without proof some well-known results about functional calculus and pseudo-differential operators. 
\begin{proposition}\label{1.1opd}
Let $a\in \Psi^{m}(\Sigma; V)$  for $m\geq 0$ be elliptic in $\Psi^{m}(\Sigma; V)$ and symmetric on $\cH(\Sigma; V)$. Then:
\ben \item $a$ is selfadjoint on $H^{m}(\Sigma; V)$,
\item  Denote by ${\rm res}(a)$ the resolvent set of $a$, with domain $H^{m}(\Sigma; V)$. Then for $z\in {\rm res}(a)$,  $(z-a)^{-1}\in \Psi^{-m}(\Sigma; V)$,
\item if $f\in S^{p}(\rr)$, $p\in \rr$,  then $f(a)$, defined by the functional calculus, belongs to $\Psi^{mp}(\Sigma; V)$. 
\item if  $f$ is elliptic in $S^{p}(\rr)$ then $\sigma_{\rm pr}(f(a))= f_{\rm pr}(\sigma_{\rm pr}(a))$.
\een
\end{proposition}

\subsection{Propagators}\label{sec1opd.5}
In this subsection we state some results about propagators, associated to elliptic operators in $\Psi^{1}(\Sigma; V)$. It is important to restrict oneself to operators with real  and {\em scalar} principal symbols.  The propagators in our presentation replace {\em Fourier integral operators}  which are often used in the literature.

Let us fix a  map $\epsilon(t)=\epsilon_{1}(t)+ \epsilon_{0}(t)$, where $\epsilon_{i}(t)\in C^{\infty}(\rr, \Psi^{i}(\Sigma; V))$ for $i=0,1$. We assume  that
\ben 
\item $\epsilon_{1}(t)$ is {\em scalar}, i.e. belongs to     $\Psi^{1}_{\rm scal}(\Sigma; V)$,
\item $\epsilon_{1}(t)$ is elliptic in $\Psi^{1}(\Sigma; V)$,
\item  $\epsilon_{1}(t)$  is symmetric on $\cH(\Sigma; V)$.
\een It follows by Prop. \ref{1.1opd} that  $\epsilon_{1}(t)$ 
is selfadjoint with domain $H^{1}(\Sigma; V)$, hence $\epsilon(t)$ with domain $H^{1}(\Sigma; V)$ is closed, with non empty resolvent set.

We denote by $\Texp(\int_{s}^{t}\i\epsilon(\sigma)d\sigma)$ the associated  propagator defined by:

\[
\left\{
\begin{array}{rl}
&\frac{\p}{\p t}\Texp(\int_{s}^{t}\i\epsilon(\sigma)d\sigma)= \i  \epsilon(t)\Texp(\int_{s}^{t}\i\epsilon(\sigma)d\sigma),\\[2mm]
&\frac{\p}{\p s}\Texp(\int_{s}^{t}\i\epsilon(\sigma)d\sigma)= -\i  \Texp(\int_{s}^{t}\i\epsilon(\sigma)d\sigma)\epsilon(s),\\[2mm]
&\Texp(\int_{s}^{s}\i\epsilon(\sigma)d\sigma)=\one.
\end{array}
\right.
\]
It is easy to see  (see e.g. \cite[Subsect. 4.6]{GW}) that  $\Texp(\int_{s}^{t}\i\epsilon(\sigma)d\sigma)$is strongly continuous in $(t,s)$ with values in $B(L^{2}(\Sigma; V))$.

\begin{definition}\label{def1}
We denote by $\Phi_{\epsilon}(t,s): \cooo{\Sigma}\to \cooo{\Sigma}$ the symplectic flow associated to  the time-dependent Hamiltonian $ -\sigma_{\rm pr}(\epsilon)(t, x, k)$. 
 \end{definition}
Clearly $\Phi_{\epsilon}(t,s)$ is  an homogeneous map of degree $0$.

We now state a version of the  {\em Egorov's theorem} for matrix-valued symbols.
\begin{proposition}\label{1.2opd}
\ben
\item  $\Texp({\textstyle\int_{s}^{t}}\i\epsilon(\sigma)d\sigma)$ is bounded on $\cH(\Sigma; V)$ hence on $\cH'(\Sigma; V)$ by duality.
\item There exists $m(t,s)\in \cinf(\rr^{2}; \Psi^{0}(\Sigma; V))$ elliptic, invertible  on $L^{2}(\Sigma; V)$ with $m^{-1}(t,s)\in \cinf(\rr^{2}; \Psi^{0}(\Sigma; V))$  such that
\[
\Texp({\textstyle\int_{s}^{t}}\i\epsilon(\sigma)d\sigma)= m(t,s)\Texp({\textstyle\int_{s}^{t}}\i\epsilon_{1}(\sigma)d\sigma).
\]
\item  Let $a\in \Psi^{m}(\Sigma; V)$. Then 
 \[
a(t,s)\defeq  \Texp({\textstyle\int_{s}^{t}}\i\epsilon(\sigma)d\sigma)a\Texp({\textstyle\int_{t}^{s}}\i\epsilon(\sigma)d\sigma) 
\]
belongs to $C^{\infty}(\rr^{2}, \Psi^{m}(\Sigma; V))$.   Moreover \[
\sigma_{\rm pr}(a)(t,s)= \sigma_{\rm pr}(a)\circ \Phi_{\epsilon}(s,t).
\] 
\een
\end{proposition}
\proof  The proposition is well-known  in the scalar case, i.e. if $\epsilon(t)= \epsilon_{1}(t)$,  see eg  \cite[Sec. 0.9]{taylor} for the proof.  It is easy to extend it to our situation.  Let us denote $\Texp({\textstyle\int_{s}^{t}}\i\epsilon(\sigma)d\sigma)$,  resp. $\Texp({\textstyle\int_{s}^{t}}\i\epsilon_{1}(\sigma)d\sigma)$ by $U(t,s)$ resp. $U_{1}(t,s)$.  Setting
\[
U(t,s)\eqdef  m(t,s)U_{1}(t,s),
\]
we obtain that $m(t,s)$ solves the equation:
\[
\left\{
\begin{array}{l}
\p_{t}m(t,s)-\i\epsilon_{0}(t,s)m(t,s)=0,\\
 m(s,s)= \one,
\end{array}
\right.
\]
for $\epsilon_{0}(t,s)\defeq U_{1}(s,t)\epsilon_{0}(t)U_{1}(t,s)$. Note that $\epsilon_{0}(t,s)\in \cinf(\rr^{2}, \Psi^{0}(\Sigma; V))$, by Egorov's theorem for  the scalar case.  The solution is
\[
 m(t,s)= \Texp({\textstyle\int_{s}^{t}}\i\epsilon_{0}(\sigma, s)d\sigma).
\]
It is easy to see that $m(t,s)\in \cinf(\rr^{2}; \Psi^{0}(\Sigma; V))$, using for example Beals criterion.  Moreover  $m(t,s): L^{2}(\Sigma; V)\to L^{2}(\Sigma; V)$ is boundedly invertible, with inverse 
\[
m^{-1}(t,s)=  \Texp({\textstyle\int_{t}^{s}}\i\epsilon_{0}(\sigma, s)d\sigma).
\] 
The same argument shows that $m^{-1}(t,s)\in  \cinf(\rr^{2}; \Psi^{0}(\Sigma; V))$, 
hence $m(t,s)$ is elliptic in $\Psi^{0}(\Sigma; V)$.   This proves (2). (1) follows from (2) and the analogous result in the scalar case. We write then
\[
a(t,s)= U_{1}(t,s) m(t,s) a m^{-1}(t,s) U_{1}(s,t)= U_{1}(t,s) \tilde{a}(t,s)U_{1}(s,t),
\]
where $\tilde{a}(t,s)= m(t,s)a m^{-1}(t,s)\in \cinf(\rr^{2}, \Psi^{m}(\Sigma; V))$ has principal symbol $\sigma_{\rm pr}(a(t,s))= \sigma_{\rm pr}(a)$. (3) follows then from Egorov's theorem for the scalar case. 
\qeds

The following two results are proved in \cite[Sect. 4]{GW} for the scalar case. By the argument outlined in the proof of Prop. \ref{1.2opd} they immediately extend to our situation.

\begin{proposition}\label{1.2b}
For $u\in \cH'(\Sigma; V)$ one has: 
\[
\WF(\Texp({\textstyle\int_{s}^{t}}\i\epsilon(\sigma)d\sigma)u)= \Phi_{\epsilon}(t,s)\WF(u),
\]
hence
\[
\WF'(\Texp({\textstyle\int_{s}^{t}}\i\epsilon(\sigma)d\sigma))= \{(x, k, x', k')  : \ (x, k)= \Phi_{\epsilon}(t,s)(x', k')\}.
\]
\end{proposition}
\begin{lemma}\label{budaopd}
 Let $\epsilon(t)\in \cinf(\rr, \Psi^{1}(\Sigma; V))$ as above,  $s_{-\infty}(t,s)\in C^{\infty}(\rr^{2}, \Psi^{-\infty}(\Sigma; V))$. Then
 \[
\Texp({\textstyle\int_{s}^{t}}\i\epsilon(\sigma)d\sigma) s_{-\infty}(t,s)\in C^{\infty}(\rr^{2}, \Psi^{-\infty}(\Sigma; V)).
\]
\end{lemma}

\section{ Some auxiliary results}\label{appB}
\subsection{ A Hardy inequality}\label{ss:hardy}
\begin{proposition}\label{prop.hard}
 There exists $C> 0$ such that 
 \beq\label{e.hardy0}
\bardels\bards\geq C\x^{-2}, \hbox{ on }L^{2}(\rr^{d}, |h|^{\12}dx)\otimes \fg.
\eeq
 \end{proposition}

\proof  
Let us denote by $M_{j}(x)\in L(\fg)$ the operator $\i^{-1}\wbar{A}_{j}(x)\wedge\,\cdot\,$ and note that $M_{j}(x)$ is selfadjoint on $(\fg, \killing)$. Let 
\[
h_{M}\defeq  \sum_{j=1}^{d}(\D_{j}+ M_{j}(x))^{2},
\]
acting on $L^{2}(\rr^{d}, dx)\otimes \fg$. 
 We claim that the proposition follows from 
 \begin{equation}
\label{e.hardy1}
h_{M}\geq C \x^{-2}.
\end{equation}
In fact  we have:
\[
h_{t}=\bardels\bards= |h|^{-\12}(x)\sum_{j,k=1}^{d}(\D_{j}+ M_{j}(x))h^{jk}(x)|h|^{\12}(x)(D_{k}+ M_{k}(x)),
\]
acting on $L^{2}(\rr^{d}, |h|^{\12}dx)\otimes \fg$. Clearly $h_{t}$ is unitarily equivalent to:
\[
\tilde{h}_{t}= |h|^{-\frac{1}{4}}(x)\sum_{j,k=1}^{d}(\D_{j}+ M_{j}(x))h^{jk}(x)|h|^{\12}(x)(D_{k}+ M_{k}(x))|h|^{-\frac{1}{4}}(x),
\]
acting on $L^{2}(\rr^{d}, dx)\otimes \fg$, by the map $U: u\mapsto |h|^{\frac{1}{4}}u$.   It suffices to prove Hardy's inequality for $\tilde{h}_{t}$. Since $c_{0}\leq |h|(x)\leq c_{0}^{-1}$ for some $c_{0}>0$, we can also replace $\tilde{h}_{t}$ by $|h|^{\frac{1}{4}}\tilde{h}_{t} |h|^{\frac{1}{4}}$. Finally since $|h|^{\frac{1}{4}}\tilde{h}_{t} |h|^{\frac{1}{4}}\geq C h_{M}$ for some $C>0$, we see that  (\ref{e.hardy1}) implies (\ref{e.hardy0}). 

Let us now prove (\ref{e.hardy1}).    From the usual Hardy inequality we know that there exists $C>0$ such that
\beq\label{e.hardy4}
-\Delta- C\x^{-2}\geq 0.
\eeq
We  use now the diamagnetic inequality:
\begin{equation}
\label{e.hardy3}
\|\e^{- t(h_{M}- C\x^{-2})}u\|\leq \e^{- t(- \Delta- C \x^{-2})}\|u\|, \ u\in L^{2}(\rr^{d}, dx)\otimes \fg, \ t\geq 0,
\end{equation}
where $\| u\|^{2}(x)= \bar{u}(x)\cdot \killing u(x)$.  The proof of (\ref{e.hardy3}) can be done as in \cite[Thm. 1.3]{CFKS}. The key fact is that 
\[
D_{j}+ \i M_{j}(x)=S^{-1}_{j}(x)D_{j}S_{j}(x)
\]
for \[
S_{j}(x)= \Texp(-\i\textstyle\int^{0}_{x_{j}}M_{j}(x_{1}, \dots, x_{j-1}, s, x_{j+1}, \dots, x_{d})ds)
\]
where $S_{j}(x)$ is unitary on $(\fg, \killing)$.  Using $a^{-1}=\int_{0}^{+\infty}\e^{- t a}dt$, we deduce from (\ref{e.hardy3}) that  for $\varepsilon>0$
\[
\begin{aligned}
(u| (h_{M}- C\x^{-2}+ \varepsilon)^{-1}u)_{L^{2}\otimes \fg}&\leq (\| u\| | (- \Delta - C\x^{-2}+\varepsilon)^{-1}\| u\|)_{L^{2}}\\[2mm]
&\leq \varepsilon^{-1}(\| u\| | \| u\| )_{L^{2}}= \varepsilon^{-1}(u| u)_{L^{2}\otimes \fg}.
\end{aligned}
\]
This implies that $h_{M}- C\x^{2}\geq 0$ and completes the proof of the proposition. \qeds

\subsection{Transition to the temporal gauge}\label{ss:tempgauge} In this section we review the transition to the temporal gauge, explained in the language of connections. 

We assume here that  $g= -\beta(t, x)dt^{2}+ h_{ij}(t, x)dx^{i}dx^{j}$, i.e. that we are in the general globally hyperbolic case.

We set:
 \[
S(t, x)\defeq  {\rm Texp}(-\textstyle\int_{t}^{0} T_{0}(s, x)ds)\in \cinf(M; L(W)),
\]
so that 
\[
\begin{cases}\p_{t}S(t, x)= S(t,x) T_{0}(t,x)\\[2mm]
S(0, x)= \one_{W}.
\end{cases}
\]
Note that $S(t, x)= S_{V}(t,x)\otimes S_{\fg}(t, x)$, for:
\[
 S_{V}(t, x)=  {\rm Texp}(-\textstyle\int_{t}^{0} \Gamma_{0}(s, x)ds), \ S_{\fg}(t, x)=  {\rm Texp}(-\textstyle\int_{t}^{0} M_{0}(s, x)ds).
\]
An easy computation  using that $ T$ is metric for $g^{-1}\otimes \killing$ shows that:
\[
g^{-1}(t, x)\otimes \killing= S^{*}(t, x)g^{-1}(0, x)\otimes \killing S(t,x). 
\]
Again if we set
\[
\tilde{T}_{a}\defeq  S \p_{a}S^{-1}+ S T_{a} S^{-1}, \quad \tilde{\rho}\defeq  S \rho S^{-1},
\]
then setting $g^{-1}_{0}(t,x)\defeq  g^{-1}(0, x)$ we have:
\[
\begin{array}{rl}
&\p_{a}g^{-1}_{0}\otimes \killing= \tilde{T}_{a}^{*}g^{-1}_{0}\otimes \killing + g^{-1}_{0}\otimes \killing \tilde{T}_{a},\\[2mm]
&\tilde{\rho}^{*} g^{-1}_{0}\otimes \killing= g^{-1}_{0}\otimes \killing \tilde{\rho},\\[2mm]
&\tilde{T}_{0}=0.
 \end{array}
\]
Setting $\widetilde{D}_1= S D_1 S^{-1}$ we have:
\[
\widetilde{D}_1= - |g|^{-\12}\nabla_{a}^{\tilde T} |g|^{\12}g^{ab}\nabla_{b}^{\tilde T}+ \tilde{\rho}.
\]
The conserved charge is:
\[
\overline{\tilde \zeta}_{1} \tilde q \tilde\zeta_{2}\defeq  \int_{\{t\}\times\Sigma}\overline{\i^{-1}\nabla_{0}^{\tilde T}\tilde \zeta_{1}}\cdot g_{0}^{-1}\otimes \killing \tilde  \zeta_{2}+ \overline{\tilde \zeta_{1}}\cdot g_{0}^{-1}\otimes \killing \ \i^{-1}\nabla_{0}^{\tilde T}\tilde \zeta_{2}|h|^{\12}dx.
\]

\subsection{Constraints for initial data of Yang-Mills equation}\label{ss:constraints}

In the main part of the text (Hypothesis \ref{as:background}, Thm. \ref{maintheo2}) we make several assumptions on the Cauchy data of the smooth solution $\wbar{A}$ of the non-linear Yang-Mills equations, used to linearize the system. To be sure that such solution $\bA$ exists, one needs to verify that the conditions on the Cauchy data are consistent with the \emph{constraint equations}. Although there is already some literature on this subject \cite{CB,CC,segal}, it does not cover directly our case, we thus briefly discuss the constraint equations below. 

 We use the framework and the notations introduced in \ref{sss:ident}, in particular we assume that the spacetime $(M,g)$ is ultra-static. We assume $\wbar{A}$ is in the temporal gauge $\wbar{A}_t\equiv 0$.

The definition $\wbar{F}=\wbar{d}\wbar{A}$ gives 
\begin{eqnarray}
&\wbar{F}_\Sig = \wbar{d}_\Sig \wbar{A}_\Sig,\label{eq:C1}\\
&\wbar{F}_t  = \partial_t \wbar{A}_\Sig.\label{eq:C2}
\end{eqnarray}

The Yang-Mills equation $\wbar{\delta}\wbar{F}=0$ reads
\begin{eqnarray}
&\bdeltas \bF_t =0, \label{eq:C3}\\
&\p_t\bF_t+\bdeltas\bF_\Sig=0.\label{eq:C4}
\end{eqnarray}

Taking the first time derivative of (\ref{eq:C1}) and using (\ref{eq:C2}) one gets
\beq\label{eq:C5}
\p_t \bF_\Sig = d_\Sig \bF_t + \bF_t \wedge \wbar{A}_\Sig.
\eeq
This allows to consider (\ref{eq:C2}), (\ref{eq:C4}) and (\ref{eq:C5}) as evolution equations
\beq\label{eq:evol}
\begin{cases}
\partial_t \wbar{A}_\Sig =\wbar{F}_t,  \\
\p_t\bF_t=-\bdeltas\bF_\Sig,\\
\partial_t \bF_\Sig=d_\Sig \bF_t + \bF_t \wedge \bA_\Sig,
\end{cases}
\eeq
subject to \emph{constraint equations} (\ref{eq:C1}) and (\ref{eq:C3}):
\beq
\begin{cases}
\bF_\Sig = \bds \bA_\Sig,\\
\bdeltas \bF_t =0.
\end{cases}
\eeq
The first constraint (\ref{eq:C1}) is not problematic in the sense that it does not restrict the set of allowed Cauchy data. It is also straightforward to see from (\ref{eq:C5}) that it is preserved by the evolution (\ref{eq:evol}).

The second constraint (\ref{eq:C3}) does significantly restrict the set of allowed Cauchy data.

First, let us check that it is preserved by the evolution (this is a known result, cf. \cite{CS} for the case of arbitrary globally hyperbolic spacetimes). Recall that $\bdeltas=\delta_\Sig + \wbar{A}_\Sig \Int\,\cdot\,$ (where for simplicity we assume the Cauchy data are real). Thus, using (\ref{eq:C2}) and (\ref{eq:C4}) one gets
\beq\label{eq:Ctemp}
\p_t \bdeltas \bF_t =  \bdeltas \p_t \bF_t + \bF_t\Int \bF_t = -\bdeltas\circ \bdeltas \bF_\Sig + \bF_t\Int \bF_t.
\eeq  
Since (\ref{eq:C1}) holds for any time slice, we have $\bds\circ\bds=\bF_\Sig \wedge\,\cdot\,$, and by taking the adjoint $\bdeltas\circ\bdeltas=\bF_\Sig \Int\,\cdot\,$. Hence (\ref{eq:Ctemp}) gives in fact
\[
\p_t \bdeltas \bF_t =  -\bF_\Sig \Int \bF_\Sig + \bF_t\Int \bF_t.
\]
Both terms identically vanish, as is easily seen by writing the expression for the interior product in an orthonormal frame. This proves that $\bdeltas \bF_t =0$ on each time slice.

\subsubsection{Existence of Cauchy data with decay at infinity}

One can construct examples of Cauchy data $\bF_t$, $\bA_\Sig$ satisfying the constraint $\bdeltas\wbar{F}_t=0$ as follows.

Let us take $\bF_t\defeq \bdeltas G$, $G\in\cE^2(\Sigma;\mathfrak{g})$. Then if we take $\bA_\Sig$ and $G$ with disjoint supports, then $\bdeltas\wbar{F}_t=0$ is trivially satisfied. If, moreover, both the supports of $\bA_\Sig$ and $G$ are compact, then the Cauchy data $\bA_\Sig,\bF_\Sig,\bF_t$ have compact support, as requested in Thm. \ref{maintheo2}.

Let now $S^m$ denote the space of $\mathfrak{g}$-valued functions (or differential forms) whose coefficients satisfy classical symbol estimates $\p_x^\alpha f(x)\in O(\bra x \ket^{m-|\alpha|})$. It suffices then to take $\bA_\Sig\in S^{-1}$ and $G\in S^{-1}$ with disjoint supports to ensure that $\bF_\Sig \in S^{-2}$ and $\bF_t\in S^{-2}$. This provides a class of examples for Hypothesis 1.4.

\subsection{Global existence of smooth space-compact solutions for non-linear Yang-Mills equations}\label{ss:global-existence}
In this subsection we explain how to deduce Prop. \ref{prop:idiotic} from the arguments  of   Chru\'sciel-Shatah \cite{CS}.

\begin{proposition}
 \ben
 \item for each $\wbar{A}\in \cE^{1}_{\rm sc}(M)\otimes{\fg}$ there exists $\wbar{A}'\in \cE^{1}_{\rm sc}(M)\otimes{\fg}$ such that $\wbar{A}'_{t}\equiv 0$ and $\wbar{A}'\sim \wbar{A}$.
 \item Assume that $\dim M \leq 4$. Let $\wbar{A}\in \cE^{1}_{\rm sc}(M)\otimes{\fg}$ be a solution of the non linear Yang-Mills equation (\ref{eq:nlYM}) near a Cauchy surface $\Sigma$.  Then there exists  $\wbar{A}'\in  \cE^{1}_{\rm sc}(M)\otimes{\fg}$ such that $\wbar{A}'\sim \wbar{A}$, $\wbar{A}'_{t}\equiv 0$ and $\wbar{A}'$ solves (\ref{eq:nlYM}) globally.   \een 
\end{proposition}

\proof 
\def\rx{{\rm x}}
(1): recall that we assumed that $G$ is represented as  a subgroup of $L(V)$ for some finite dimensional vector space $V$.
The gauge transformation generated by  the map $M\ni x\mapsto \mathcal{G}(x)\in  G$  is:
\[
\wbar{A}_{\mu}\mapsto \wbar{A}_{\mu}'=\mathcal{G}^{-1}\wbar{A}_{\mu}\mathcal{G}+ \mathcal{G}^{-1}\p_{\mu}\mathcal{G}. 
\]
Writing $M= \rr_{t}\times \Sigma_{x}$,  we obtain $\wbar{A}_{t}'\equiv 0$ if  $\p_{t}\mathcal{G}+ \wbar{A}_{t}\mathcal{G}=0$. 
This can be solved by
\[
\mathcal{G}(t, x)= {\rm Texp}(\textstyle\int_{0}^{t}-\wbar{A}_{t}(s, x)ds).
\]
Since $\wbar{A}_{\mu}\in \cinf_{\rm sc}(M)\otimes{\fg}$, we obtain that $\mathcal{G}-\one\in \cinf_{\rm sc}(M; G)$, hence $\wbar{A}_{\mu}'\in  \cinf_{\rm sc}(M)\otimes{\fg}$.

(2):   By (1) we can assume that $\wbar{A}_{t}\equiv 0$, i.e. that $\wbar{A}$ is in the temporal gauge.  We recall  the form of the Yang-Mills equations in the temporal gauge, recalled in \cite[Sect. 4]{CS}. Denoting by $\wbar{F}_{\mu\nu}$ the curvature, we obtain the equations:
\beq\label{equato}
\begin{cases}\p_{t}\wbar{A}_{i}= \wbar{F}_{0i}, \\[2mm]
\mathcal{D}_{t}\wbar{F}_{ij}= \mathcal{D}_{j}\wbar{F}_{i0}- \mathcal{D}_{i}\wbar{F}_{j0},\\[2mm]
\mathcal{D}_{t}\wbar{F}^{0i}= \mathcal{D}_{j}\wbar{F}^{ji},
\end{cases}
\eeq
where $\mathcal{D}_{\mu}= \nabla_{\mu}+ [\wbar{A}_{\mu}, \cdot]$, and $\mathcal{D}_{t}= \mathcal{D}_{0}$. 

 Another fact is that if $G_{\mu\nu}\defeq  \wbar{F}_{\mu\nu}- \p_{\mu} \wbar{A}_{\nu}+ \p_{\nu}\wbar{A}_{\mu}- [\wbar{A}_{\mu}, \wbar{A}_{\nu}]$ vanishes at $t=0$ and (\ref{equato}) holds in some region $I\times \mathcal{O}$  where $I$  is a time interval, then $G_{\mu\nu}$ vanishes identically in $I\times \mathcal{O}$, hence $\wbar{F}=\wbar{d}\wbar{A}$. 
 
 By \cite[Thm. 1.1]{CS}  the local in time solution $(\wbar{A}_{i}, \wbar{F}_{ij}, \wbar{F}_{0j})$ of (\ref{equato}) extends globally as a smooth solution. Moreover since (\ref{equato}) is a symmetric hyperbolic semi-linear system of equations (see eg  the proof of \cite[Prop. 4.1]{CS}), its solutions satisfy Huygens' principle, which implies that the global solution of (\ref{equato}) belongs to $\cE^{1}_{\rm sc}(M)\otimes{\fg}$. 
 Note that \cite{CS} deals with the most difficult case $\dim M=4$. It is easy to extend the result to lower dimensions. In fact if $\dim M=n<4$, we consider $\tilde{M}=M\times\rr^{4-n}_{y}$ with metric $g+ dy^{2}$. A $1-$form $A= A_{\mu}(x)dx^{\mu}\in \cE^{1}(M)\otimes{\fg}$ is extended to $\tilde{A}= A_{\mu}(x)dx^{\mu}\in \cE^{1}(\tilde{M})\otimes{\fg}$. It is easy to see that $A$ satisfies the Yang-Mills equation on $M$ iff $\tilde{A}$ satisfies the YM equation on $\tilde{M}$. It follows that the Cauchy problem can be globally solved for smooth Cauchy data in $M$. The fact that a local space-compact solution extends as a global space-compact solution follows by the same argument based on Huygens' principle.   \qed

{ 

}


\begin{thebibliography}{aaaa}
\small
\bibitem[AS]{araki} Araki, H., Shiraishi, M.: {\em On quasi-free states of canonical commutation relations I}, Publ. RIMS Kyoto Univ. 7 (1971/72), 105-120.

\bibitem[B]{benini} Benini, M.: {\em Optimal space of linear classical observables for Maxwell $k$-forms via spacelike and timelike compact de Rham cohomologies}, \texttt{arXiv:1401.7563} (2014).

\bibitem[BF]{BF} B\"ar, C. (ed.), Fredenhagen, K. (ed.): {\em Quantum Field Theory on Curved spacetimes}, Lect. Notes Phys. 786 (2009).

\bibitem[BGP]{BGP} B\"ar, C., Ginoux, N., Pf\"affle, F.: {\em Wave equation on Lorentzian
Manifolds and Quantization}, ESI Lectures in Mathematics and Physics, EMS 2007.

\bibitem[C-B]{CB} Choquet-Bruhat, Y.: {\em Yang-Mills Fields on Lorentzian Manifolds}, Mechanics, Analysis and Geometry: 200 Years After Lagrange, North-Holland Delta Ser., North Holland, 1991, 289-313.
 
\bibitem[CC]{CC} Choquet-Bruhat, Y., Christodoulou, D.: {\em Existence of global solutions of the Yang-Mills, Higgs and spinor fields equations in $3+1$ dimensions}, Ann. Sci. \'Ecole Norm. Sup. 14 (1981), 481-506.

\bibitem[CFKS]{CFKS} Cycon, H.L., Froese, R., Kirsch, W., Simon, B.:
{\em Schr\"odinger Operators with applications to Quantum Mechanics and Global Geometry}, Springer 1987.

\bibitem[BG]{BG} B\"ar, C., Ginoux, N.: {\em Classical and quantum fields on Lorentzian manifolds.}, Global Differential Geometry, 359-400, Springer Berlin Heidelberg, 2012.

\bibitem[CS]{CS} Chru\'sciel, P.T.,  Shatah, J.: {\em Global existence of solutions of the Yang-Mills equations on globally hyperbolic four dimensional Lorentzian manifolds}, Asian Jour. Math. 1 (1997), 530-548.

\bibitem[Der]{derezinski} Derezi\'nski, J.: {\em Quantum fields with classical perturbations}, \texttt{arXiv:1307.1162} (2013)

{ \bibitem[DF]{DF} D\"utsch, M., Fredenhagen, K.: {\em A Local (perturbative) construction of observables in gauge theories: The Example of QED}, Comm.\ Math.\ Phys.\  {\bf 203} (1999) 71.}

\bibitem[DG]{derger} Derezi\'nski, J., G\'erard, C.: {\em Mathematics of Quantization and Quantum Fields}, Cambridge Monographs in Mathematical Physics, Cambridge University Press 2013.

\bibitem[Dim]{dimock} Dimock, J.: \textsl{Dirac quantum fields on a manifold}. Tran. Amer. Math. Soc., 269 (1) (1982), 133--147.

\bibitem[Dim2]{dimock2} Dimock, J.: {\em Quantized electromagnetic field on a manifold}, Rev. Math. Phys., 4(02) (1992), 223-233.

\bibitem[DH]{DH} Duistermaat, J.J., H\"{o}rmander, L.:  \textsl{Fourier integral
operators II}, Acta Math. {\bf 128}  (1972), 183--269.

\bibitem[DHK]{SDH} Dappiaggi, C., Hack, T.-P., Sanders, K.: {\em Electromagnetism, local covariance, the Aharonov-
Bohm effect and Gauss' law}, \texttt{arXiv:1211.6420} (2012).

\bibitem[DS]{DS} Dappiaggi C., Siemssen D.: {\em Hadamard States for the Vector Potential on Asymptotically Flat
spacetimes}, Rev. Math. Phys. 25, 1350002 (2013).

\bibitem[FNW]{FNW}Fulling, S.A., Narcowich, F.J., Wald, R.M.: \textsl{Singularity structure of the two-point function in quantum field theory in curved spacetime, II}, Annals of Physics,  {\bf 136} (1981), 243-272.

\bibitem[FP]{FP} Fewster, C.J., Pfenning, M.J.: \textsl{A quantum weak energy inequality for spin-one fields in curved spacetime}, J. Math. Phys., 44, 4480 (2003).

\bibitem[FS]{FS} Finster, F., Strohmaier, A.: \textsl{Gupta-Bleuler quantization of the Maxwell field in globally hyperbolic space-times}, \texttt{arXiv:1307.1632} (2013).

\bibitem[Fur]{furlani0} Furlani, E.P.: \textsl{Quantization of the electromagnetic field on static spacetimes}, J. Math. Phys. 36 (1995), no. 3, 1063-1079.

\bibitem[Fur2]{furlani} Furlani, E.P.: \textsl{Quantization of massive vector fields in curved spacetime}, J. Math. Phys. 40, 2611 (1999).

\bibitem[GW]{GW} G\'erard, C., Wrochna, M.: \textsl{Construction of Hadamard states by pseudo-differential calculus}, Comm. Math. Phys. \textbf{325} (2) (2014), 713-755.

\bibitem[HS]{HS} Hack, T.-P., Schenkel, A.: \textsl{Linear bosonic and fermionic quantum gauge theories on curved spacetimes}, General Relativity and Gravitation,  (2012), 1--34.
\bibitem[Hol]{hollands} Hollands, S.: \textsl{{The Hadamard Condition for Dirac Fields and Adiabatic
  States on Robertson-Walker spacetimes}}, \newblock Comm. Math. Phys. \textbf{ 216} (2001), 635--661. 

\bibitem[Hol2]{hollands2} Hollands, S.: \textsl{Renormalized quantum Yang-Mills fields in curved spacetime}, Rev. Math. Phys., 20(09)  (2008), 1033-1172.
\bibitem[H\"{o}r]{H1} H\"ormander, L.: {\em The analysis of linear partial differential
  operators I. Distribution Theory and Fourier Analysis},  Springer, Berlin Heidelberg New York, 1985.



\bibitem[J]{junker}Junker, W.: {\em Adiabatic Vacua and Hadamard States for Scalar Quantum Fields on Curved Spacetime}, PhD thesis, University of Hamburg 1995. %ArXiv preprint hep-th/9507097v1.

\bibitem[K]{khavkine} Khavkine, I.: {\em Characteristics, conal geometry and causality in locally covariant field theory}, \texttt{arXiv:1211.1914} (2012).

%\bibitem[Mel2]{melrose2} Melrose, R.: \textsl{Lectures on Pseudodifferential operators}, unpublished lecture notes (2006).

%\bibitem[Mel]{melrose} Melrose, R.: \textsl{Introduction to Microlocal Analysis}, unpublished lecture notes (2007).

\bibitem[MM]{MM} Marathe, K.B., Martucci, G: \textsl{Mathematical foundations of gauge theories}, Studies in Mathematical Physics, 5, North-Holland 1992.

\bibitem[M\"uh]{muehlhoff} M\"uhlhoff, R.: \textsl{Cauchy problem and Green's functions for first order differential operators and algebraic quantization}, J. Math. Phys., 52, 022303 (2011).

\bibitem[M\"ul]{olaf} M\"uller, O.: \textsl{Asymptotic flexibility of globally hyperbolic manifolds}, R. Math. Acad. Sci. Paris 350, no. 7-8 (2012), 421-423.

\bibitem[P]{pfenning} Pfenning, M. J.: \textsl{Quantization of the Maxwell field in curved spacetimes of arbitrary dimension}, Classical and Quantum Gravity, 26(13), 135017 (2009).

\bibitem[Rad]{radzikowski} Radzikowski, M.: {\em Micro-local approach to the Hadamard condition in quantum field theory on curved space-time}, Comm. Math. Phys. {\bf 179} (1996), 529--553.

\bibitem[Rej]{rejzner} Rejzner, K.: {\em Remarks on local gauge invariance in perturbative algebraic quantum field theory}, \texttt{arXiv:1301.7037} (2013).

%\bibitem[RT]{ruzhansky} Ruzhansky, M., Turunen, V.: \textsl{Pseudo-differential Operators and Symmetries: Background Analysis and advanced topics}, Birkhauser (2009)

\bibitem[Seg]{segal} Segal, I.: {\em The Cauchy problem for the Yang-Mills equations}, J. Funct. Anal., 33(2) (1979), 175-194.

\bibitem[Shu]{shubin} Shubin, M.A.: \textsl{Pseudodifferential Operators and Spectral Theory}, Springer 2001.

{ \bibitem[SV]{SV} Sahlmann, H., Verch, R.: {\em Microlocal spectrum condition and Hadamard form for vector-valued quantum fields in curved spacetime}, Rev. Math. Phys., 13(10) (2001), 1203-1246.}

\bibitem[T]{taylor}Taylor, M.: {\em Pseudo-differential Operators and Nonlinear PDE}, Birkh\"{a}user, 1991.

\bibitem[W]{static}Wrochna, M.: {\em Quantum Field Theory in Static External Potentials and Hadamard States}, Ann. Henri Poincar\'e, vol. 13, no. 8 (2012), 1841--1871.

\bibitem[W2]{wrothesis} Wrochna, M.: \textsl{Singularities of two-point functions in Quantum Field Theory}, PhD thesis, University of G\"ottingen 2013.

{ \bibitem[WZ]{WZ} Wrochna, M., Zahn, J.: {\em Classical phase space and Hadamard states in the BRST formalism for gauge field theories on curved spacetime}, \texttt{arXiv:1407.8079} (2014).}

\bibitem[Z]{zahn} Zahn, J.: {\em The renormalized locally covariant Dirac field.} Rev. Math. Phys. 26, 1330012 (2014).
\end{thebibliography}
\end{document}